\documentclass{article}
\usepackage{amsmath,amsthm}

\newtheorem{theorem}{Theorem}[section]
\newtheorem{lemma}[theorem]{Lemma}
\newtheorem{definition}[theorem]{Definition}
\newtheorem{corol}[theorem]{Corollary}

\newenvironment{remark}%
  {\par\medbreak\refstepcounter{theorem}%
    \noindent\textbf{Remark~\thetheorem. }}%
  {\par\medskip}

\newcommand{\vz}[1]{\ensuremath{\mathbb{#1}}}

\newcommand{\R}{{\vz R}}

\newcommand{\C}{{\vz C}}
\newcommand{\monod}{\delta_m}
\newcommand{\bid}{\delta_b}
\newcommand{\mlength}{\nu}
\newcommand{\mrelu}{\varrho}
\newcommand{\mrelv}{\varsigma}
\newcommand{\mreluv}{\mu}
\newcommand{\blength}{\upsilon}
\newcommand{\brel}{\zeta}
\newcommand{\matrixB}{\overset{\maltese}{B}}
\DeclareMathOperator{\supp}{supp}
\def\pref#1{(\ref{#1})}

\long\def\drop#1{}

\let\e\varepsilon
\let\epsilon\varepsilon
\def\HNo{\mathcal H^{N-1}}
\def\zvec{0}

\def\XXint#1#2#3{{\setbox0=\hbox{$#1{#2#3}{\int}$}
     \vcenter{\hbox{$#2#3$}}\kern-.5\wd0}}

\usepackage{a4wide}
\usepackage{a4}
\usepackage{latexsym}
\usepackage{amssymb}
\usepackage{amsfonts}
\usepackage{bm}
\usepackage{epsfig}
\usepackage{bbm}
\usepackage{amsmath}
\usepackage{graphicx}

\usepackage{psfrag}
\usepackage{color}
\usepackage{multicol}
\usepackage{subfig}

\title{Stability of monolayers and bilayers in a copolymer-homopolymer blend model}
\author{Yves van Gennip\footnote{
Dept. of Mathematics and Computer Science,
Technische Universiteit Eindhoven,
PO Box 513, 5600 MB  Eindhoven, The Netherlands
(e-mail: y.v.gennip@tue.nl, m.a.peletier@tue.nl)}, Mark A. Peletier\footnotemark[\value{footnote}]}
\date{\today}

\begin{document}
\maketitle
\begin{abstract}
We study the stability of layered structures in a variational model for diblock copolymer-homopolymer blends with respect to perturbations of their interfaces. The main step consists of calculating the first and second derivatives of a sharp-interface Ohta-Kawasaki energy for straight mono- and bilayers and determining the latter's sign. By developing the interface perturbations in a Fourier series we fully characterise the stability of the structures in terms of the energy parameters. Both for the monolayer and for the bilayer there exist parameter regions where these structures are unstable. For strong repulsive interaction between the monomer types in the diblock copolymer the bilayer is always stable with respect to interface perturbations, irrespective of the domain size. The monolayer is only stable for small domain size.

In the course of our computations we also give a Green's function for the Laplacian on a two-dimensional periodic strip.
\\[2\jot]
\textbf{Keywords:} block copolymers, copolymer-homopolymer blends, pattern formation, variational model, partial localisation, Green's function for Laplacian on a strip\\[2\jot]
\textit{Mathematics Subject Classification (2000):} 49N99, 82D60
\end{abstract}

\section{Introduction}

\subsection{Localised and partially localised patterns}

Localised patterns are observed in a wide variety of systems, including
experimental systems such as the Belusov-Zabotinsky reaction~\cite{VanagEpstein04}, nonlinear optics~\cite{TaranenkoStaliunasWeiss97,StaliunasSanchezMorcillo98}, vertically shaken granular media~\cite{UmbanhowarMeloSwinney96,TsimringAranson97}, and Bose-Einstein condensates~\cite{StreckerPartridgeTruscottHulet02}, and also in idealised systems such as the Swift-Hohenberg equation~\cite{CrossHohenberg93,SakaguchiBrand96,SakaguchiBrand98,TlidiMandelLefever94} or networks of reacting cells~\cite{MooreHorsthemke05}. More recently objects have been observed that are only \emph{partially} localised: structures in two dimensions, for instance,  that are `thin' in one spatial direction and `long' in the other. Such \emph{partially localised patterns} have been observed in Nonlinear Schr\"odinger equations~\cite{DAprile00,BadialeDAprile02,AmbrosettiMalchiodiNi02,AmbrosettiMalchiodiNi03,AmbrosettiMalchiodiNi04}, Gierer-Meinhardt-type systems~\cite{DoelmanVanderPloeg02}, and even in scalar nonlinear elliptic equations~\cite{MalchiodiMontenegro02,MalchiodiMontenegro03,Malchiodi05}. In addition, the membrane that surrounds each living cell, for instance, is such a structure~\cite{Lipowski98,BlomPeletier04,PeletierRoeger08}.

In this paper we study an example of \emph{energy-driven} partial localisation, arising in the study of mixtures of \emph{diblock copolymers} with \emph{homopolymers}. Such mixtures feature two opposing forces: a repelling force between different monomer types favours separation into homogeneous phases, while covalent bonds between some of the repelling monomers impose an upper limit on the separation length. As a result a wide variety of patterns are observed (both in physical and in numerical experiments), ranging from spheres~\cite{KoizumiHasegawaHashimoto94,OhtaNonomura97,UneyamaDoi05,ZhangJinMa05}, cylinders~\cite{KinningWineyThomas88}, dumbbells~\cite{OhtaIto95}, helices~\cite{HashimotoMitsumuraYamaguchiTakenakaMoritaKawakatsuDoi01}, `labyrinths' and
`sponges'~\cite{LoewenhauptSteurerHellmannGallot94,Ito98,OhtaIto95}, `ball-of-thread'~\cite{LoewenhauptSteurerHellmannGallot94}, layered structures~\cite{KinningWineyThomas88,KoizumiHasegawaHashimoto94,OhtaIto95,ZhangJinMa05}, and many more.

Our focus is on \emph{layered patterns}, consisting of two or more parallel layers of roughly uniform thickness. In each layer the composition is dominated by one of the polymer types, and in the separation into layers one can recognise a phase separation phenomenon triggered by the repelling forces between polymer types. In addition to their interest as particular patterns in copolymer-homopolymer blends, such layered structures are examples of energy-driven partial localisation.

The main goal of this article is to understand the (in)stability of such layered structures in this simple model of copolymer-homopolymer blends.

\subsection{Diblock copolymers and blends}
Diblock copolymers are linear polymer molecules that consist of two parts (blocks) called the U-part and the V-part in this paper, with corresponding volume fractions given by the functions~$u$ and~$v$. Each part contains monomers of a single type only, U or V. As described above, the interaction between the two types of monomers is the net result of two opposing influences. On the one hand the U- and V-parts repel each other, leading to a tendency of the U-and V-phases to separate; on the other hand the U-and V-parts are chemically bonded together in a single diblock copolymer molecule, forcing both parts to remain close to each other. As a result of these two types of interaction, the separation between the U-and V-phases is restricted to length scales of the order of the molecule size.

We consider systems that contain, in addition to the diblock copolymers, some species of \emph{homopolymer}, that we call the 0-phase. A homopolymer is made up of a single type of monomers, here named 0. The system therefore contains three phases, and because of an assumption of incompressibility we can use the functions $u$ and $v$ to describe the distributions of the three phases.

In~\cite{ChoksiRen05} the following energy is derived:
\[
\mathcal{F}(u, v) = \left\{ \begin{array}{ll} \displaystyle
 c_0 \int_{S_L} |\nabla (u + v)| + c_u \int_{S_L} |\nabla u| + c_v \int_{S_L} |\nabla v|
  \hspace{0.3cm}+  \|u - v\|_{H^{-1}(S_L)}^2
  & \mbox{ if $(u, v) \in \mathcal{K}$,}\\
\infty &\mbox{ otherwise,} \end{array} \right.
\]
where the coefficients $c_i$ are nonnegative (and not all equal to zero), $S_L$ is a periodic strip $\vz{T}_L \times \R$ (where $\vz{T}_L$ is the one-dimensional torus of length $L$), and the set of admissible functions is given by
\[
\mathcal{K} := \left\{ (u, v) \in \left(\text{BV}\left(S_L\right)\right)^2 :
   u(x), v(x) \in \{0, 1\} \text{ a.e., } uv = 0 \text{ a.e., and } \int_{S_L} u = \int_{S_L} v \; \right\}.
\]
Since unconstrained minimisation will lead to the trivial structure $u\equiv v\equiv 0$, the natural problem to look at here is minimisation under constrained mass, i.e. with the constraint $\int_{S_L} u = \int_{S_L} v = M$ for some $M>0$.

Under the extra restriction $u+v\equiv 1$---no 0-phase---the functional $\mathcal F$ is a well-known sharp-interface model for diblock copolymer melts~\cite{RenWei00,ChoksiRen03}. The sharp-interface character of this model, known in the physics literature as the strong-segregation limit, is recognizable in the fact that the variables $u$ and $v$ are characteristic functions, implying that at each point in space only one phase is present. The underlying diffuse-interface model is well studied~\cite{NishiuraOhnishi95,FifeKowalczyk99,EscherMayer01,Muratov02,RenWei02,ChoksiRen03,RenWei03b,RenWei05,RenWei06a,RenWei06b,Tzoukmanis06,RoegerTonegawa07} because of the interesting pattern formation phenomena it exhibits.

The first three terms of $\mathcal F$ can be recognised as the sharp-interface manifestation of the repelling forces between the U-, V-, and 0-monomers. The last term, the $H^{-1}$-norm, is a remainder of the chemical bond between the U- and V-parts and penalises large-scale separation of the U- and V-phases.

The functional $\mathcal{F}$ resembles the energy functional used to model triblock copolymers, i.e. block copolymers consisting of three chemically bonded parts, \cite{RenWei03c, RenWei03d}. The interface penalisation part is present in that functional as well, and the long range interaction term includes interaction between the third phase (the phase corresponding to the third part of the triblock copolymers) and the other two phases in addition to the interaction between the first two phases present in the functional $\mathcal{F}$ above.

For a more extensive review of the modelling of diblock copolymers and diblock copolymer-homopolymer blends and its study in mathematics, we refer to \cite[Chapter 2]{vanGennip08}.

\subsection{From one-dimensional to two-dimensional structures}\label{sec:1D2D}
A layered structure with perfectly straight layers can be described by functions $u$ and $v$ of one spatial variable. In a companion paper~\cite{vanGennipPeletier07a} (see also~\cite{ChoksiRen05}) we study this one-dimensional case and give a full characterisation of global minimisers.

\begin{figure}[ht]
\subfloat[Monolayer1D][A one-dimensional monolayer]
{
    \psfrag{y}{$x_2$}
    \psfrag{u}{U}
    \psfrag{v}{V}
    \psfrag{0}{$0$}
    \psfrag{1}{$1$}
    \includegraphics[width=0.35\textwidth]{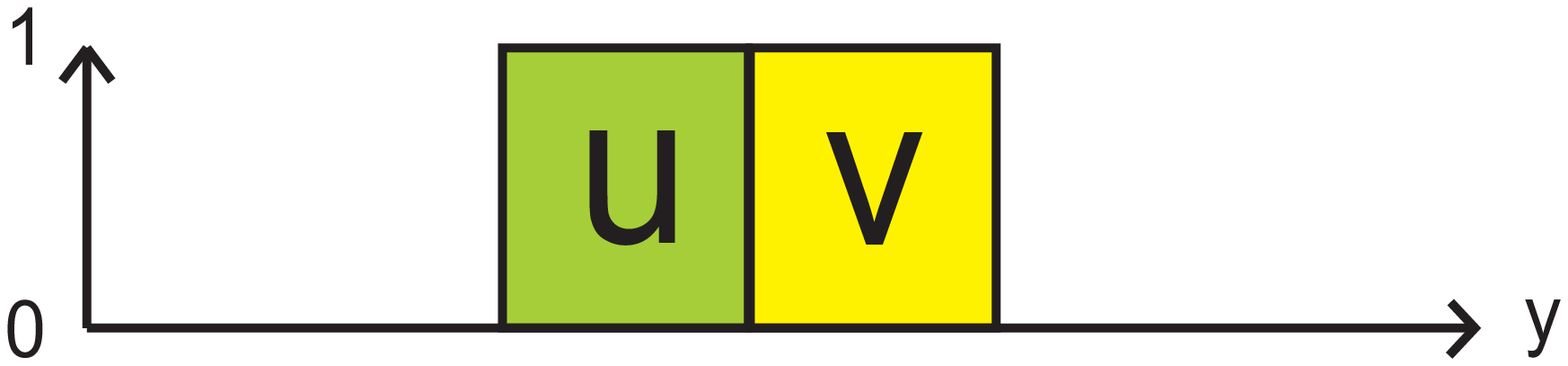}\\
    \label{fig:monolayer1D}
}
\hspace{0.15\textwidth}
\subfloat[Bilayer1D][A one-dimensional bilayer]
{
    \psfrag{y}{$x_2$}
    \psfrag{u}{U}
    \psfrag{v}{V}
    \psfrag{0}{$0$}
    \psfrag{1}{$1$}
    \includegraphics[width=0.35\textwidth]{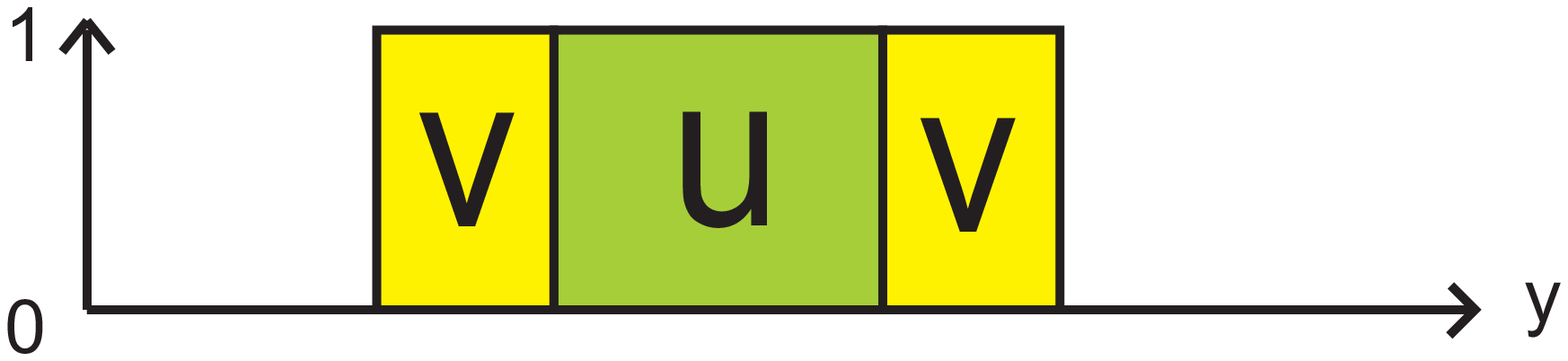}\\
    \label{fig:bilayer1D}
}
\hspace{0.1\textwidth}
\subfloat[Monolayer2D][A straight monolayer on the periodic strip $S_L$]
{
    \psfrag{x}{$x_1$}
    \psfrag{y}{$x_2$}
    \psfrag{u}{U}
    \psfrag{v}{\hspace{0.05cm}V}
    \psfrag{L}{$L$}
    \psfrag{0}{$0$}
    \psfrag{1}{$1$}
    \psfrag{d}{$\delta_m$}
    \includegraphics[width=0.35\textwidth]{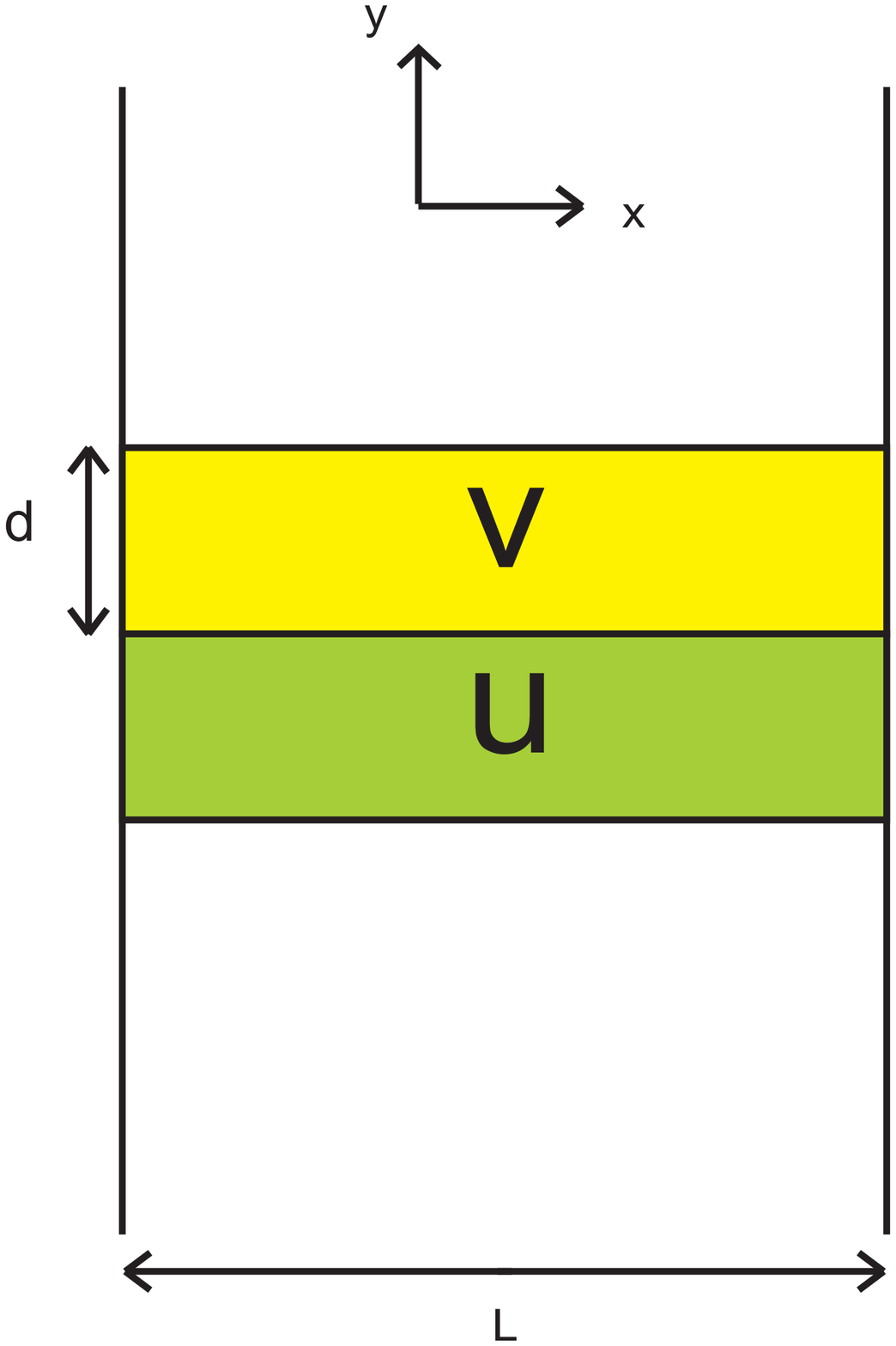}\\
    \label{fig:monolayer2D}
}
\hspace{0.25\textwidth}
\subfloat[Bilayer2D][A straight VUV bilayer on the periodic strip $S_L$]
{
    \psfrag{x}{$x_1$}
    \psfrag{y}{$x_2$}
    \psfrag{u}{U}
    \psfrag{v}{\hspace{0.05cm}V}
    \psfrag{L}{$L$}
    \psfrag{0}{$0$}
    \psfrag{1}{$1$}
    \psfrag{d}{$\delta_b$}
    \includegraphics[width=0.35\textwidth]{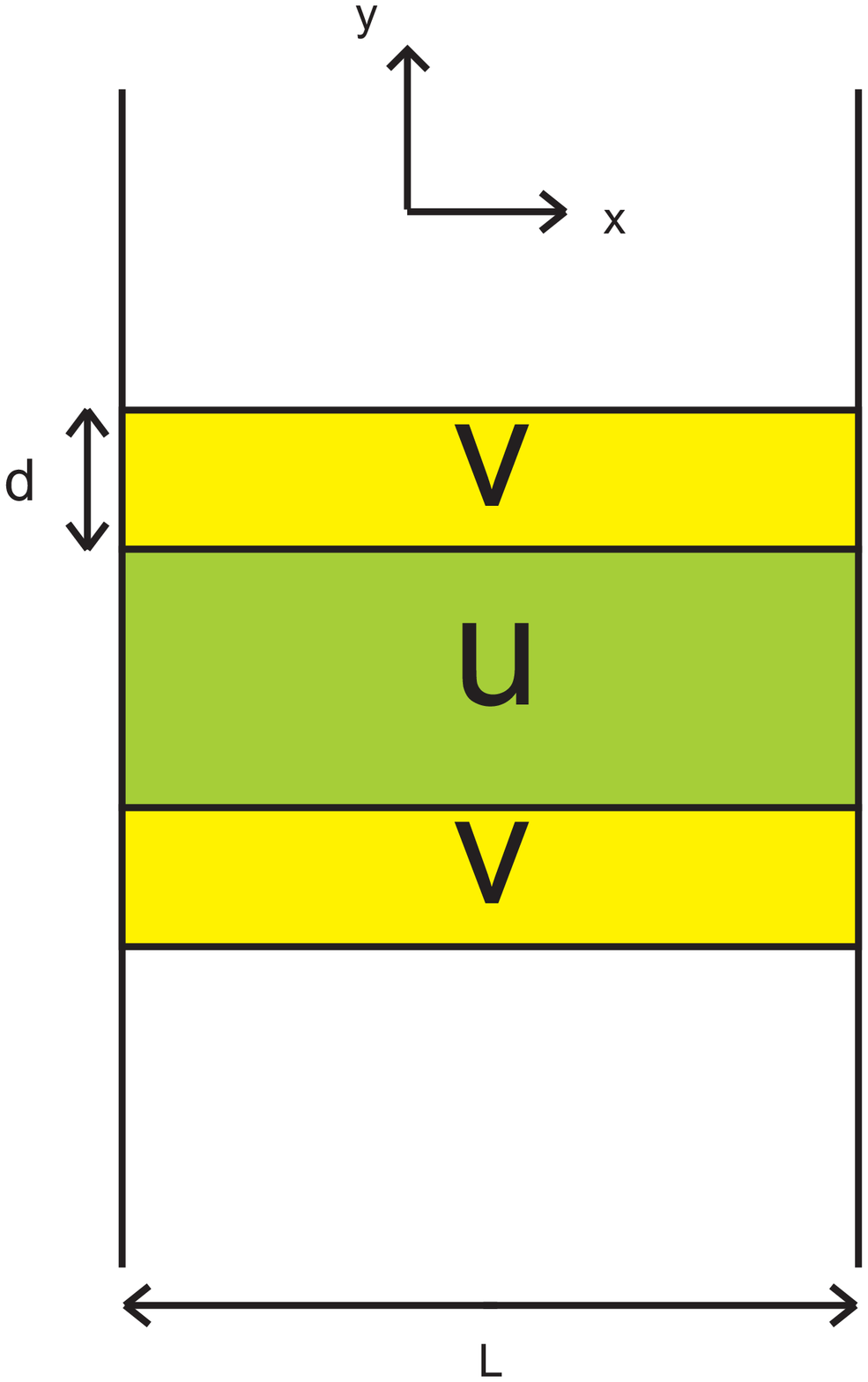}\\
    \label{fig:bilayer2D}
}
\caption{Mono- and bilayers on a strip as trivial extensions of one-dimensional structures. We assume the 0-phase to fill up the rest of the domain where there is no U- or V-phase. We do not indicate this in our pictures.}
\label{fig:2d-multilayers}
\end{figure}

One of the results in that paper is that, for generic parameter values, every constrained-mass global minimiser on $\R$ is a \emph{concatenation of equal-width monolayers}. A monolayer is shown in Figure~\ref{fig:2d-multilayers}: a structure, described by a pair of functions $(u,v)$, in which the supports of $u$ and $v$ are adjacent intervals of equal length---or, in the higher-dimensional context, adjacent layers of equal width (see Figures~\ref{fig:monolayer1D} and~\ref{fig:monolayer2D}).

For small constrained mass, the global minimiser in one dimension is a monolayer. For slightly larger constrained mass, the global minimiser switches to a \emph{bilayer}, a pair of monolayers joined back to back (Figures~\ref{fig:bilayer1D} and~\ref{fig:bilayer2D}). As the constrained mass further increases the global minimiser switches to structures of increasing numbers of monolayers (see~\cite{vanGennipPeletier07a}).

In the present paper we are interested in the stability properties under $\mathcal{F}$ of a particular subset of two-dimensional mono- and bilayer structures:
\begin{itemize}
\item For both mono- and bilayers we assume that the layer thickness is such that the energy-to-mass ratio $\mathcal F/\int u$ is minimal among all such layers;
\item For monolayers we assume that $c_u=c_v$, i.e. that the interface penalisation is the same for U-0 and V-0 interfaces.
\end{itemize}
Both restrictions arise from our interest in thin, partially localised structures in $\R^2$, as is explained in detail in Appendices~\ref{sec:energy-per-unit-mass} and~\ref{sec:cu=cv}.

The optimal widths $\monod$ and $\bid$ for which the energy-to-mass ratio is minimal for the mono- and bilayer respectively are indicated in Figure~\ref{fig:2d-multilayers} and defined in (\ref{eq:monod}) and (\ref{eq:bid}).

\subsection{Stability of mono- and bilayers in two dimensions}

The aim of this paper is to investigate the stability of these mono- and bilayers in two dimensions. Since the functions $u$ and $v$ are forced to be characteristic functions of sets, the only admissible perturbations are changes in the supports of these functions. In this paper we only consider \emph{local} stability with respect to perturbation of the position of the interfaces; other perturbations, such as those that change the topology of the structure are disregarded (see the discussion in Section~\ref{sec:concdisc}).

Specifically, we consider perturbations of the interfaces that are periodic with period $L$ along the length of the layer, and therefore we assume a domain that is periodic in one direction ($x_1$) and unbounded in the other (see Figure~\ref{fig:2d-multilayers}). Because of this periodicity each perturbation of an interface is given by a periodic function $p:\vz{T}_L \to \R^3$ (for the monolayer) or $p:\vz{T}_L \to \R^4$ (for the bilayer), where each component is the lateral displacement of one of the interfaces. By expanding the perturbations in Fourier modes, and using the usual vanishing of cross terms of different frequency, the positivity of the second derivative of the energy reduces to the positivity
on each Fourier mode.%
\footnote{First and second derivatives of similar functionals have been calculated by Muratov and Choksi \& Sternberg~\cite{Muratov02,ChoksiSternberg06}. Our calculations differ in the number of phases (three instead of two) and in the early adoption of a Fourier framework.} 


Fourier modes have a natural scale invariance: the $k^\mathrm{th}$ Fourier mode on the interval of length~$L$ is equivalent to the $1^\mathrm{st}$ Fourier mode on an interval of length $L/k$. This allows us to establish the stability with respect to the first Fourier mode as a function of $L$, rescale for the stability properties of the $k^\mathrm{th}$ mode, and aggregate the results.

Using this approach we show in Section~\ref{sec:stability} that the monolayer of optimal width $\monod$ is linearly stable with respect to mode-$1$ perturbations iff
\[
\frac{c_u}{2c_u+c_0}  \geq f_1(L/\monod),
\]
where $f_1$ is an explicit function given in~\pref{def:f1}.
By combining all Fourier modes we find

\begin{figure}[ht]
\hspace{0.025\textwidth}
\subfloat[Monolayer][The monolayer; plotted are the curves $L/\monod\mapsto f_1(L/(k\monod))$ for $k=1\dots 20$]
{
    \psfrag{a}{$L/\monod$}
    \psfrag{b}{$f_1(L/(k\monod))$}
    \includegraphics[width=0.45\textwidth]{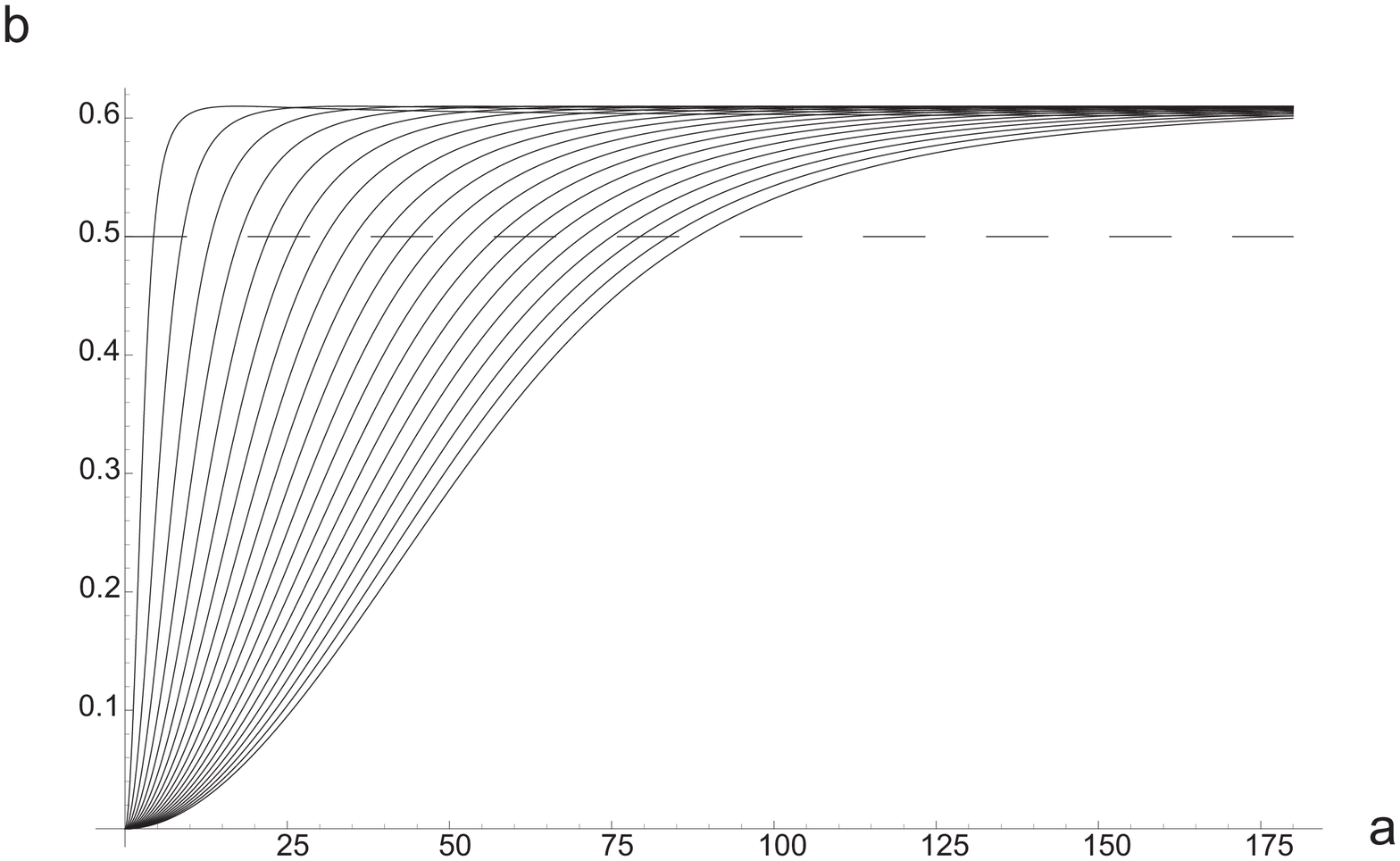}\\
    \label{fig:negeigmonolay}
}
\hspace{0.05\textwidth}
\subfloat[Bilayer][The bilayer; plotted are the curves $L/\bid\mapsto g_1(L/(k\bid))$ for $k=1\dots 20$]
{
    \psfrag{a}{$L/\bid$}
    \psfrag{b}{$g_1(L/(k\bid))$}
    \includegraphics[width=0.45\textwidth]{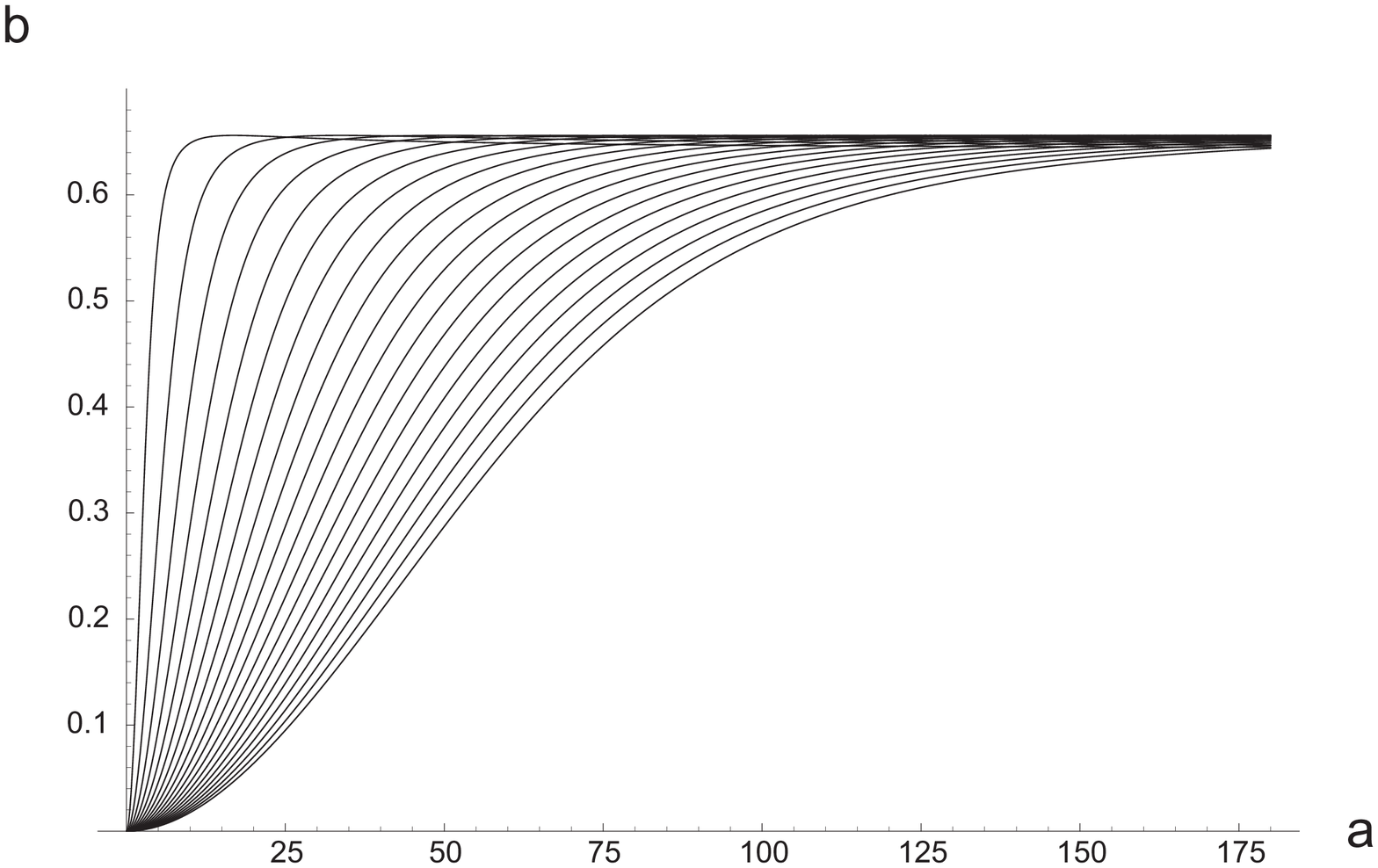}\\
    \label{fig:negeigbilay}
}
\caption{The graphs of the functions 
$x\mapsto f_1(x/k)$ and $x\mapsto g_1(x/k)$ ($k=1, \ldots, 20$) portray the curves in parameter space that separate the parts where the first twenty Fourier modes of the second variation for the monolayer (Figure~\ref{fig:negeigmonolay}) and bilayer (Figure~\ref{fig:negeigbilay}) are positive and negative. If $c_u/(2c_u+c_0) < f_1(L/(k\monod))$ the $k$th Fourier mode is negative for the monolayer, if the reverse inequality holds the mode is positive. Similarly for the bilayer the $k$th Fourier mode is negative if $(c_u+c_v)/(c_0+c_u+2c_v) < g_1(L/(k\bid))$. The leftmost curve in each figure corresponds to the first order Fourier mode, the order increases towards the right. Note that the positivity of the parameters $c_u$ and $c_0$ implies that $c_u/(2c_u+c_0) \leq \frac12$ as indicated in Figure~\ref{fig:negeigmonolay} by the dashed line.}\label{fig:negeig}
\end{figure}

\begin{theorem}
\label{th:stab-monolayer}
Assume $c_u=c_v$. The monolayer of optimal width is linearly stable iff
\begin{equation}
\label{cond:stab-monolayer}
\frac{c_u}{2c_u+c_0} \geq f(L/\monod)
\end{equation}
where
\[
f(\ell) := \sup_{k\geq1}  f_1(\ell/k).
\]
\end{theorem}
\noindent
The graphs of the functions
$x\mapsto f_1(x/k)$ for different values of $k$ are shown in Figure~\ref{fig:negeigmonolay}.

\medskip
For a bilayer of optimal width $\bid$ a similar result holds:

\begin{theorem}
\label{th:stab-bilayer}
The VUV bilayer of optimal width is linearly stable iff
\begin{equation}
\label{cond:stab-bilayer}
\frac{c_u+c_v}{c_0+c_u+2c_v} \geq g(L/\bid)
\end{equation}
where
\[
g(\ell) := \sup_{k\geq1}  g_1(\ell/k)
\]
and $g_1$ is given by~\pref{def:g_1}.
\end{theorem}
\noindent
In Figure~\ref{fig:negeigbilay}
the graphs of the functions $x\mapsto g_1(x/k)$ are shown for different values of $k$.

\begin{figure}[ht]
\hspace{0.025\textwidth}
\subfloat[Monolayerzoom][Blow-up of Figure~\ref{fig:negeigmonolay}]
{
    \psfrag{a}{$L/\monod$}
    \psfrag{b}{$\mreluv$}
    \includegraphics[width=0.45\textwidth]{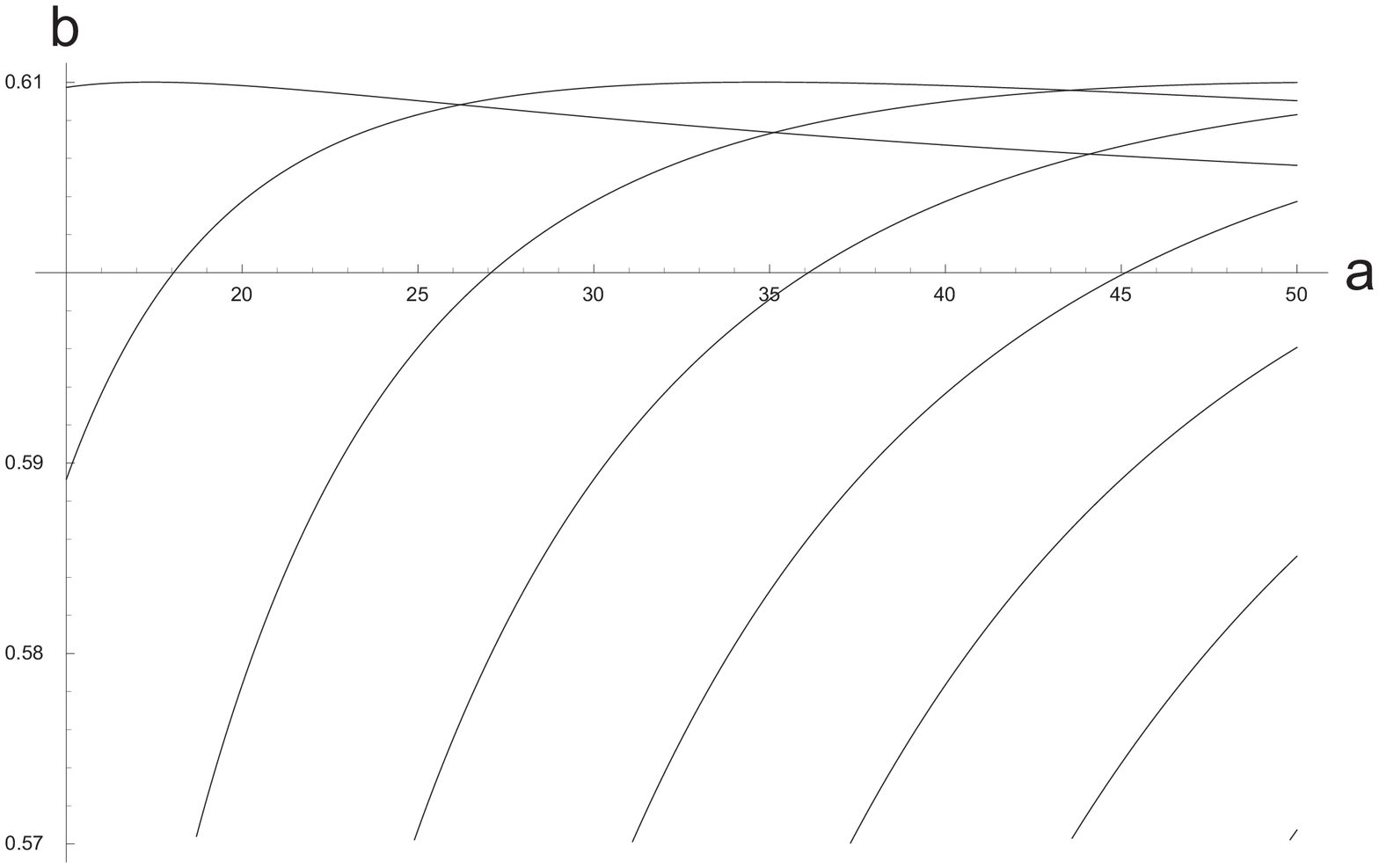}\\
    \label{fig:negeigmonolayzoom}
}
\hspace{0.05\textwidth}
\subfloat[Bilayerzoom][Blow-up of Figure~\ref{fig:negeigbilay} ]
{
    \psfrag{a}{$L/\bid$}
    \psfrag{b}{$\brel$}
    \includegraphics[width=0.45\textwidth]{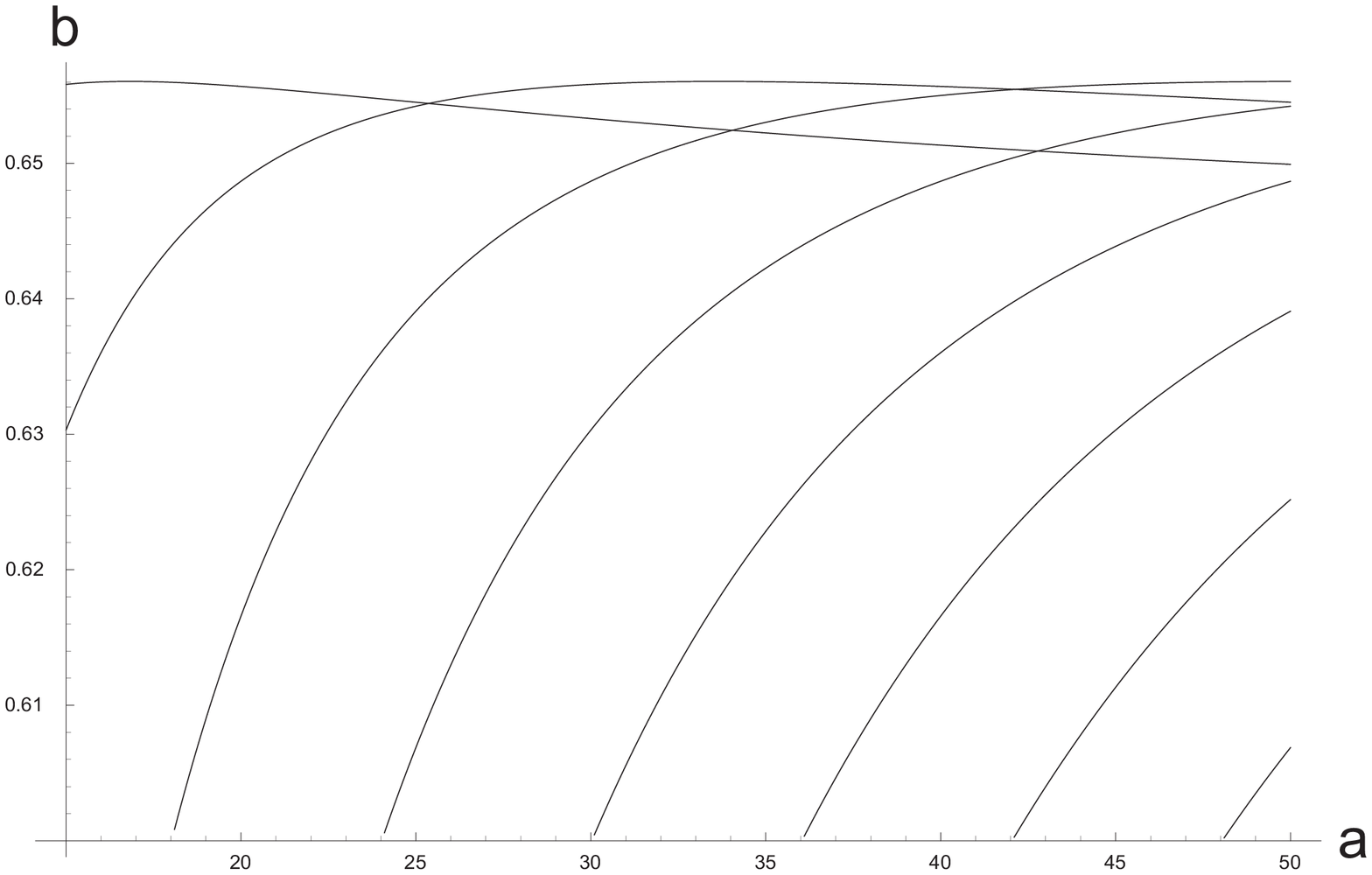}\\
    \label{fig:negeigbilayzoom}
}
\caption{A blow-up of the graphs in Figure~\ref{fig:negeig}. Curves corresponding to different Fourier modes clearly cross. Here $\mreluv=\frac{c_u+c_v}{2 (c_0+c_u+c_v)}$ and $\brel=\frac{c_u+c_v}{c_0+c_u+2c_v}$, see also (\ref{eq:upslamb}) and (\ref{eq:yeahgoaheadanddefinemreluv}).}\label{fig:negeigzoom}
\end{figure}

From Figures~\ref{fig:negeigmonolay} and~\ref{fig:negeigbilay} one might think that curves belonging to higher orders remain below curves of lower orders. The blow-ups in Figure~\ref{fig:negeigzoom} however show that this is not the case. However, it is true that only the first Fourier mode is of importance for determining the stability of the monolayer. This can be recognised by noting that the left-hand side of~\pref{cond:stab-monolayer} cannot reach values larger than $1/2$, and Figures~\ref{fig:negeigmonolay} and~\ref{fig:negeigmonolayzoom} show that the non-monotonicity for the monolayer plays a role only for values above $1/2$.

\bigskip

\begin{figure}[ht]
\hspace{0.1\textwidth}
\subfloat[Monolayer]
{
    \psfrag{a}{$\mlength$}
    \psfrag{b}{$\mreluv$}
    \psfrag{c}{\scriptsize $+$}
    \psfrag{d}{\color{white}$+/-$}
    \includegraphics[width=0.35\textwidth]{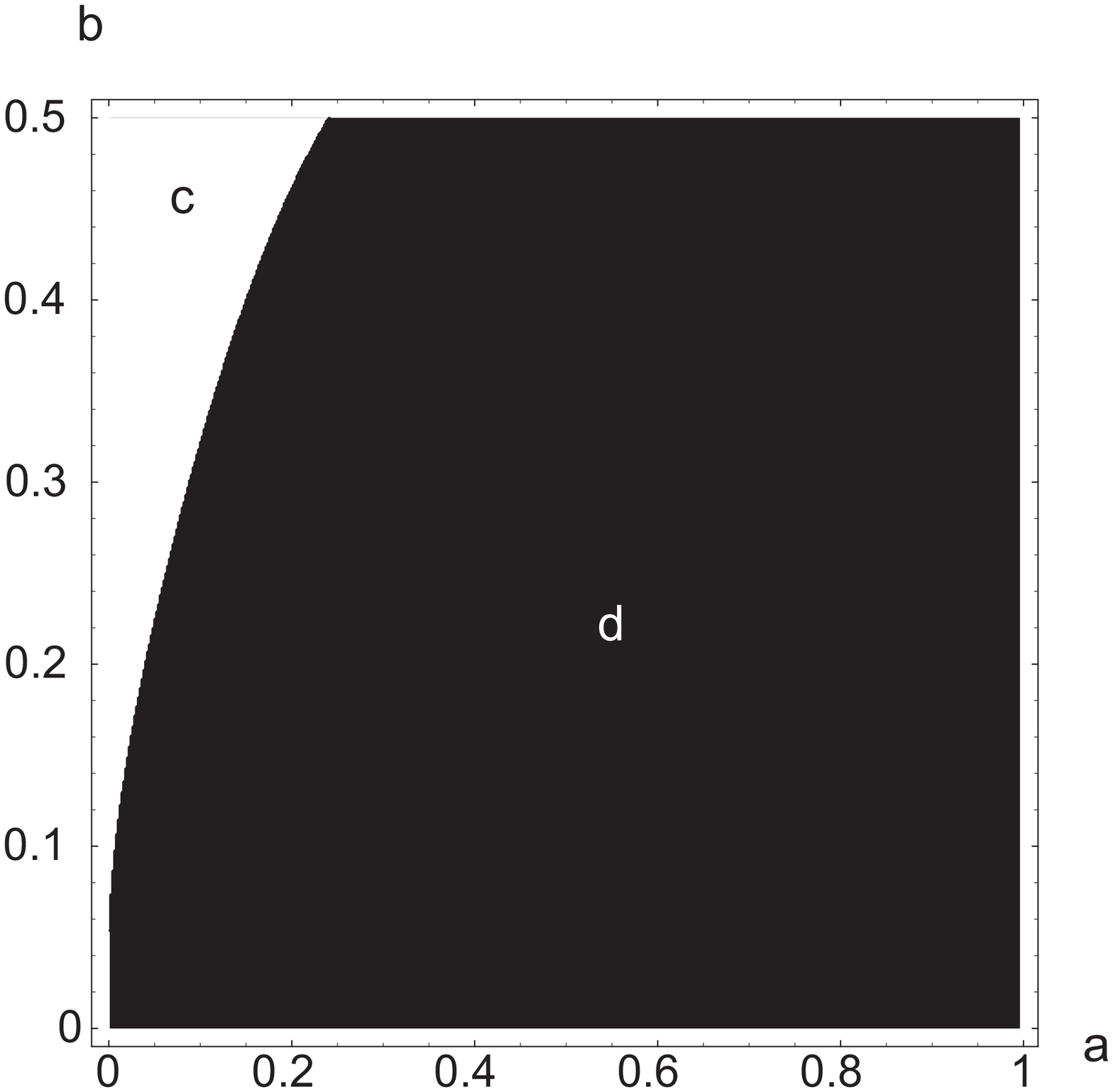}\label{fig:signs-intro-mono}\\

}
\hspace{0.1\textwidth}
\subfloat[Bilayer]
{
    \psfrag{a}{$\blength$}
    \psfrag{b}{$\brel$}
    \psfrag{c}{$+$}
    \psfrag{d}{\color{white} $+/-$}
    \includegraphics[width=0.35\textwidth]{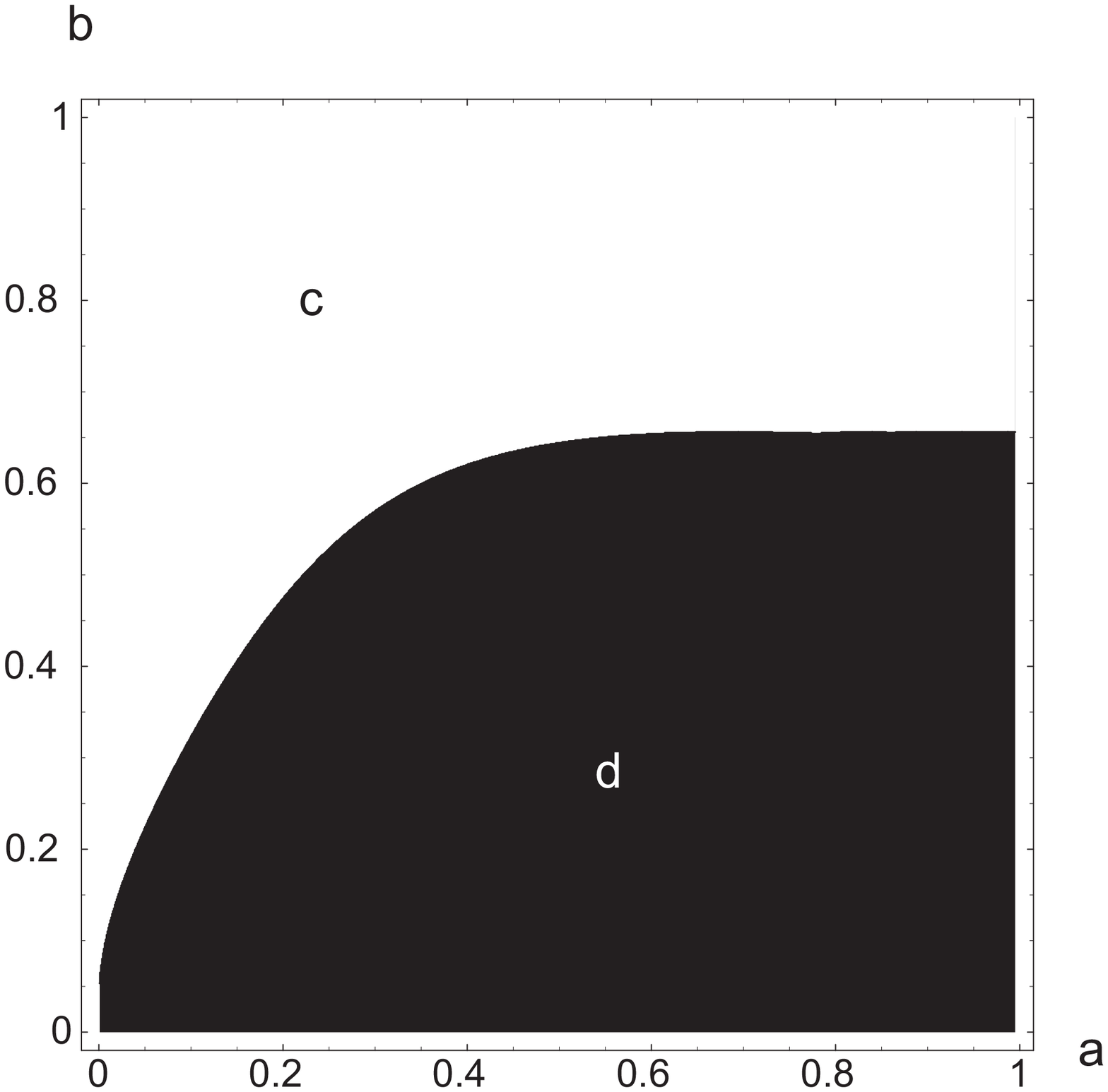}
    \label{fig:signs-intro-bi}\\
}
\caption{The sign of the second derivative operator for the mono- and bilayer of optimal width. $+/-$ indicates indeterminate sign, due to the negativity of one or more eigenvalues. Along the horizontal axes are plotted $\mlength = e^{-{2 \pi \monod}/L}$ and $\blength = e^{-2 \pi \bid/L}$. The vertical axes show $\mreluv=c_u/(2c_u+c_0)$ and $\brel=(c_u+c_v)/(c_0+c_u+2c_v)$. These figures are based on a calculation involving Fourier modes up to and including order 100.}\label{fig:signs-intro}
\end{figure}

Figure~\ref{fig:signs-intro} summarises the stability properties of both the mono- and the bilayer. In Figure~\ref{fig:signs-intro-mono} the vertical axis is restricted to the interval $[0,1/2]$ to reflect the value set of the left-hand side of~\pref{cond:stab-monolayer}.
This implies that monolayers can only be stable if $L$ is sufficiently small, and even then only for a subset of the coefficients $c_0$, $c_u$, and $c_v$; for sufficiently large $L$ the monolayer is unstable  for all choices of interface penalisation.

For the bilayer the situation is different: here the condition~\pref{cond:stab-bilayer} allows for both stability and instability at all values of $L$. The function $g$ is bounded from above (away from 1), implying that a threshold $\alpha$ exists such that
\[
\frac{c_u+c_v}{c_0+c_u+2c_v} \geq \alpha
\qquad \Longrightarrow\qquad
\text{Bilayer is stable for all $L$}.
\]
From Figure~\ref{fig:signs-intro-bi} we estimate that $\alpha \approx 0.65$.

\subsection{Directions of instability}
For the functional $\mathcal{F}$ one may imagine a number of different evolution problems, such as gradient flows based on the $L^2$, $H^{-1}$, or Wasserstein metrics. Under such an evolution the straight mono- and bilayer structures are stationary. If they are unstable, the evolution will amplify small deviations and move away from the straight configurations. While the perturbations are still small, the main contribution of the evolution will be in the directions of the eigenvectors of the second variation\footnote{For each Fourier mode the bilinear form that is the second variation can be identified with a bilinear form on $\R^3$ (monolayer) or $\R^4$ (bilayer) whose eigenvalues and eigenvectors can be studied. Details can be found in Sections~\ref{subsec:bilayerstability}--\ref{subsec:monolayerstability}.} belonging to the (most) negative eigenvalues.

\begin{figure}[ht]
\hspace{0.1\textwidth}
\subfloat[Monolayerunstable2][The (possibly) unstable deformation direction]
{
    \psfrag{u}{U}
    \psfrag{v}{V}
    \includegraphics[width=0.2\textwidth]{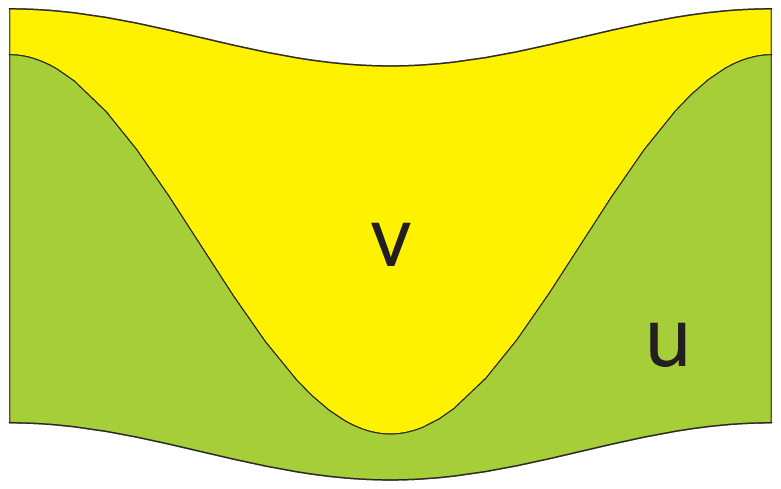}\\
    \label{fig:monounstab2}
}
\hspace{0.1\textwidth}
\subfloat[Monolayerstable1][One of the stable directions]
{
    \psfrag{u}{U}
    \psfrag{v}{V}
    \includegraphics[width=0.2\textwidth]{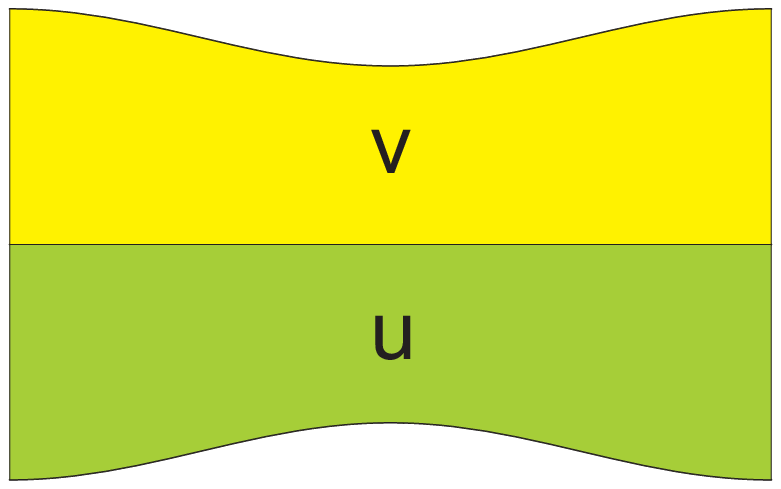}\\
    \label{fig:monostab1}
}
\hspace{0.1\textwidth}
\subfloat[Monolayerstable3][The other stable direction]
{
    \psfrag{u}{U}
    \psfrag{v}{V}
    \includegraphics[width=0.2\textwidth]{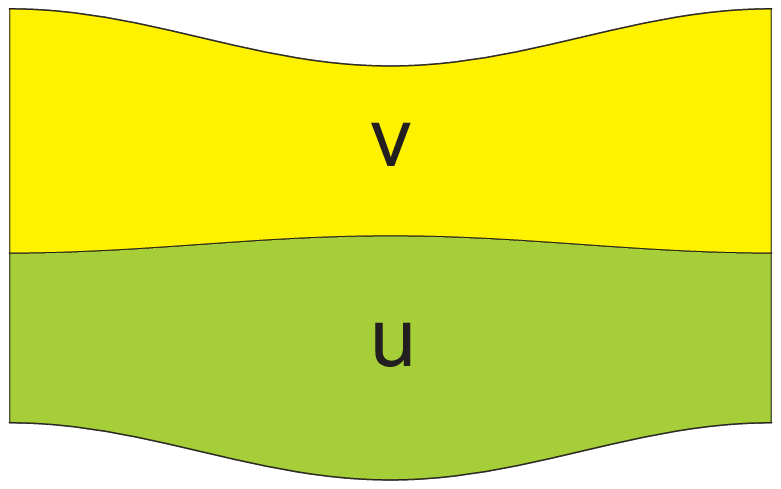}\\
    \label{fig:monostab3}
}
\hspace{0.1\textwidth}
\caption{One (possibly) unstable and two stable first order Fourier modes of deformation for the monolayer; see Section~\ref{sec:stability}, in particular Remark~\ref{rem:stableunstablemodesmonolayer}.}
\end{figure}

For the monolayer there is, for each Fourier mode, one eigenvalue that can become negative (for the first Fourier mode: $E_3$ in Lemma~\ref{lem:M1negeig}; other modes follow by rescaling as above) and there are two which are always positive. Each component of the corresponding eigenvectors is associated with the deformation of one of the interfaces in the layer.
A cartoon of the (possibly) unstable deformation direction is given in Figure~\ref{fig:monounstab2}, the two stable directions are shown in Figures~\ref{fig:monostab1} and~\ref{fig:monostab3}.

\begin{figure}[ht]
\hspace{0.1\textwidth}
\subfloat[SigneigenvaluecorrespondingtoG2][The sign in parameter space of the eigenvalue corresponding to the eigenvalue $G_+$ of the reduced matrix $\tilde B_1$ in the proof of Lemma~\ref{lem:B1negeig}]
{
    \psfrag{x}{$L/\bid$}
    \psfrag{y}{$\brel$}
    \psfrag{a}{$\color{white}-$}
    \psfrag{b}{$+$}
    \includegraphics[width=0.35\textwidth]{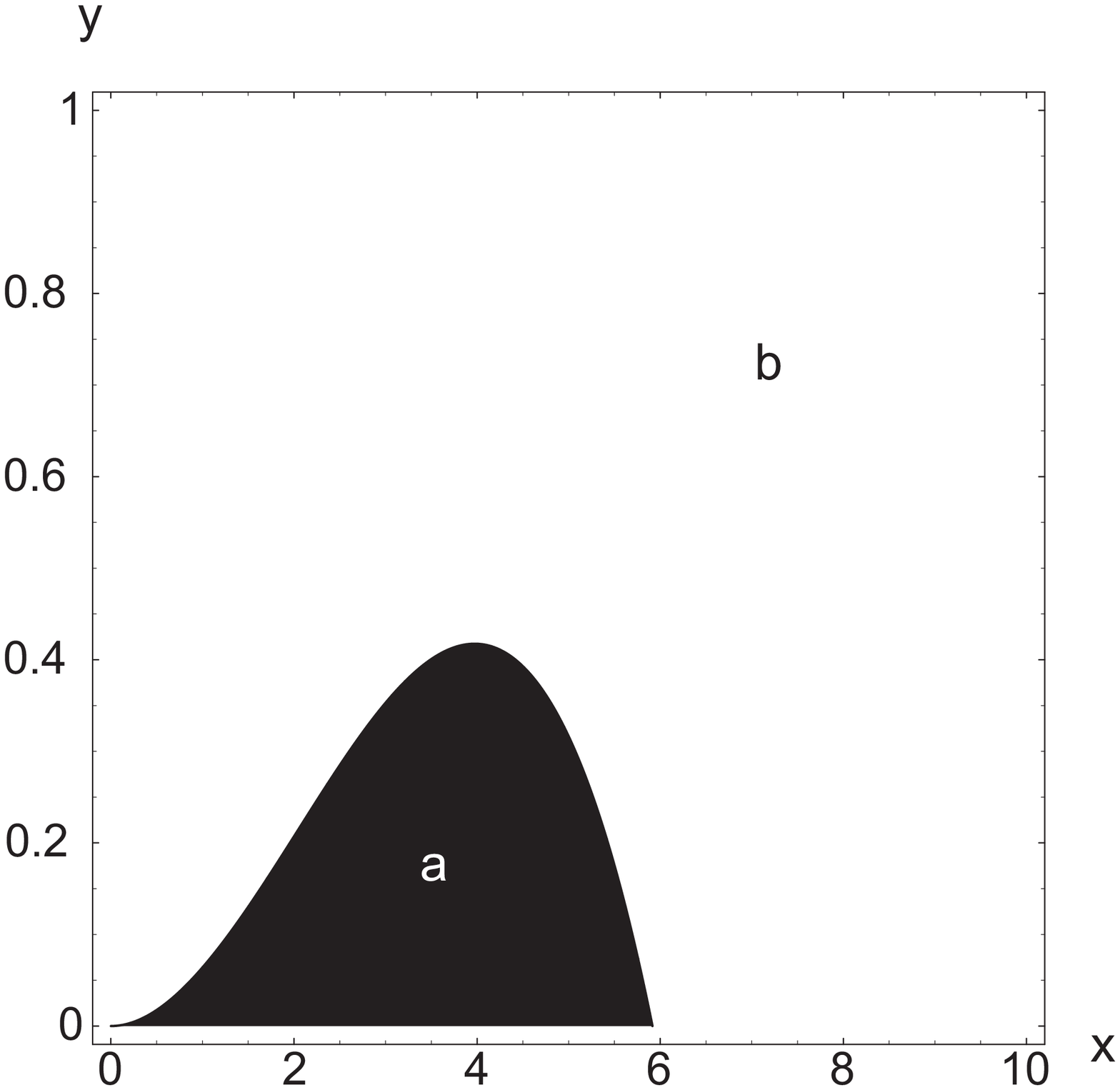}\\
    \label{fig:signeigenvaluecorrespondingtoG2}
}
\hspace{0.1\textwidth}
\subfloat[SigneigenvaluecorrespondingtoG1][The sign in parameter space of the eigenvalue corresponding to the eigenvalue $G_-$ of the reduced matrix $\tilde B_1$ in the proof of Lemma~\ref{lem:B1negeig}]
{
    \psfrag{x}{$L/\bid$}
    \psfrag{y}{$\brel$}
    \psfrag{a}{$\color{white}-$}
    \psfrag{b}{$+$}
    \includegraphics[width=0.35\textwidth]{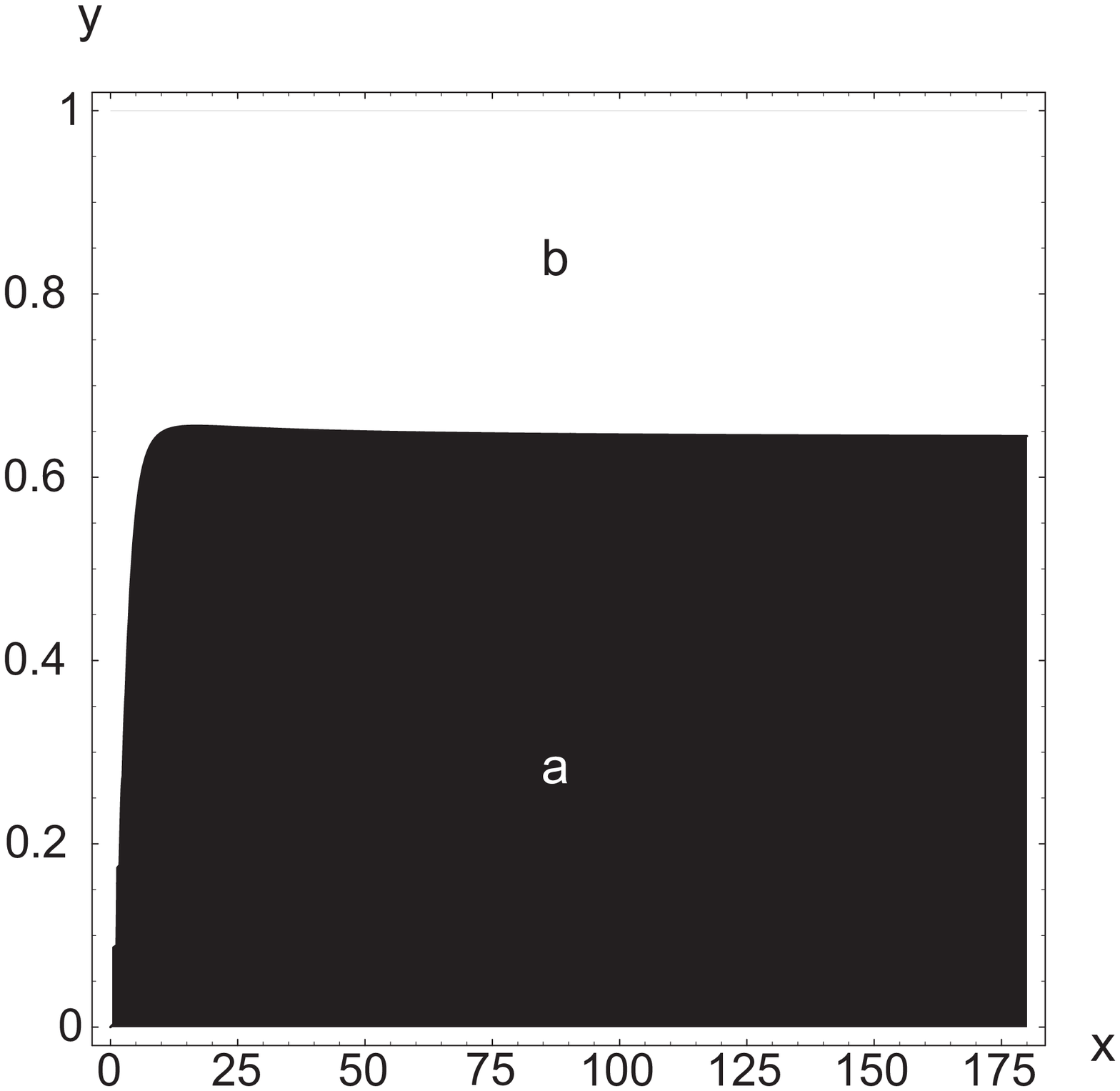}\\
    \label{fig:signeigenvaluecorrespondingtoG1}
}
\caption{The black patches in parameter space indicate where two of the eigenvalues of the first Fourier order second variation operator for the bilayer become negative.}
\end{figure}

For the bilayer two eigenvalues are always positive, and two eigenvalues may also become negative. For the first Fourier mode the dependence of the sign of the latter two on the parameters $L/\bid$ and $\brel$ is given in Figures~\ref{fig:signeigenvaluecorrespondingtoG2} and~\ref{fig:signeigenvaluecorrespondingtoG1}. We recognise in the second figure the first order curve ($k=1$) from Figure~\ref{fig:negeigbilay}; a similar curve for the first figure would always stay below the curve from the latter one, which is why its influence is not recognisable in Figure~\ref{fig:negeigbilay}.

\begin{figure}[h]
\hspace{0.2\textwidth}
\subfloat[Bilayerunstable1][One of the (possibly) unstable deformation directions]
{
    \psfrag{u}{U}
    \psfrag{v}{V}
    \includegraphics[width=0.2\textwidth]{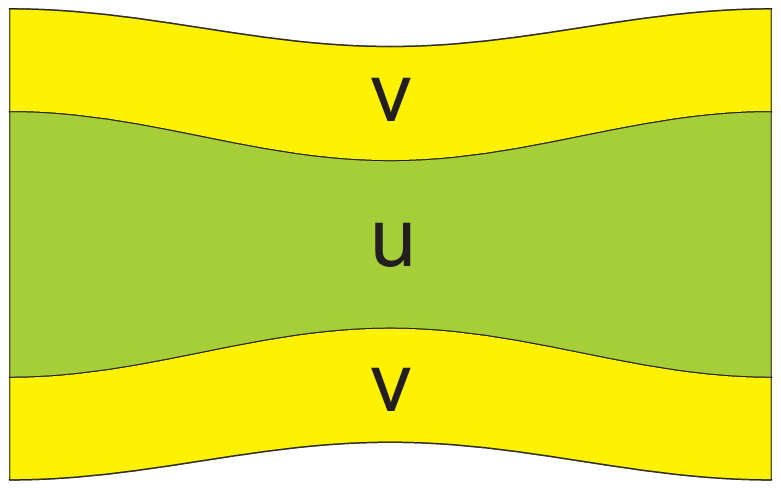}\\
    \label{fig:biunstab1}
}
\hspace{0.2\textwidth}
\subfloat[Bilayerunstable3][The other (possibly) unstable deformation direction]
{
    \psfrag{u}{U}
    \psfrag{v}{V}
    \includegraphics[width=0.2\textwidth]{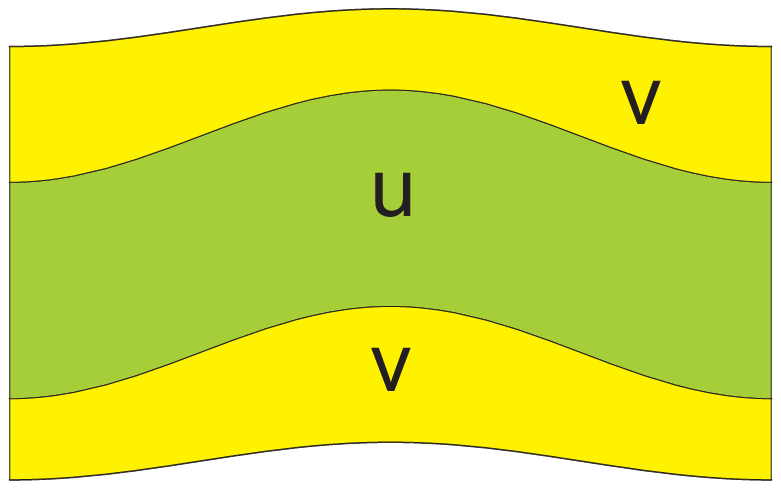}\\
    \label{fig:biunstab3}
}
\\
\vspace{0.5cm}\hspace{0.2\textwidth}
\subfloat[Bilayerstable2][One of the stable directions]
{
    \psfrag{u}{U}
    \psfrag{v}{V}
    \includegraphics[width=0.2\textwidth]{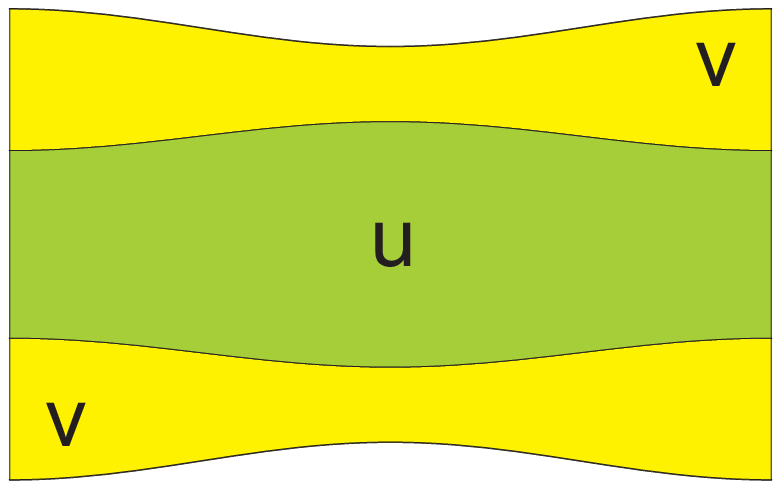}\\
    \label{fig:bistab2}
}
\hspace{0.2\textwidth}
\subfloat[Bilayerstable4][The other stable direction]
{
    \psfrag{u}{U}
    \psfrag{v}{V}
    \includegraphics[width=0.2\textwidth]{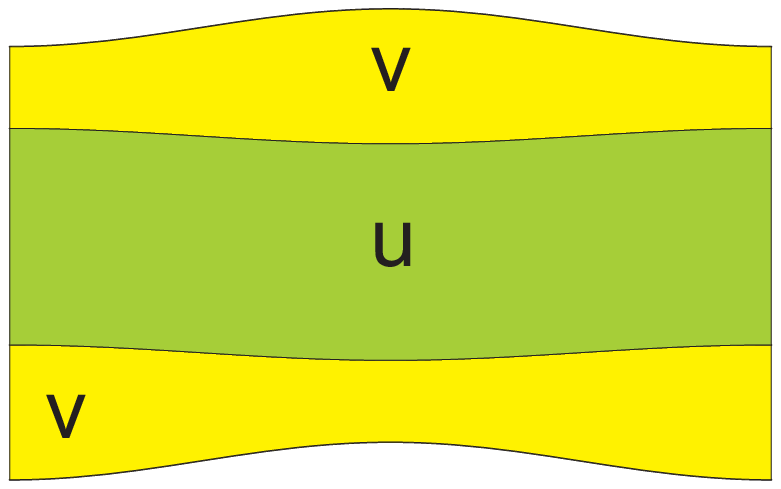}\\
    \label{fig:bistab4}
}

\caption{Two (possibly) unstable and two stable first-order Fourier modes of deformation for the bilayer. The (possibly) unstable deformation in Figure~\ref{fig:biunstab1} corresponds to the eigenvalue in Figure~\ref{fig:signeigenvaluecorrespondingtoG2} and the deformation in Figure~\ref{fig:biunstab3} to the eigenvalue in Figure~\ref{fig:signeigenvaluecorrespondingtoG1}. For details see the discussion in Section~\ref{sec:stability}, in particular Remark~\ref{rem:stableunstablemodesbilayer}.}
\end{figure}

The (possibly) unstable deformation directions are shown in Figures~\ref{fig:biunstab1} and~\ref{fig:biunstab3}, corresponding to the eigenvalues in Figures~\ref{fig:signeigenvaluecorrespondingtoG2} and~\ref{fig:signeigenvaluecorrespondingtoG1}, the stable ones in Figures~\ref{fig:bistab2} and~\ref{fig:bistab4}.

These results all show that depending on the parameters in the model the monolayer and bilayer structures can be unstable. This mirrors closely the results in~\cite{Muratov02}, \cite{RenWei03b}, and \cite{RenWei05}, where it is shown that in the pure diblock case `wriggled' lamellar structures bifurcate off the straight lamellar pattern if the spacing between the lamellae becomes too large. In Section~\ref{sec:wriggledlamellar} we discuss the relation with these results in more detail.

\subsection{Structure of this paper}

We start in Section~\ref{sec:defcon} by defining the functional under consideration and clarifying some of the notation that is used throughout the paper. In Section~\ref{sec:perstrip} we prove via a calculation of the first variation of $\mathcal{F}$ that the monolayer and bilayer are both stationary points of $\mathcal{F}$ with respect to mass preserving perturbations of the interfaces. We then proceed to compute the second variation for both these structures. Since this calculation for the monolayer is similar to that for the bilayer, we only give the details in the latter case and even there we defer most of the computational details to Appendix~\ref{app:proofbilayersecondvar}. Section~\ref{sec:stability} is dedicated to computing the sign of the second variations for the monolayer and bilayer in order to determine the parameter regions of stability and instability. Much of the work in the proofs is again of a calculational nature, some of which we have also moved to the back of the paper in Appendix~\ref{app:details}. Finally Section~\ref{sec:greens} gives a Green's function of $-\Delta$ on the periodic strip. This Green's function is heavily used in this paper and since the authors could not trace a previous appearance of it in the literature a section on its validity closes this paper.

\section{Definitions and conventions}\label{sec:defcon}

\subsection{Problem setting}\label{subsec:problemsetting}

The domain of definition is the strip $S_L := \vz{T}_L \times \vz{R}$, where $\vz{T}_L$ is the one-dimensional torus of length $L$, i.e. the interval $[0, L]$ with the endpoints identified. For functions on $S_L$ the $H^{-1}$-norm is defined by convolution:

\begin{definition}\label{def:H-1norm}
For $f\in L^{\infty}(S_L)$ with compact support satisfying $\int_{S_L} f  = 0$
we define
\[
\|f\|_{H^{-1}(S_L)}^2 := \int_0^L \int_\R f(x_1, x_2) G*f(x_1, x_2)\, dx_2dx_1,
\]
where $G$ is the Green's function of the operator $-\Delta$ on $S_L$, i.e. it satisfies $-\Delta G = \delta$ in the sense of distributions ($\delta$ is the Dirac delta distribution).
\end{definition}
Note that $\phi_f := G*f$ satisfies $-\Delta \phi_f = f$ on $S_L$. Also note that while the Green's function is only unique up to addition of an affine function of $x_2$,  this non-uniqueness is irrelevant for the definition above.

We repeat the definition of $\mathcal{F}$ and $\mathcal{K}$ for convenience.

\begin{definition}\label{def:functional}
Let $c_0$, $c_u$, and $c_v$ be real numbers.
Define
\[
\mathcal{F}(u, v) = \left\{ \begin{array}{ll}
\displaystyle c_0 \int_{S_L} |\nabla (u + v)| + c_u \int_{S_L} |\nabla u|
+ c_v \int_{S_L} |\nabla v| \hspace{0.3cm} + \|u - v\|_{H^{-1}(S_L)}^2
& \mbox{ if $(u, v) \in \mathcal{K}$,}\vspace{0.25cm}\\
\infty &\mbox{ otherwise,} \end{array} \right.
\]
where the admissible set is given by
\[
\mathcal{K} := \left\{ (u, v) \in \left(\text{BV}\left(S_L\right)\right)^2 :
   u(x), v(x) \in \{0, 1\}, \ uv = 0 \text{ a.e., and } \int_{S_L}
u = \int_{S_L} v\right\}.
\]
\end{definition}
We will require that all $c_i$ are non-negative and at least one of them is positive.

Another, equivalent, form of the functional will be useful, in which the
penalisation of the three types of interface U-0, V-0, and U-V, is given explicitly by surface tension
coefficients $d_{kl}$:
\begin{lemma}
\label{lemma:d_ij}
Let the \emph{surface tension coefficients} be given by
\begin{align*}
d_{u0} &:= c_u+c_0,\\
d_{v0} &:= c_v+c_0,\\
d_{uv} &:= c_u+c_v.
\end{align*}

Non-negativity of the $c_i$ is equivalent to the conditions
\footnote{The indices $j, k, l$ take values in $\{u, v, 0\}$ and the $d_{kl}$
are taken symmetric in their indices, i.e. $d_{vu} = d_{uv}$ etc.}
\begin{equation}\label{eq:ddemands}
0 \leq d_{kl} \leq d_{kj} + d_{jl} \qquad\text{for each } k\not=l\not=j\not=k.
\end{equation}

Then
\[
\mathcal{F}(u, v) = \left\{ \begin{array}{ll}
  d_{u0}\HNo(S_{u0}) + d_{v0}\HNo(S_{v0}) + d_{uv}\HNo(S_{uv})
+  \|u - v\|_{H^{-1}(S_L)}^2
& \mbox{ if $(u, v) \in \mathcal{K}$,}\\ \infty &\mbox{ otherwise.} \end{array} \right.
\]
where $S_{kl}$ is the interface between the phases $k$ and $l$:
\begin{align*}
&S_{u0} = \partial^* \supp u \setminus \partial^* \supp v,\\
&S_{v0} = \partial^* \supp v \setminus \partial^*\supp u,\\
&S_{uv} = \partial^* \supp u \cap \partial^* \supp v,
\end{align*}
and $\partial^*$ is the essential boundary of a set.
\end{lemma}
\noindent
The essential boundary of a set consists of all points in the set that have a density other than $0
$ or $1$ in the set; see e.g.~\cite[Chapter 3.5]{AmbrosioFuscoPallara00}.

\begin{proof}[Proof of Lemma~\ref{lemma:d_ij}]
The main step in recognising the equivalence of both forms of $\mathcal{F}$ is noticing
that, for characteristic functions of a set, such as $u, v$ and $u+v$, the equality
\[
\int_\Omega |\nabla u| = \HNo(\partial^* \supp u \cap \Omega)
\]
(see \cite[Theorem 4.4]{Giusti84}, \cite[Theorems 3.59, 3.61]{AmbrosioFuscoPallara00}).
\end{proof}

Note the different interpretations of the coefficients $c_i$ and the surface tension coefficients $d_{kl}$. The latter have a direct physical interpretation (and can be related to material parameters, see \cite{ChoksiRen05}): they determine the mutual repulsion between the different constituents of the diblock copolymer-homopolymer blend. For example, the value of $d_{uv}$ (as compared to the values of $d_{u0}$, $d_{v0}$ and $1$, the coefficient in front of the $H^{-1}$-norm) determines the energy penalty associated with close proximity of U- and V-polymers. In particular, if one of these surface tension coefficients is zero, the corresponding polymers do not repel each other and many interfaces between their respective phases in the model can be expected. On the other hand the coefficients $c_i$, when taken separately, do not convey complete information about the penalisation of the boundary of a phase. If for instance $c_u=0$, but $c_v\neq0$, the part of the U-phase interface that borders on the V-phase still receives a penalty, because $d_{uv}=c_v$. For this reason the use of surface tension coefficients makes more sense from a physical point of view. For the mathematics it is often easier to use the formulation in terms of $c_i$.

If we consider the functional $\mathcal{F}$ on three-dimensional `physical space' we can also see from dimensional considerations that the name ``surface tension coefficients'' for the $d_{kl}$ is justified. Since $\mathcal{H}^2(S_{kl})$ measures surface area, if $\mathcal{F}$ has the dimension of energy then the coefficients $d_{kl}$ have the dimension of energy per unit area, which is also the dimension of surface tension.

\begin{remark}
The conditions (\ref{eq:ddemands}) can be understood in several ways. If, for
instance, $d_{uv}>d_{u0}+d_{v0}$, then the U-V type interface, which is penalised
with a weight of $d_{uv}$, is unstable, for the energy can be reduced by
slightly separating  the U and V regions and creating a thin zone of 0
inbetween. A different way of seeing the necessity of~\pref{eq:ddemands} is by
remarking that the equivalent requirement of non-negativity of the $c_i$ is
necessary for $\mathcal{F}$ to be lower semicontinuous in e.g. the $L^1$ topology (see e.g. \cite[Theorem 1.9]{Giusti84}). Our assumption that at least one $c_i$ is positive is equivalent to requiring at least two $d_{kl}$ to be positive.
\end{remark}

\subsection{Fourier transformation}\label{subsec:fourtfm}

To clarify the notation we use, we will explicitly define the Fourier series we are using. For future reference we will also state some results we will need.

\begin{definition}\label{def:fourser}
Let $f \in L^2\left(\vz{T}_L\right)$, then we will denote by $\hat f \in L^2(\vz{Z}; \C)$, the \emph{Fourier transform of~$f$}:
\[
\hat f(k) := \frac{1}{\sqrt{L}} \int_0^L f(x) e^{-2 \pi i x k / L} \, dx,
\]
and by $a_j$ and $b_j$, $j \in \vz{N}$, the Fourier coefficients of $f$ with respect to the normalised basis of cosines and sines:
\begin{align*}
a_0 &:= \frac1{\sqrt{L}} \int_0^L f(x)\, dx,\\
a_j &:= \sqrt{\frac2L} \int_0^L f(x) \cos\left(\frac{2 \pi x j}{L}\right) \, dx,\\
b_j &:= \sqrt{\frac2L} \int_0^L f(x) \sin\left(\frac{2 \pi x j}{L}\right) \, dx,
\end{align*}
\end{definition}

For easy reference we give here the relations between $\hat f(j)$ and $a_j, b_j$: $\hat f(0) = a_0$ and, for $j \geq 1$, $\hat f(j) = \frac{1}{\sqrt{2}} (a_j - i b_j), \hat f(-j) = \frac{1}{\sqrt{2}} (a_j + i b_j), a_j = \frac{1}{\sqrt{2}} \left( \hat f(j) + \hat f(-j) \right)$ and $b_j = \frac{i}{\sqrt{2}} \left(\hat f(j) - \hat f(-j)\right)$.

Furthermore we have
\begin{align*}
f(x) &= \frac{a_0}{\sqrt{L}} + \sqrt{\frac2L} \sum_{j=1}^{\infty} a_j \cos\left(2 \pi x j / L \right) + \sqrt{\frac2L} \sum_{j=1}^{\infty} b_j \sin\left(2 \pi x j /L\right),\\
f(x) &= \frac{1}{\sqrt{L}} \sum_{q \in \vz{Z}} \hat f(q) e^{2 \pi i x q / L},
\end{align*}
where the convergence is in the $L^2$ topology.
Finally Parseval's theorem takes the form
\begin{align}
\int_0^L f(x) g(x) \, dx &= 
\hat f(0) \hat g(0) + 2 \text{Re}\, \sum_{q=1}^{\infty} \hat f(q) \overline{\hat g(q)}\nonumber\\
&= 
a_{f,0} a_{g,0} + \sum_{j=1}^{\infty} \bigl[a_{f,j} a_{g,j} + b_{f,j} b_{g,j}\bigr],\label{eq:Parseval}
\end{align}
and as a consequence we have for $p_1, p_2, p_3 \in L^2(\vz{T}_L)$
\begin{equation}\label{eq:parsconv}
\int_{\vz{T}_L} \int_{\vz{T}_L} p_1(x) p_2(x - y) p_3(y) \, dx  dy = L^{1/2} \sum_{q \in \vz{Z}} \hat p_1(q) \overline{\hat p_2(q)} \overline{\hat p_3(q)}.
\end{equation}

\section{Geometrical derivatives of the energy}
\label{sec:perstrip}

In the following two sections we will take a look at the stability of two-dimensional periodic monolayer and bilayer configurations. First we need to determine under which conditions these structures are stationary points of the functional $\mathcal{F}$. For the bilayer this will be done in Section~\ref{subsec:bilaystat}, after which we compute the second variation for a bilayer in Section~\ref{subsec:secvarbilay}. We will give analogous results for the monolayer in Section~\ref{subsec:varmonolay}. In Section~\ref{sec:stability} we will use these results to derive the explicit stability criteria of Theorems~\ref{th:stab-monolayer} and~\ref{th:stab-bilayer}.

Of the two possible bilayer structures---UVU and VUV---we only discuss the VUV structure. The results for the UVU structure follow from exchanging the roles of $u$ and $v$.

\subsection{Bilayer: admissible perturbations and stationarity}\label{subsec:bilaystat}

The VUV bilayer of optimal width is a structure given by functions $(u_0,v_0)\in\mathcal{K}$ with
\begin{equation}
\label{def:bilayer-optimal}
u_0 := \chi^{}_{\vz{T}_L\times[-\bid,\bid]}
\qquad\text{and}\qquad
v_0 := \chi^{}_{\vz{T}_L\times([-2\bid,-\bid]\cup[\bid,2\bid])},
\end{equation}
where~\cite{vanGennipPeletier07a}
\begin{equation}\label{eq:bid}
\bid := \sqrt[3]{\frac34 (c_0+c_u+2c_v)},
\end{equation}
and $\chi_A$ is the characteristic function of the set $A$. The set of admissible boundary perturbations of this structure is only restricted by regularity and the equal-mass constraint:
\begin{definition}
The set of admissible perturbations is characterised by
\begin{equation}
\label{def:Pb}
\mathcal{P}_b := \left\{p \in \left(W^{1,2}(\vz{T}_L)\right)^4:
  2\int_{\vz{T}_L}(p_1+p_3) = \int_{\vz{T}_L} (p_2+p_4)\right\}.
\end{equation}
For $p\in \mathcal{P}_b$ and $\e>0$ we define a perturbed structure  $(u_\e,v_\e)$,
\begin{align*}
u_\e(x_1, x_2) &= \left\{ \begin{array}{ll} 1 & \text{if } x_2 \in \bigl(-\bid-\e p_3(x_1), \bid + \e p_1(x_1)\bigr),\\ 0 & \text{otherwise,} \end{array} \right.\\
v_\e(x_1, x_2) &= \left\{ \begin{array}{ll} 1 & \text{if } x_2 \in \bigl(-2\bid -\e p_4(x_1), -\bid - \e p_3(x_1)\bigr) \cup \bigl(\bid+\e p_1(x_1), 2\bid+\e p_2(x_1)\bigr),\\ 0 & \text{otherwise.} \end{array} \right.
\end{align*}
We also introduce the subset of perturbations that conserve mass:
\begin{equation}
\label{ass:mass_conservation}
\mathcal{P}_b^M := \left\{p \in \mathcal{P}_b:
  \int_{\vz{T}_L}(p_1+p_3) = \int_{\vz{T}_L}(p_2+p_4) = 0 \right\}
\end{equation}
\end{definition}
\noindent
Note that since $W^{1,2}(\vz{T}_L)$ is imbedded in $L^\infty(\vz{T}_L)$ by the Sobolev imbedding theorem, the pair $(u_\e,v_\e)$ belongs to $\mathcal{K}$ for sufficiently small $\e$.

\begin{figure}[ht]
\centering
{
    \psfrag{1}{$\bid + \epsilon p_1(x_1)$}
    \psfrag{2}{$2\bid + \epsilon p_2(x_1)$}
    \psfrag{3}{$-\bid - \epsilon p_3(x_1)$}
    \psfrag{4}{$-2\bid - \epsilon p_4(x_1)$}
    \psfrag{x}{$x_1$}
    \psfrag{y}{$x_2$}
    \psfrag{u}{U}
    \psfrag{v}{V}
    \includegraphics[width=0.35\textwidth]{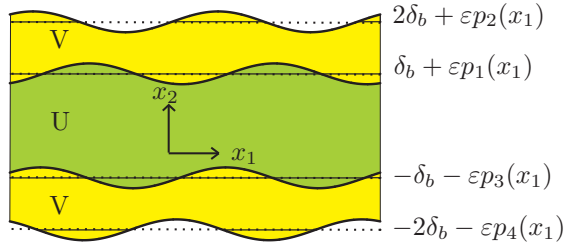}
}
\caption{The bilayer of optimal width with perturbations}
\label{fig:perturbationsbilayer}
\end{figure}

A picture of a bilayer of optimal width with perturbations $p$ is shown in Figure~\ref{fig:perturbationsbilayer}.

\begin{remark}\label{rem:differentmassconstraints}
We should stress the difference between the two mass constraints~\pref{def:Pb} and~\pref{ass:mass_conservation}. The constraint~\pref{def:Pb} is equivalent to the condition that $u_\e$ and $v_\e$ have the same mass. 
This property is a basic element of the model of block copolymers, via the set of admissible functions $\mathcal{K}$.

The additional condition~\pref{ass:mass_conservation} expresses the requirement that $\int u_\e$ and $\int v_\e$ both equal the mass $\int u_0$ of the unperturbed bilayer; perturbations without this property are meaningful in a situation where the joint mass of $u_\e$ and $v_\e$ may change. The bilayer of optimal width is a stationary point of the functional  $\mathcal{F}$ under mass-preserving changes (see Lemma~\ref{lemma:bilayer_is_stationary} below); but as can be inferred from equation~\pref{eq:bilayer_deriv_HMO}, the functional is \emph{not} stationary under perturbations that do change the mass.
\end{remark}

\begin{definition}\label{def:stationarity}
We say that the bilayer of optimal width is stationary with respect to the admissible perturbations $\mathcal{P}_b$ (or $\mathcal{P}_b^M$) if, for all $p \in \mathcal{P}_b$ (or all $p \in \mathcal{P}_b^M$),
\[
\left.\frac{d}{d \epsilon} \mathcal{F}(u_{\epsilon}, v_{\epsilon})\right|_{\epsilon=0} = 0.
\]
\end{definition}

\begin{lemma}
\label{lemma:bilayer_is_stationary}
The VUV bilayer of optimal width is stationary with respect to all $p\in \mathcal{P}_b^M$. \end{lemma}
\begin{proof}
Choksi and Sternberg calculate the first and second variations of a related functional~\cite{ChoksiSternberg06}, and their method can be adapted without much difficulty to the functional $\mathcal{F}$. Here we give a self-contained proof.

Since the interfaces of the bilayer are straight, the derivative of the interfacial terms with respect to the perturbation is zero for all $p\in \mathcal{P}_b$:
\begin{align}
&\hspace{.4cm}\left. \frac{d}{d \epsilon} \left(c_0 \int_{S_L} |\nabla (u_\epsilon + v_\epsilon)| + c_u \int_{S_L} |\nabla u_\epsilon| + c_v \int_{S_L} |\nabla v_\epsilon| \right)\right|_{\epsilon = 0}  \notag \\
&=\left. \frac{d}{d \epsilon} \left[d_{uv} \int_0^L \left( \sqrt{1 + \epsilon^2 {p_1'}^2} + \sqrt{1 + \epsilon^2 {p_3'}^2} \right) \, dx
+ d_{v0} \int_0^L \left( \sqrt{1 + \epsilon^2 {p_2'}^2} + \sqrt{1 + \epsilon^2 {p_4'}^2} \right) \, dx\right]\right|_{\e=0}\notag\\
&= 0.
\label{eq:inteps}
\end{align}
For the derivative of the $H^{-1}$-norm, let $\eta \in C(\R)$ and compute
\begin{align}
\frac{d}{d \epsilon} \int_{S_L} \eta(x_2) u_{\epsilon}(x) \, dx\Bigr|_{\e=0}
 &= \int_0^L \frac{d}{d \epsilon} \int_{-\bid - \epsilon p_3(x_1)}^{\bid + \epsilon p_1(x_1)} \eta(x_2) \, dx_2 \Bigr|_{\e=0} dx_1 \nonumber\\
&= \int_0^L \Bigl( p_1(x_1)\, \eta(\bid + \epsilon p_1(x_1)) + p_3(x_1)\, \eta(-\bid - \epsilon p_3(x_1)) \Bigr) \, dx_1\Bigr|_{\e=0} \notag\\
&= \eta(\bid)\int_0^L p_1 + \eta(-\bid)\int_0^L p_3.
\label{eq:weakudir}
\end{align}
Similarly,
\begin{align}
\frac{d}{d \epsilon} \int_{S_L} \eta(x_2) v_{\epsilon}(x) \, dx\Bigr|_{\e=0}
&= - \eta(\bid)\int_0^L p_1 + \eta(2\bid)\int_0^L p_2
        - \eta(-\bid)\int_0^L p_3 + \eta(-2\bid)\int_0^L p_4.
\label{eq:weakvdir}
\end{align}
Let $G$ be the Green's function of $-\Delta$ on $S_L$ from Theorem~\ref{thm:gf}, then
\begin{align*}
\left. \frac{d}{d \epsilon} \|u_{\epsilon} - v_{\epsilon}\|_{H^{-1}(S_L)}^2\right\vert_{\epsilon=0}
&= \left.  \frac{d}{d \epsilon} \int_{S_L}
  |\nabla G\ast (u_\e-v_\e) |^2\, dx\,\right|_{\e=0}\\
&= 2\int_{S_L} \nabla G\ast (u_0 - v_0)
   \left[ \frac{d}{d \epsilon} \nabla G \ast (u_\e - v_\e)\right]_{\epsilon=0}
     dx\\
&= 2\left. \frac{d}{d \epsilon}\int_{S_L} \nabla G\ast (u_0 - v_0) \cdot
    \nabla G \ast (u_\e - v_\e)\, dx \right|_{\epsilon=0}
    \\
&=  2\left. \frac{d}{d \epsilon}\int_{S_L} \bigl[G\ast (u_0 - v_0)\bigr]
    (u_\e - v_\e)\, dx \right|_{\epsilon=0}
\end{align*}
Setting $\eta(x_2) := [G\ast (u_0-v_0)](x_1,x_2)$ (which is independent of $x_1$, because $u_0-v_0$ is independent of $x_1$) we calculate by the Fourier series~\pref{eq:greenfour} (or by remarking that this is a one-dimensional situation) that
\[
\eta(x_2) = -\frac12 \int_{\vz{R}} |x_2 - y| (u_0-v_0)(0, y) \,dy,
\]
from which it follows that $\eta(\bid) = \eta(-\bid)$ and $\eta(\pm2\bid)=0$. Therefore $\eta\in C(\R)$ and thus we obtain from~(\ref{eq:weakudir}) and (\ref{eq:weakvdir}) that
\begin{equation}
\label{eq:bilayer_deriv_HMO}
\left. \frac{d}{d \epsilon} \|u_{\epsilon} - v_{\epsilon}\|_{H^{-1}(S_L)}^2
  \right|_{\e=0}
 = 4L\eta(\bid)\int (p_1+p_3)
 \stackrel{\pref{ass:mass_conservation}}= 0.
\end{equation}
\end{proof}

\begin{remark}\label{rem:whyoptimalwidth}
Note that in Lemma~\ref{lemma:bilayer_is_stationary} we nowhere use the specific definition of $\bid$ in (\ref{eq:bid}). As we explain in Appendix~\ref{sec:energy-per-unit-mass} the optimal width is relevant when considering energy per unit mass. If we define the mass functional as
\[
\mathcal{M}(u,v) := \int_{S_L} u
\]
for $(u,v)\in \mathcal{K}$, then the following calculations show that the bilayer of optimal width is a stationary point of $\mathcal{F}/\mathcal{M}$ with respect to all perturbations in $\mathcal{P}_b$ (so not only the mass preserving ones) in the sense of Defintion~\ref{def:stationarity} (with the functional $\mathcal{F}$ replaced by $\mathcal{F}/\mathcal{M}$). Thus, let now $p\in \mathcal{P}_b$.

We first compute that $\eta(\bid) = \frac12 \bid^2$, where $\eta$ is such as chosen at the end of the proof of Lemma~\ref{lemma:bilayer_is_stationary}. Then with the help of (\ref{eq:bilayer_deriv_HMO}) and the computations for the one-dimensional case in \cite{vanGennipPeletier07a} we find that
\begin{align*}
\left. \frac{d}{d\e} \mathcal{F}(u_\e, v_\e)\right|_{\e=0} &= 2 \bid^2 \int_0^L (p_1+p_3),\\
\left. \frac{d}{d\e} \mathcal{M}(u_\e, v_\e)\right|_{\e=0} &= \int_0^L (p_1+p_3),\\
\mathcal{F}(u_0,v_0) &= 2 L (c_0+c_u+2c_v) + \frac43 L \bid^3,\\
\mathcal{M}(u_0,v_0) &= 2 L \bid.
\end{align*}
Now we conclude
\begin{align*}
\left.\frac{d}{d\e} \mathcal{F}/\mathcal{M}(u_\e, v_\e)\right|_{\e=0} &= \mathcal{M}(u_0,v_0)^{-2} \Biggl( \mathcal{M}(u_0,v_0)\left. \frac{d}{d\e} \mathcal{F}(u_\e, v_\e)\right|_{\e=0}+\\&\hspace{2.8cm} - \mathcal{F}(u_0,v_0)\left. \frac{d}{d\e} \mathcal{M}(u_\e, v_\e)\right|_{\e=0} \Biggr)\\
&= \frac12 L^{-1} \bid^{-2} \int_0^L (p_1+p_3) \Biggl( \frac43 \bid^3 - (c_0+c_u+2c_v) \Biggr)\\
&\hspace{-0.145cm}\stackrel{\pref{eq:bid}}= 0.
\end{align*}
We see that the optimal width condition \pref{eq:bid} is not necessary for stationarity under $\mathcal{F}/\mathcal{M}$ with respect to the mass preserving perturbations in $\mathcal{P}_b^M$, but it is for stationarity with respect to perturbations in $\mathcal{P}_b\setminus\mathcal{P}_b^M$.
\end{remark}

\subsection{Second variation for a bilayer}\label{subsec:secvarbilay}

We express the components $p_i$ of a given perturbation $p\in \mathcal{P}_b$ as a Fourier series (see Section~\ref{subsec:fourtfm}):
\begin{equation}
\label{eq:Fourier-pi}
p_i(x) = \frac{a_{i,0}}{\sqrt{L}} + \sqrt{\frac2L} \sum_{j=1}^{\infty} a_{i,j} \cos\left( \frac{2 \pi x j}{L} \right) + \sqrt{\frac2L} \sum_{j=1}^{\infty} b_{i,j} \sin\left( \frac{2 \pi x j}{L} \right).
\end{equation}
The equal-mass condition in~\pref{def:Pb} translates into
\begin{equation}\label{eq:masscondition}
2 \left( a_{1,0} + a_{3,0}\right) = a_{2,0} + a_{4,0}.
\end{equation}
We also write
\[
\mathfrak{a}_j := \left(a_{1,j}, a_{2,j}, a_{3,j}, a_{4,j}\right) \qquad \text{and} \qquad \mathfrak{b}_j := \left(b_{1,j}, b_{2,j}, b_{3,j}, b_{4,j}\right).
\]

\begin{theorem}\label{thm:bilayersecondvar}
Using the notation introduced above, the second variation of $\mathcal F$ at the VUV bilayer of optimal width~\pref{def:bilayer-optimal} in the direction $p\in\mathcal{P}_b$ is given by
\[
\left.\frac{d^2}{d \epsilon^2} \mathcal{F}(u_\e,v_\e) \right\vert_{\epsilon=0} = B_0\left(\mathfrak{a}_0, \bid\right) + \sum_{j=1}^{\infty} B_j\left(\mathfrak{a}_j, \mathfrak{b}_j, d_{uv}, d_{v0}, L\right),
\]
where
\begin{align*}
&B_0\left(\mathfrak{a}_0, \bid\right) :=\\ &4\bid \left\{- a_{1,0}^2 - a_{3,0}^2 + a_{1,0} a_{2,0} + a_{3,0} a_{4,0} - 4 a_{1,0} a_{3,0} + 3 a_{2,0} a_{3,0} + 3 a_{1,0} a_{4,0} - 2 a_{2,0} a_{4,0}\right\},
\end{align*}
and, for $j \in \vz{N}_{>0}$,
\begin{align*}
&B_j\left(\mathfrak{a}_j, \mathfrak{b}_j, d_{uv}, d_{v0}, L\right) :=\\ &\frac{4 \pi^2 j^2}{L^2} \left[ d_{uv} \left\{  a_{1,j}^2 + a_{3,j} ^2 + b_{1,j} ^2 + b_{3,j}^2 \right\}
  + d_{v0} \left\{  a_{2,j} ^2 +  a_{4,j} ^2 +  b_{2,j} ^2 +  b_{4,j} ^2 \right\} \right] \Biggr.\\
&+ \left.\frac{L}{ \pi j}\right.  \left. \biggl[2 \left(1 - \frac{2 \pi \bid j}{L} \right) \left\{ a_{1,j} ^2 + a_{3,j}^2 + b_{1,j}^2 + b_{3,j}^2 \right\}\biggr.\right.\\
&\quad \quad \quad \quad \quad+ \left.\left. \frac12 \left\{ a_{2,j}^2 + a_{4,j}^2 + b_{2,j}^2 + b_{4,j}^2 \right\}\right.\right.\\
&\quad \quad \quad \quad \quad-2 \left. \left. \left\{ a_{1,j} a_{2,j} + a_{3,j} a_{4,j} + b_{1,j} b_{2,j} + b_{3,j} b_{4,j} \right\} e^{-2 \pi \bid j / L} \right.\right.\\
&\quad \quad \quad \quad \quad+4 \left. \left. \left\{ a_{1,j} a_{3,j} + b_{1,j} b_{3,j} \right\} e^{-4 \pi \bid j / L}\right.\right.\\
&\quad \quad \quad \quad \quad-2 \left. \left. \left\{ a_{1,j} a_{4,j} + a_{2,j} a_{3,j} + b_{1,j} b_{4,j} + b_{2,j} b_{3,j} \right\} e^{-6 \pi \bid j / L}\right.\right.\\
&\quad \quad \quad \quad \quad+ \Biggl. \biggl. \left\{ a_{2,j} a_{4,j} + b_{2,j} b_{4,j}\right\} e^{-8 \pi \bid j / L}\biggr].
\end{align*}
\end{theorem}

The proof is given in Appendix~\ref{app:proofbilayersecondvar}.

\subsection{Variations for a monolayer}\label{subsec:varmonolay}

Analogous results also hold for monolayers as defined below. In the current subsection we will state them. Since the proofs are completely analogous to the proofs for bilayers, we will not write them out here.

\begin{figure}[ht]
\centering
{
    \psfrag{0}{$\epsilon p_1(x_1)$}
    \psfrag{1}{$\monod + \epsilon p_2(x_1)$}
    \psfrag{2}{$2\monod + \epsilon p_3(x_1)$}
    \psfrag{x}{$x_1$}
    \psfrag{y}{$x_2$}
    \psfrag{u}{U}
    \psfrag{v}{V}
    \includegraphics[width=0.35\textwidth]{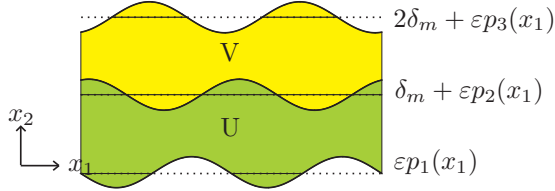}\\
}
\caption{The monolayer of optimal width with perturbations}
\label{fig:perturbationsmonolayer}
\end{figure}

The monolayer of optimal width is a structure given by functions $(u_0,v_0)\in\mathcal{K}$ with
\begin{equation}
\label{def:monolayer-optimal}
u_0 := \chi^{}_{\vz{T}_L\times [0,\monod]}
\qquad\text{and}\qquad
v_0 := \chi^{}_{\vz{T}_L\times [\monod,2\monod]},
\end{equation}
where~\cite{vanGennipPeletier07a}
\begin{equation}\label{eq:monod}
\monod := \sqrt[3]{\frac32 (c_0+c_u+c_v)}.
\end{equation}
The set of admissible boundary perturbations of this structure is again restricted by regularity and the equal-mass constraint:
\begin{definition}
The set of admissible perturbations is characterised by
\[
\mathcal{P}_m := \left\{p \in \left(W^{1,2}(\vz{T}_L)\right)^3:
  \int_{\vz{T}_L}(p_2-p_1) = \int_{\vz{T}_L} (p_3-p_2)\right\}.
\]
For $p\in \mathcal{P}_m$ and $\e>0$ we define a perturbed structure  $(u_\e,v_\e)$,
\begin{align*}
u_\e(x_1, x_2) &= \left\{
  \begin{array}{ll}
    1 & \text{if } x_2 \in \bigl(\e p_1(x_1), \monod + \e p_2(x_1)\bigr),\\
    0 & \text{otherwise,} \end{array} \right.\\
v_\e(x_1, x_2) &= \left\{
  \begin{array}{ll}
    1 & \text{if } x_2 \in \bigl(\monod + \e p_2(x_1), 2\monod + \e p_3(x_1)\bigr),\\ 0 & \text{otherwise.} \end{array} \right.
\end{align*}
We also define the subset of mass preserving perturbations:
\begin{equation}\label{assum:monolayermassconservation}
\mathcal{P}_m^M := \left\{p \in \mathcal{P}_m : \int_{\vz{T}_L}(p_2-p_1) = \int_{\vz{T}_L}(p_3-p_2) = 0\right\}.
\end{equation}
\end{definition}
A picture of a monolayer of optimal width with perturbations $p$ is shown in Figure~\ref{fig:perturbationsmonolayer}.

Stationarity for the monolayer of optimal width is defined analogously to stationarity for the bilayer, see Definition~\ref{def:stationarity}.

\begin{lemma}\label{lem:monolayerstationary}
The monolayer of optimal width is stationary with respect to all $p\in \mathcal{P}_m^M$.
\end{lemma}
\begin{proof}
Analogous to the proof of Lemma~\ref{lemma:bilayer_is_stationary} we find that the first variation of the interfaces with respect to all $p \in \mathcal{P}_m$ is zero.

For any $\eta \in C(\R)$ we compute
\begin{align*}
\frac{d}{d \epsilon} \int_{S_L} \eta(x_2) u_{\epsilon}(x) \, dx\Bigr|_{\e=0} &= \eta(\monod) \int_0^L p_2 - \eta(0) \int_0^L p_1,\\
\frac{d}{d \epsilon} \int_{S_L} \eta(x_2) v_{\epsilon}(x) \, dx\Bigr|_{\e=0} &= \eta(2\monod) \int_0^L p_3 - \eta(\monod) \int_0^L p_2.
\end{align*}

With $G$ the Green's function of $-\Delta$ on $S_L$ from Theorem~\ref{thm:gf} we make the choice $\eta(x_2) = G\ast(u_0-v_0)(x_1, x_2)$, which is independent of $x_1$. Using $\eta(\monod) = 0$ and $\eta(0) = -\eta(2 \monod) > 0$, we compute, as in the above mentioned proof,
\begin{equation}\label{eq:stationaritymonolayer}
\frac{d}{d\epsilon} \|u_{\epsilon}-v_{\epsilon}\|_{H^{-1}(S_L)}^2 = 2 \eta(0) \int_0^L(p_3-p_1) \stackrel{\pref{assum:monolayermassconservation}}= 0.
\end{equation}
\end{proof}
Note that by equation~(\ref{eq:stationaritymonolayer}) the monolayer of optimal width is not stable with respect to perturbations that are allowed to change the total mass, i.e with respect to $p \in \mathcal{P}_m\setminus \mathcal{P}_m^M$.

\begin{remark}
In complete analogy to Remark~\ref{rem:whyoptimalwidth} we see that the optimal width condition (\ref{eq:monod}) plays no role in the stationarity of the monolayer under the functional $\mathcal{F}$, but it plays an important role when considering the stationarity of the monolayer under $\mathcal{F}/\mathcal{M}$, the energy per unit mass functional. Appendix~\ref{sec:energy-per-unit-mass} also argues the relevance of optimal width when considering energy per unit mass.  Using, \cite{vanGennipPeletier07a}, 
\[
\mathcal{F}(u_0,v_0) = 2 L (c_0+c_u+c_v) + \frac23 L \monod^3,
\]
a computation analogous to that in Remark~\ref{rem:whyoptimalwidth} shows that the optimal width condition \pref{eq:monod} is not necessary for stationarity under $\mathcal{F}/\mathcal{M}$ with respect to the mass preserving perturbations in $\mathcal{P}_m^M$, but it is for stationarity with respect to perturbations in $\mathcal{P}_m\setminus\mathcal{P}_m^M$.\end{remark}

Similar to~\pref{eq:Fourier-pi} we express a $p\in \mathcal P_m$ in terms of its Fourier modes $a_{i,j}$ and $b_{i,j}$ and introduce the notation
\[
\mathfrak{a}_j := \left(a_{1,j}, a_{2,j}, a_{3,j}\right) \qquad \text{and} \qquad \mathfrak{b}_j := \left(b_{1,j}, b_{2,j}, b_{3,j}\right).
\]

\begin{theorem}\label{thm:monolayersecond}
Using the notation given above, the second variation of $\mathcal{F}$ at $(u_0, v_0)$ in the direction of $p\in\mathcal{P}_m$ is given by
\[
\left. \frac{d^2}{d \epsilon^2} \mathcal{F}(u_{\epsilon}, v_{\epsilon}) \right\vert_{\epsilon=0} = M_0\left(\mathfrak{a}_0, \monod\right)
+ \sum_{j=1}^{\infty} M_j\left(\mathfrak{a}_j, \mathfrak{b}_j, d_{u0}, d_{uv}, d_{v0}, L\right),
\]
where
\[
M_0\left(\mathfrak{a}_0, \monod\right) := \monod \left(a_{1,0} - a_{3,0}\right)^2,
\]
and, for $j \in \vz{N}$,
\begin{align*}
&M_j\left(\mathfrak{a}_j, \mathfrak{b}_j, d_{u0}, d_{uv}, d_{v0}, L\right) :=\\
&\frac{4 \pi^2 j^2}{L^2} \left[ d_{u0} \left\{ a_{1,j}^2 + b_{1,j}^2 \right\} + d_{uv} \left\{ a_{2,j}^2 + b_{2,j}^2 \right\} + d_{v0} \left\{ a_{3,j}^2 + b_{3,j}^2 \right\} \right] \Biggr.\\
&+ \left.\frac{L}{\pi j}\right.  \left. \biggl[2 \left(1 - \frac{2 \pi \monod j}{L} \right) \left\{ a_{2,j}^2 + b_{2,j}^2 \right\}\biggr.\right.\\
&\quad \quad \quad \quad \quad+ \left.\left. \frac12 \left\{ a_{1,j}^2 + a_{3,j}^2 + b_{1,j}^2 + b_{3,j}^2 \right\}\right.\right.\\
&\quad \quad \quad \quad \quad-2 \left. \left. \left\{ a_{1,j} a_{2,j} + a_{2,j} a_{3,j} + b_{1,j} b_{2,j} + b_{2,j} b_{3,j} \right\} e^{-2 \pi \monod j / L} \right.\right.\\
&\quad \quad \quad \quad \quad+ \Biggl. \biggl. \left\{ a_{1,j} a_{3,j} + b_{1,j} b_{3,j} \right\} e^{-4 \pi \monod j / L} \biggr].
\end{align*}
\end{theorem}

\begin{proof}
Analogous to the proof of Theorem~\ref{thm:bilayersecondvar}.
\end{proof}

\section{Stability}\label{sec:stability}

In this section we study stability of monolayers and bilayers with respect to the admissible perturbations. The bilayer will be treated in Section~\ref{subsec:bilayerstability}, the monolayer in Section~\ref{subsec:monolayerstability}.

\subsection{Preliminary definitions and results}

In this paper we only consider \emph{linear} stability---whenever we use the words \emph{stable} or \emph{unstable}, this refers to the sign of the second derivative:

\begin{definition}\label{def:stability}
Using the notation of Section~\ref{sec:perstrip}, the VUV bilayer (monolayer) of optimal width $(u_0,v_0)$ is called \emph{stable} iff
\[
\left.\frac{d^2}{d\epsilon^2}\mathcal{F}(u_{\epsilon}, v_{\epsilon})\right|_{\epsilon=0} \geq 0,
\]
for every $p\in \mathcal{P}_b^M$ ($\mathcal{P}_m^M$), and unstable otherwise.
\end{definition}

The following property simplifies the study of stability of the bilayers and monolayers.

\begin{lemma}\label{lem:oneforall}
Using the notation from Theorem~\ref{thm:bilayersecondvar} we have, for any $x, y\in\vz{R}^4$ and for $j \geq 1$,
\begin{align*}
B_j\left(x, y, d_{uv}, d_{v0}, L\right) &= B_1\left(x, y, d_{uv}, d_{v0}, L/j\right),\\
B_j\left(x, \zvec, d_{uv}, d_{v0}, L\right) &= B_j\left(\zvec, x, d_{uv}, d_{v0}, L\right),\\
B_j\left(x, y, d_{uv}, d_{v0}, L\right) &= B_j\left(x, \zvec, d_{uv}, d_{v0}, L\right) + B_j\left(\zvec, y, d_{uv}, d_{v0}, L\right).
\end{align*}

Similarly, in the notation from Theorem~\ref{thm:monolayersecond} we have, for any $x, y\in\vz{R}^3$ and for $j \geq 1$,
\begin{align*}
M_j\left(x, y, d_{u0}, d_{uv}, d_{v0}, L\right) &= M_1\left(x, y, d_{u0}, d_{uv}, d_{v0}, L/j\right),\\
M_j\left(x, \zvec, d_{u0}, d_{uv}, d_{v0}, L\right) &= M_j\left(\zvec, x, d_{u0}, d_{uv}, d_{v0}, L\right),\\
M_j\left(x, y, d_{u0}, d_{uv}, d_{v0}, L\right) &= M_j\left(x, \zvec, d_{u0}, d_{uv}, d_{v0}, L\right) + M_j\left(\zvec, y, d_{u0}, d_{uv}, d_{v0}, L\right).
\end{align*}
\end{lemma}
\begin{proof}
These properties follow from the definitions of $B_j$ in Theorem~\ref{thm:bilayersecondvar} and $M_j$ in Theorem~\ref{thm:monolayersecond}.
\end{proof}

\subsection{Stability of the bilayer}\label{subsec:bilayerstability}

Throughout this subsection we will use the notation as introduced in Section~\ref{subsec:secvarbilay}. Lemma~\ref{lem:oneforall} provides us with a simpler characterisation of stability:
\begin{corol}\label{cor:bilayerstability}
The VUV bilayer is stable iff
\begin{enumerate}
\item $B_0(\mathfrak{a}_0, \bid)\geq 0$ for all $\mathfrak{a}_0\in \R^4$ satisfying \pref{eq:masscondition}, and
\item $B_1(x, \zvec, d_{uv}, d_{v0}, L/j)
  \geq 0$ for all $x\in \R^4$ and all $j\geq 1$.
\end{enumerate}
\end{corol}
We therefore study $B_0$ and $B_1$ as quadratic forms on $\R^4$ subject to~\pref{eq:masscondition} and investigate their sign. Note that $B_0$ and $B_1$ can be identified with symmetric $4\times 4$ matrices, and we will continuously make this identification. Among other things that means we can speak of eigenvalues of $B_0$ and $B_1$, and relate the sign of the quadratic forms to the signs of their eigenvalues.

\begin{lemma}
\label{lemma:B0_positive}
$B_0(\mathfrak{a}, \bid)\geq 0$ for all $\bid>0$ and for all $\mathfrak{a}_0\in \R^4$ satisfying~\pref{eq:masscondition}.
\end{lemma}

\begin{proof}
The Lemma follows immediately from writing $B_0$ as
\[
\frac{1}{4\bid}B_0(\mathfrak{a}_0, \bid) =
  -\frac12\left(2a_{1,0}-a_{2,0} + 2a_{3,0} - a_{4,0}\right)^2
  + \frac12\left(a_{1,0} - a_{2,0} - a_{3,0} + a_{4,0}\right)^2
  + \frac12\left(a_{1,0} + a_{3,0}\right)^2.
\]
\end{proof}
\begin{lemma}\label{lem:B1negeig}
Two of the four eigenvalues of $B_1$ are nonnegative for all $d_{uv}$, $d_{v0}$, and $L$; the other two do not have a definite sign. Denote the smallest eigenvalue by $\lambda_1^b(d_{uv}, d_{v0}, L)$. Define
\begin{equation}
\label{eq:upslamb}
\blength := e^{-2 \pi \bid/L}, \hspace{2cm}
\brel := \frac{d_{uv}}{d_{uv}+d_{v0}} = \frac{c_u + c_v}{c_0 + c_u + 2 c_v}.
\end{equation}
There exists a function $\brel_1 \in C([0, 1])$ (see~\pref{eq:brel1}) such that
\begin{align*}
&\lambda_1^b(d_{uv}, d_{v0}, L) \geq 0 \Longleftrightarrow \brel \geq \brel_1(\blength).
\end{align*}
\end{lemma}

\begin{proof}
Note that $\blength \in (0, 1)$ and, by conditions (\ref{eq:ddemands}), \mbox{$\brel \in \left[\frac12 - \frac{c_u + c_0}{2(c_0 + c_u + 2 c_v)}, \frac12 + \frac{c_u + c_0}{2(c_0 + c_u + 2 c_v)}\right] \subset [0,1]$}.
Let $x\in\vz{R}^4$. We now write
\[
B_1\left(x, \zvec, d_{uv}, d_{v0}, L\right) = \frac{2L}{\pi} \tilde B_1\left(x, \brel, \blength\right),
\]
where
\begin{align}
\tilde B_1\left(x, \brel, \blength\right) &:= -\frac13 \log^3 \blength\, \Bigl( \brel \left(x_1^2 + x_3^2 \right) + (1 - \brel) \left( x_2^2 + x_4^2 \right) \Bigr)\label{eq:tildeB1}\\
&\hspace{0.5cm} + (1 + \log \blength) \left(x_1^2 + x_3^2\right) + \frac14 \left( x_2^2 + x_4^2 \right)\nonumber\\
&\hspace{0.5cm} - \left( x_1 x_2 + x_3 x_4 \right) \blength + 2 x_1 x_3 \blength^2 - \left(x_1 x_4 + x_2 x_3\right) \blength^3 + \frac12 x_2 x_4 \blength^4.\nonumber
\end{align}
Note that when $x_1=x_3=0$,
\[
\tilde B_1\left(x, \brel, \blength\right)
= (1-\brel)\left( x_2^2 + x_4^2 \right)
  + \frac14 \left( x_2^2 + x_4^2 \right)
  + \frac12 x_2 x_4 \blength^4
\geq 0,
\]
so that by the max-min characterisation of the third eigenvalue $\lambda_3^b$, for fixed $\brel, \blength$, we have
\[
\lambda^b_3 \;= \;\max_{\dim L = 2}\; \min_{\substack{x\in \R^4/L\\|x|=1}}\; \tilde B_1(x,\brel,\blength)
  \;\geq \;\min_{\substack{x_1=x_3=0\\|x|=1}} \;\tilde B_1(x,\brel,\blength) \geq0,
\]
implying that the largest two eigenvalues are always non-negative.

We now turn to the question of existence of admissible $x$ such that $\tilde B_1$ is negative, and we simplify the problem by minimizing $\tilde B_1$ with respect to $x_2$ and $x_4$ under fixed $x_1$ and $x_3$. The stationarity conditions $\frac{\partial}{\partial x_2} \tilde B_1\left(x, \brel, \blength\right) = 0$ and $\frac{\partial}{\partial x_4} \tilde B_1\left(x, \brel, \blength\right) = 0$ lead to the equations
\[
\left(\begin{array}{c} x_2^{\text{opt}} \\ x_4^{\text{opt}} \end{array}\right) = \frac1{\det A(\brel, \blength)} A(\brel, \blength) \left(\begin{array}{cc} \blength & \blength^3 \\ \blength^3 & \blength \end{array}\right) \left(\begin{array}{c} x_1 \\ x_3 \end{array}\right),
\]
where
\[
A(\brel, \blength) := \left(\begin{array}{cc} \frac12 - \frac23 (1 - \brel) \log^3 \blength & -\frac12 \blength^4 \\ -\frac12 \blength^4 & \frac12 - \frac23 (1 - \brel) \log^3 \blength \end{array}\right).
\]
Inserting these results into $\tilde B_1$ gives
\[
\tilde B_1\left(x_1, x_2^{\text{opt}}, x_3, x_4^{\text{opt}}, \brel, \blength\right) = \left(x_1, x_3\right) \matrixB(\brel, \blength) \left(x_1, x_3\right)^T,
\]
where the matrix entries of $\matrixB$ are given by
\begin{align*}
\matrixB_{11}(\brel, \blength) = \matrixB_{22}(\brel, \blength) &= \log \blength - \frac13 \brel \log^3 \blength\\
&\hspace{0.47cm} - \frac{\Bigl(3 (-1 + \blength^2) - 4 (-1 + \brel) \log^3 \blength\Bigr) \Bigl(3 (-1 + \blength^6) - 4 (-1 + \brel) \log^3 \blength\Bigr)}{9 \Bigl(-1 + \blength^8\Bigr) + 8 \Bigl(-1 + \brel\Bigr) \Bigl(-3 - 2 (-1 + \brel) \log^3 \blength\Bigr) \log^3 \blength},\\
\matrixB_{12}(\brel, \blength) = \matrixB_{21}(\brel, \blength) &= -\frac{\Bigl( 3 \blength (-1 + \blength^2) - 4 \blength (-1 + \brel) \log^3 \blength\Bigr)^2}{9 \Bigl(-1 + \blength^8\Bigr) + 8 \Bigl(-1 + \brel\Bigr) \Bigl(-3 -2 (-1 + \brel) \log^3 \blength\Bigr) \log^3 \blength}.
\end{align*}
The eigenvalues of $\matrixB$ are
\begin{align*}
G_-(\brel, \blength) &:= 1 - \blength^2 + \log \blength - \frac13 \brel \log^3 \blength + \frac{3 \blength^2 (-1 + \blength^2)^2}{3 (-1 + \blength^4) - 4 (-1 + \brel) \log^3 \blength}\\
    &= \Bigl(3 (-1 + \blength^4) - 4 (-1 + \brel) \log^3 \blength\Bigr)^{-1} h_-(\brel, \blength),\\
G_+(\brel, \blength) &:= 1 + \blength^2 + \log \blength - \frac13 \brel \log^3 \blength - \frac{3 \blength^2 (1 + \blength^2)^2}{3 (1 + \blength^4) + 4 (-1 + \brel) \log^3 \blength}\\
    &= \Bigl(3 (1 + \blength^4) + 4 (-1 + \brel) \log^3 \blength\Bigr)^{-1} h_+(\brel, \blength),
\end{align*}
with
\begin{align*}
h_-(\brel, \blength) &:= \left( \frac43 \log^6 \blength \right) \brel^2 + \left( -\frac43 \log^6 \blength - 4 \log^4 \blength + (-3 + 4 \blength^2 - \blength^4) \log^3 \blength \right) \brel\\
&\hspace{.5cm}   - 3 (1 - \blength^2)^2 + 3 (-1 + \blength^4) \log \blength + 4 (1 - \blength^2) \log^3 \blength + 4 \log^4 \blength,\\
h_+(\brel, \blength) &:= -\left( \frac43 \log^6 \blength\right) \brel^2 + \left(\frac43 \log^6 \blength + 4 \log^4 \blength + (3 + 4 \blength^2 - \blength^4) \log^3 \blength \right) \brel\\
&\hspace{.5cm} + 3 (1 - \blength^4) + 3 (1 + \blength^4) \log \blength - 4 (1 + \blength^2) \log^3 \blength - 4 \log^4 \blength .
\end{align*}
Note that $G_- < G_+$, since for $\blength \in (0, 1), \brel \in [0, 1]$,
\[
3(1 + \blength^4) + 4 (-1 + \brel) \log^3 \blength > 0,      \qquad      3(-1 + \blength^4) - 4 (-1 + \brel) \log^3 \blength < 0,
\]
and thus
\[
G_+(\brel, \blength) - G_-(\brel, \blength) = \frac{-2 \blength^2 \Bigl(3 (-1+\blength^2)-4(-1+\brel) \log^3\blength\Bigr)^2}{\Bigl(3(1 + \blength^4) + 4 (-1 + \brel) \log^3 \blength\Bigr)\Bigl(3(-1 + \blength^4) - 4 (-1 + \brel) \log^3 \blength\Bigr)} > 0.
\]
We have now the following equivalences:
\begin{align*}
\forall x\in\vz{R}^4, B_1\left(x, \zvec, d_{uv}, d_{v0}, L\right) \geq 0 &\Longleftrightarrow \forall x\in\vz{R}^4, \tilde B_1\left(x, \brel, \blength\right) \geq 0\\
&\Longleftrightarrow \matrixB \left(\brel, \blength\right) \geq 0\\
&\Longleftrightarrow G_-\left(\brel, \blength\right) \geq 0 .
\end{align*}
We prove the following characterisation of the sign of $G_-$:
\begin{equation}
G_-(\brel, \blength) \geq 0 \Longleftrightarrow \brel \geq \brel_1(\blength), \label{eq:F1neg}
\end{equation}
where
\begin{align}
\brel_1(\blength) &= (8 \log^3 \blength)^{-1} \Biggl(9 - 12 \blength^2 +  3 \blength^4 + (4 \log \blength) (3 + \log^2 \blength)\Bigr.\nonumber\\
    &\hspace{2.3cm}\left.+ \Bigl\{ 225 - 504 \blength^2 + 342 \blength^4 - 72 \blength^6 + 9 \blength^8 + (360 - 288 \blength^2 - 72 \blength^4) \log \blength \right.\nonumber\\
    &\hspace{2.8cm}\Bigl. + 144 \log^2 \blength + (-120 + 96 \blength^2 + 24 \blength^4) \log^3 \blength - 96 \log^4 \blength + 16 \log^6 \blength \Bigr\}^{\frac12} \Biggr).\label{eq:brel1}
\end{align}
The details of this calculation can be found in Appendix~\ref{app:details}. This concludes the proof.
\end{proof}
The function $g_1$ mentioned in Theorem~\ref{th:stab-bilayer} in the introduction is related to $\brel_1$, given in (\ref{eq:brel1}), by
\begin{equation}
\label{def:g_1}
g_1(\ell) := \brel_1\bigl(e^{2\pi/\ell}\bigr).
\end{equation}

\begin{figure}[ht]
\hspace{0.1\textwidth}
\subfloat[G1][The sign of $G_-$]
{
    \psfrag{a}{$\blength$}
    \psfrag{b}{$\brel$}
    \psfrag{c}{$G_- > 0$}
    \psfrag{d}{\color{white} $G_- < 0$}
    \includegraphics[width=0.35\textwidth]{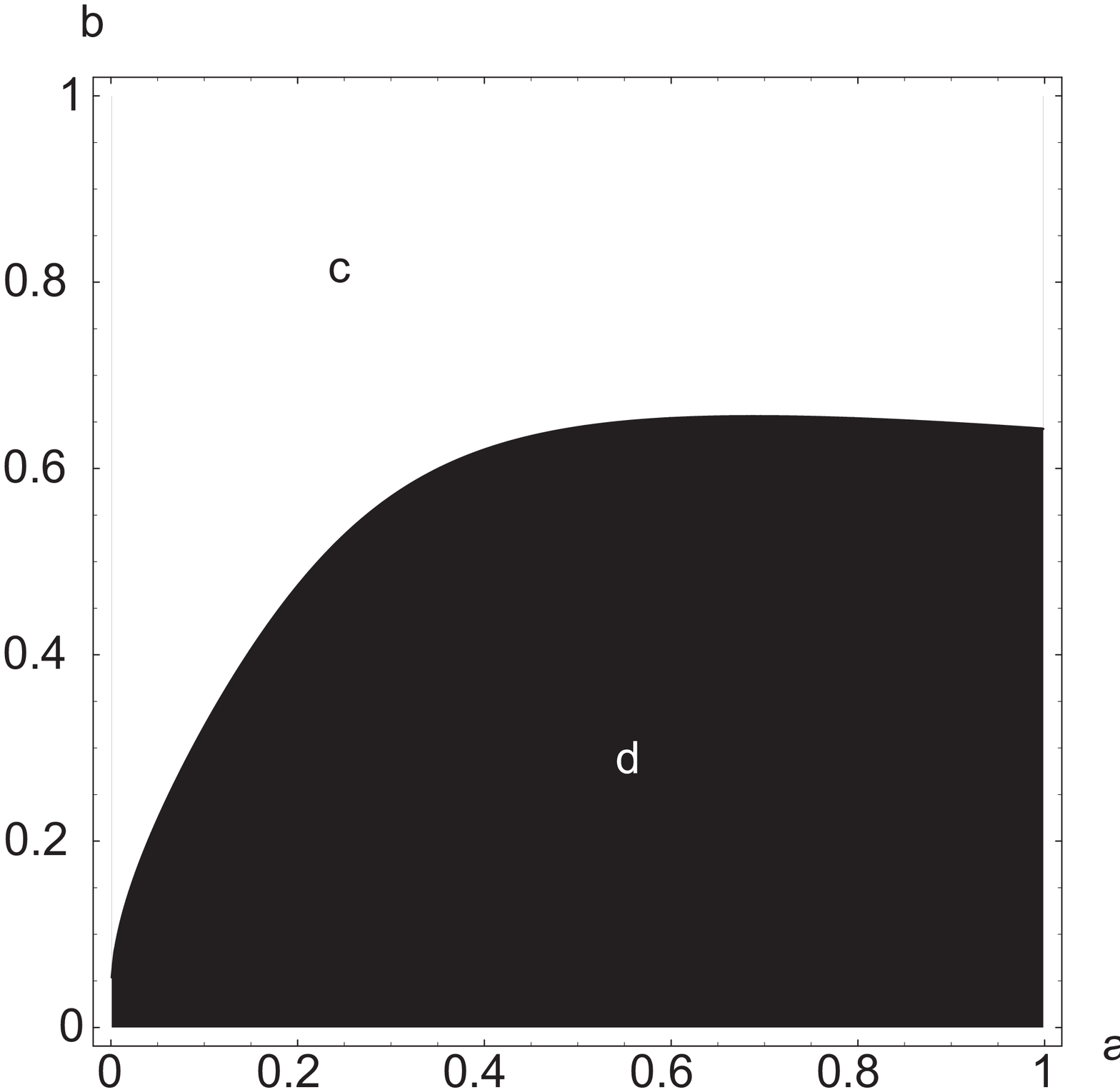}\\

}
\hspace{0.1\textwidth}
\subfloat[G2][The sign of $G_+$]
{
    \psfrag{a}{$\blength$}
    \psfrag{b}{$\brel$}
    \psfrag{c}{$G_+ > 0$}
    \psfrag{d}{\color{white} $G_+ < 0$}
    \includegraphics[width=0.35\textwidth]{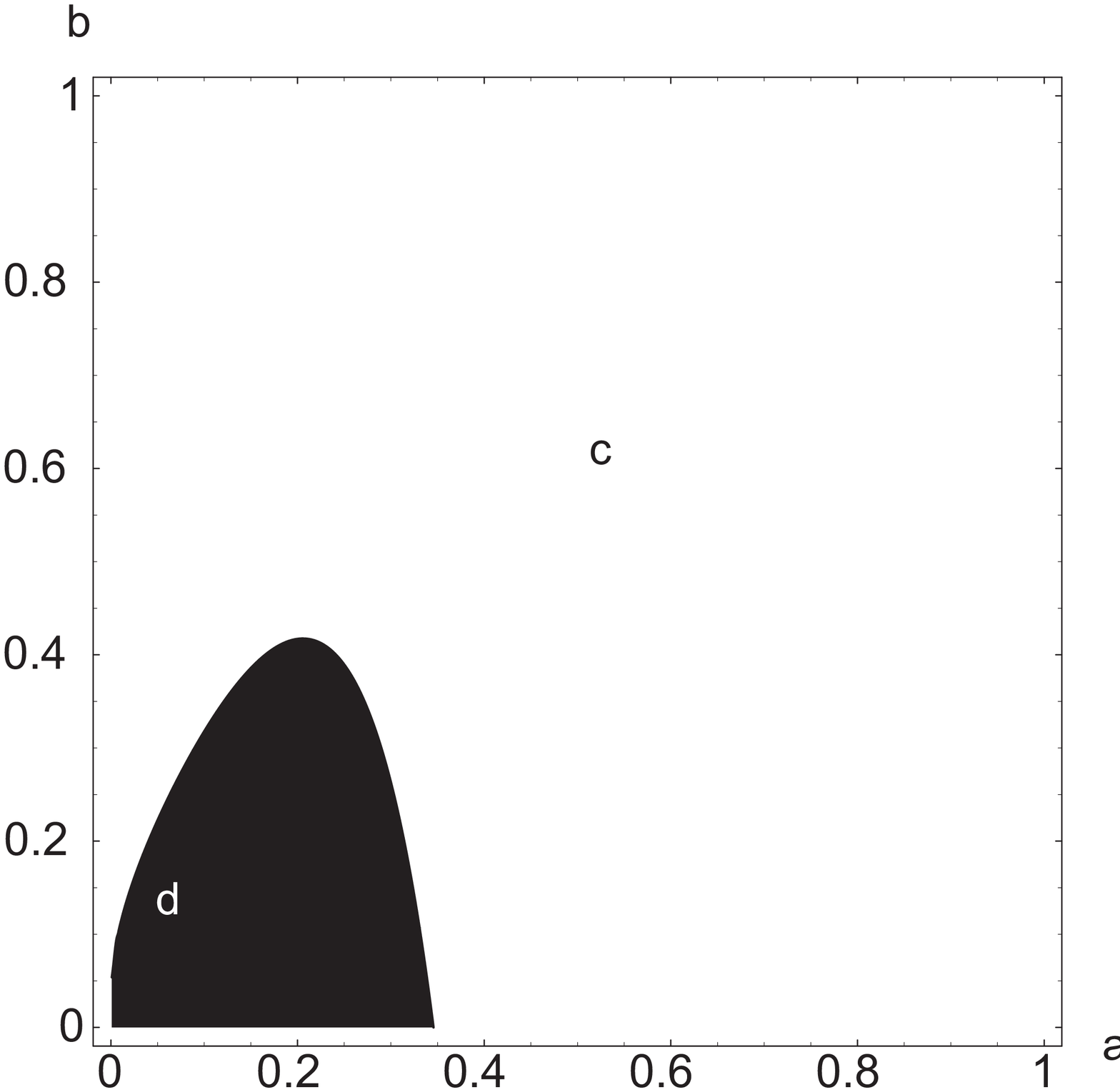}\\

}
\caption{The sign in parameter space of the eigenvalues $G_-<G_+$. The boundary between the two regions in the left-hand figure is given by $\brel=\brel_1(v)$.}
\label{fig:bilayerev}
\end{figure}

\begin{remark}
The four eigenvalues of $\tilde B_1$ from the proof of Lemma~\ref{lem:B1negeig} are
\begin{align*}
&\frac1{72} \Biggl( 45 - 36 \blength^2 - 9 \blength^2 + 36 \log\blength - 12 \log^3\blength \\
&\hspace{0.6cm}\pm \biggl\{ \Bigl(-45 + 36 \blength^2 + 9 \blength^4 - 36 \log\blength + 12 \log^3\blength\Bigr)^2\\
&\hspace{1.3cm}- 144 \Bigl(9 - 18 \blength^2 + 9 \blength^4 + 9 \log\blength - 9 \blength^4 \log\blength - 12 \log^3\blength + 12 \blength^2 \log^3\blength + 9 \brel \log^3 \blength\\
&\hspace{2.5cm}- 12 \blength^2 \brel \log^3\blength + 3 \blength^4 \brel \log^3\blength - 12 \log\blength^4 + 12 \brel \log^4\blength + 4 \brel \log^6\blength - 4 \brel^2 \log^6\blength\Bigr)\biggr\}^{\frac12}\Biggr),
\end{align*}
and
\begin{align*}
&\frac1{72} \Biggl( 45 + 36 \blength^2 + 9 \blength^2 + 36 \log\blength - 12 \log^3\blength \\
&\hspace{0.6cm}\pm \biggl\{ \Bigl(45 + 36 \blength^2 + 9 \blength^4 + 36 \log\blength - 12 \log^3\blength\Bigr)^2\\
&\hspace{1.3cm}+ 144 \Bigl(- 9 + 9 \blength^4 - 9 \log\blength - 9 \blength^4 \log\blength + 12 \log^3\blength + 12 \blength^2 \log^3\blength - 9 \brel \log^3\blength \\
&\hspace{2.5cm}- 12 \blength^2 \brel \log^3\blength + 3 \blength^4 \brel \log^3 \blength + 12 \log\blength^4 - 12 \brel \log^4\blength - 4 \brel \log^6\blength + 4 \brel^2 \log^6\blength)\biggr\}^{\frac12}\Biggr).
\end{align*}
Plotting the areas where these eigenvalues are negative shows that the eigenvalues with the plus sign chosen for $\pm$ are positive everywhere for $\blength \in (0, 1)$ and $\brel \in [0, 1]$. The plots for the other two eigenvalues correspond with those in Figure~\ref{fig:bilayerev}.
\end{remark}

\medskip

Collecting Lemmas~\ref{lemma:B0_positive} and~\ref{lem:B1negeig} we can summarise the stability properties with the use of Corollary~\ref{cor:bilayerstability} as follows:

\begin{theorem}\label{thm:stabilitybilayer}
Let $\brel$, $\blength$, and $\brel_1$ be as in Lemma~\ref{lem:B1negeig}. Define the functions $\underline{\brel}_j$, $j\geq 1$, and $\tilde \brel$ by
\[
\underline{\brel}_j(\blength) :=\brel_1(\blength^j), \quad \tilde \brel(\blength) := \sup_{j \geq 1} \underline{\brel}_j(\blength).
\]
Then the VUV bilayer of optimal width~\pref{def:bilayer-optimal} is stable with respect to all (mass-conserving) perturbations in $\mathcal{P}_b^M$ iff
\[
\brel \geq \tilde \brel(\blength).
\]
\end{theorem}
This is Theorem~\ref{th:stab-bilayer} from the introduction. Its implications are illustrated in Figure~\ref{fig:signs-intro-bi}.

\begin{remark}\label{rem:whichperturbationsbilayer}
Note that the statement in Theorem~\ref{thm:stabilitybilayer} about the positivity of the second variation also holds true if we allow the perturbations to come from the larger set of perturbations $\mathcal{P}_b$, instead of $\mathcal{P}_b^M$. However, as stated in Remark~\ref{rem:differentmassconstraints}, the bilayer of optimal width is not stationary under perturbations that do not preserve mass.
\end{remark}

We next show that $\tilde \brel$ is bounded from above away from $1$. Therefore there is a threshold $\alpha$ (as mentioned in the introduction) such that the bilayer is stable if $\brel \geq \alpha$.

\begin{lemma}\label{lem:brel1boundedawayfrom1}
Let $\tilde \brel$ be as in Theorem~\ref{thm:stabilitybilayer}, then there exists $\alpha\in(0,1)$ such that for all $\blength \in (0, 1)$,
\[
\tilde \brel(\blength) < \alpha < 1.
\]
\end{lemma}
\begin{proof}
First note that per definition of $\tilde \brel$ it suffices to show that there exists a $c \in (0,1)$, such that for all $\blength \in (0, 1)$,
\[
\brel_1(\blength) < c < 1.
\]
Since $\brel_1$ is continuous on the interval $(0, 1)$ and goes to zero for $\blength \downarrow 0$ and to $\frac52 - \frac12\sqrt{\frac{69}5}$ for $\blength \uparrow 1$ (see Remark~\ref{rem:limitsbrel}), this is equivalent to
\[
(8 \log^3\blength) (\brel_1(\blength) - 1) > 0.
\]
By (\ref{eq:lotsofinfoinhere}) we know that
\begin{align*}
0 &< \Biggl(\Bigl(9 - 12 \blength^2 +  3 \blength^4 + (4 \log \blength) (3 + \log^2 \blength)\Bigr) - 8 \log^3 \blength\Biggr)^2\\
&< 225 - 504 \blength^2 + 342 \blength^4 - 72 \blength^6 + 9 \blength^8 + (360 - 288 \blength^2 - 72 \blength^4) \log \blength \\
&\hspace{0.5cm}+ 144 \log^2 \blength + (-120 + 96 \blength^2 + 24 \blength^4) \log^3 \blength - 96 \log^4 \blength + 16 \log^6 \blength.
\end{align*}
Taking square roots completes the proof.
\end{proof}

\begin{remark}\label{rem:stableunstablemodesbilayer}
To find out the stable and unstable first-order Fourier modes of deformation for the bilayer, we compute the eigenvectors belonging to the positive and (potentially) negative eigenvalues of $\tilde B_1$ from (\ref{eq:tildeB1}). For the stable directions we find
\begin{align*}
\mathfrak{a}_1^{s_1}(\brel, \blength) &:= \left(\frac1{12 \blength( 1 + \blength^2)} \left(f_1(\brel, \blength) - \sqrt{f_4(\brel, \blength)}\right), 1, \frac2{\blength} \frac{f_2(\brel, \blength) - \sqrt{f_4(\brel, \blength)}}{f_3(\brel, \blength) + \sqrt{f_4(\brel, \blength)}}, 1 \right),\\
\mathfrak{a}_1^{s_2}(\brel, \blength) &:= \left(\frac1{12 \blength(-1 + \blength^2)} \left(g_1(\brel, \blength) - \sqrt{g_4(\brel, \blength)}\right), -1, \frac2{\blength} \frac{g_2(\brel, \blength) + \sqrt{g_4(\brel, \blength)}}{g_3(\brel, \blength) - \sqrt{g_4(\brel, \blength)}}, 1 \right),
\end{align*}
where
\begin{align*}
f_1(\brel, \blength) &:= -9 - 12 \blength^2 + 3 \blength^4 - 12 \log\blength + (-4 + 8 \brel) \log^3\blength\\
f_2(\brel, \blength) &:= -9 - 3 \blength^2 (6 + \blength^2) - 12 \log\blength + (-4+8\brel) \log^3\blength\\
f_3(\brel, \blength) &:= 15+3\blength^2 (4+\blength^2) - 12 \log\blength + (-4+8 \brel) \log^3\blength\\
f_4(\brel, \blength) &:= 9 \left( 9 + 40 \blength^2 + 42 \blength^4 + 8 \blength^6 + \blength^8\right)\\ &\hspace{0.4cm} + 8 \log\blength \Bigl( -3 + (-1+2 \brel) \log^2\blength\Bigr) \Bigl(3 (-3 - 4\blength^2 + \blength^4) - 6 \log\blength + (-2+4 \brel) \log^3\blength\Bigr),\\
g_1(\brel, \blength) &:= -9 + 12 \blength^2 - 3 \blength^4 - 12 \log\blength + (-4 + 8 \brel) \log^3\blength\\
g_2(\brel, \blength) &:= 9 - 3 \blength^2 (2 + \blength^2) + 12 \log\blength + (4-8\brel) \log^3\blength\\
g_3(\brel, \blength) &:= -15+3\blength^2 (4+\blength^2) + 12 \log\blength + (4-8 \brel) \log^3\blength\\
g_4(\brel, \blength) &:= 9 (-1+\blength^2)^2 (1+\blength^2) (9+\blength^2)\\&\hspace{0.4cm} + 8 \log\blength \Bigl(-3 + (-1+2\brel) \log^2\blength\Bigr) \Bigl(-3 (3-4\blength^2+\blength^4) - 6\log\blength + (-2+4\brel)\log^3\blength\Bigr).
\end{align*}
The directions belonging to the eigenvalues that can become negative, corresponding to the eigenvalues $G_+$ and $G_-$ of the reduced matrix $\matrixB$ in the proof of Lemma~\ref{lem:B1negeig}, are
\begin{align*}
\mathfrak{a}_1^{u_1}(\brel, \blength) &:= \left(\frac1{12 \blength( 1 + \blength^2)} \left(f_1(\brel, \blength) + \sqrt{f_4(\brel, \blength)}\right), 1, \frac2{\blength} \frac{f_2(\brel, \blength) + \sqrt{f_4(\brel, \blength)}}{f_3(\brel, \blength) - \sqrt{f_4(\brel, \blength)}}, 1 \right),\\
\mathfrak{a}_1^{u_2}(\brel, \blength) &:= \left(\frac1{12 \blength(-1 + \blength^2)} \left(g_1(\brel, \blength) + \sqrt{g_4(\brel, \blength)}\right), -1, \frac2{\blength} \frac{g_2(\brel, \blength) + \sqrt{g_4(\brel, \blength)}}{g_3(\brel, \blength) + \sqrt{g_4(\brel, \blength)}}, 1 \right),
\end{align*}
respectively.

The direction of the perturbation $\mathfrak{a}_1^{u_1}$ is depicted in Figure~\ref{fig:biunstab1}. Here we have chosen the values $d_{uv}=0.7, d_{v0}=0.3, L=5$ and $\epsilon=0.25$. Similarly we get Figures~\ref{fig:biunstab3}, \ref{fig:bistab2}, and ~\ref{fig:bistab4} using perturbations $\mathfrak{a}_1^{u_2}$, $\mathfrak{a}_1^{s_1}$, and $\mathfrak{a}_1^{s_2}$ respectively.
\end{remark}

\subsection{Stability of the monolayer}
\label{subsec:monolayerstability}

We now redo the arguments for the monolayer of optimal width~\pref{def:monolayer-optimal}.
Throughout this subsection we use the notation of Section~\ref{subsec:varmonolay}.

We can simplify $M_1$ a bit by writing
\[
\mlength := e^{-{2 \pi \monod}/L}, \qquad
\mrelu := \frac{d_{u0}}{d_{u0} + d_{uv} + d_{v0}}
        =\frac{c_u + c_0}{2 (c_0 + c_u + c_v)}, \qquad
\mrelv := \frac{d_{v0}}{d_{u0} + d_{uv} + d_{v0}}
        =\frac{c_v + c_0}{2 (c_0 + c_u + c_v)}.
\]
Note the slightly different definition of $\mlength$ than for the bilayer~\pref{eq:upslamb}.
Then, for all $x\in\vz{R}^3$,
\[
M_1\left(x, \zvec, d_{u0}, d_{uv}, d_{v0}, L\right)
  = \frac L{\pi}
    \tilde M_1\left(x, \mrelu, \mrelv, \mlength\right),
\]
where
\begin{align*}
\tilde M_1\left(x, \mrelu, \mrelv, \mlength\right) &:= -\frac23 \log^3\mlength \left(\mrelu x_1^2 + (1 - \mrelu - \mrelv) x_2^2 + \mrelv x_3^2\right)\\
&\hspace{0.5cm} + 2 (1 + \log \mlength) x_2^2 + \frac12 \left(x_1^2 + x_3^2 \right)\\
&\hspace{0.5cm} - 2\left(x_1 + x_3 + x_2 x_3 \right) \mlength + x_1 x_3 \mlength^2.
\end{align*}
We now can write
\[
\tilde M_1\left(x, \mrelu, \mrelv, \mlength\right)
  = x^T \hat M (\mrelu, \mrelv, \mlength) \,x,
\]
with
\[
\hat M (\mrelu, \mrelv, \mlength) := \left(\begin{array}{ccc} -\frac23 \mrelu \log^3\mlength + \frac12 & -\mlength & \frac12 \mlength^2\\ -\mlength & -\frac23 (1 - \mrelu - \mrelv) \log^3 \mlength + 2 (1 + \log \mlength) & -\mlength\\ \frac12 \mlength^2 & -\mlength & -\frac23 \mrelv \log^3 \mlength + \frac12 \end{array} \right).
\]
This matrix is well defined for all $\mrelu, \mrelv \in \vz{R}$, $\mlength > 0$, but note that the nonnegativity of the parameters $c_0$, $c_u$, and $c_v$---or equivalently conditions (\ref{eq:ddemands})---translates into
\begin{equation}\label{eq:conditionsonchikappa}
0 \leq \mrelu \leq \frac12, \hspace{2cm} 0 \leq \mrelv \leq \frac12, \hspace{2cm} \mrelu + \mrelv \geq \frac12,
\end{equation}
and furthermore $\mlength \in (0, 1)$ by definition.

\begin{remark}
As explained in the introduction and Appendix~\ref{sec:cu=cv}, we assume throughout the paper that for the monolayer the interfaces U-0 and V-0 are penalised equally strongly, i.e. $d_{u0} = d_{v0}$ or equivalently $c_u=c_v$. Under this assumption $\chi=\kappa$, and the inequalities in (\ref{eq:conditionsonchikappa}) imply that $\chi$ and $\kappa$ take values in $[\tfrac14,\tfrac12]$.
\end{remark}

\begin{lemma}\label{lem:M1negeig}
Let $c_u = c_v$. Two of the three eigenvalues of $\hat M (\mrelv, \mrelv, \mlength)$ are nonnegative for all $\mlength \in (0, 1)$ and $\mrelv \in [\frac14, \frac12]$. The third eigenvalue is given by
\[
E_3(\mrelv, \mlength) := \frac1{12} \left( e_1(\mrelv, \mlength) - \sqrt{e_2(\mrelv, \mlength)}\right),
\]
where $\mlength \in (0, 1), \mrelv \in [\frac14, \frac12]$ and $e_1$ and $e_2$ are given in (\ref{eq:e1}) and (\ref{eq:e2}).
The sign of $E_3$ is characterised by
\begin{equation}\label{eq:signofE2conditions}
E_3(\mrelv, \mlength) \geq 0 \Longleftrightarrow \mrelv \leq \mrelv_2(\mlength)\end{equation}
with $\mrelv_2$ as given in (\ref{eq:mrelv2}).
\end{lemma}

\begin{figure}[ht]
\hspace{0.1\textwidth}
\subfloat[mrelv1zoom][Plot of $\mrelv_1$ away from $\mlength=1$]
{
    \psfrag{a}{$\mlength$}
    \psfrag{b}{$\mrelv_1$}
    \includegraphics[width=0.35\textwidth]{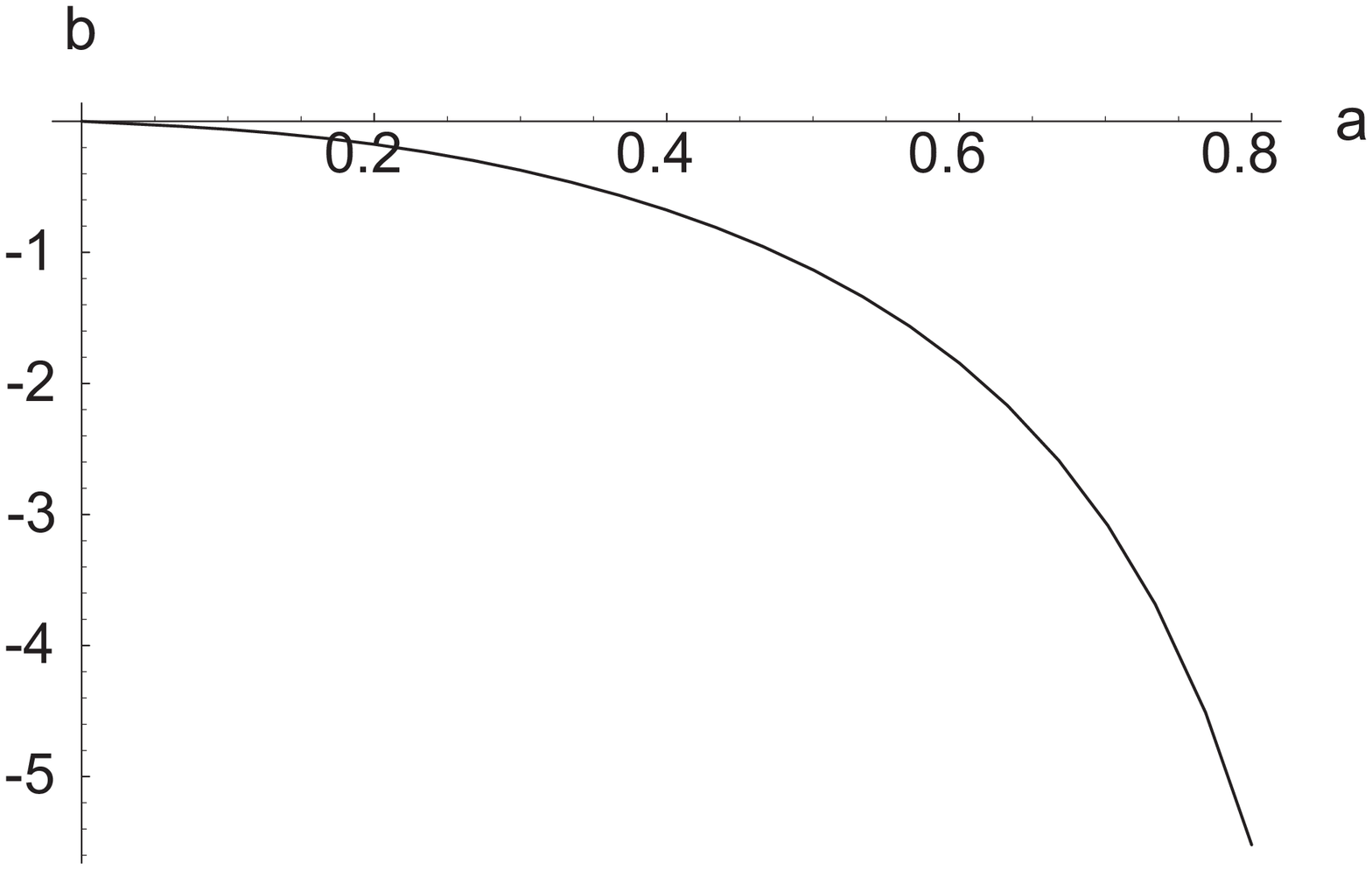}\\

}
\hspace{0.1\textwidth}
\subfloat[mrelv2][Plot of $\mrelv_2$, with the admissible range $\mrelv_2 \in \lbrack\frac14, \frac12 \rbrack$ indicated]
{
    \psfrag{a}{$\mlength$}
    \psfrag{b}{$\mrelv_2$}
    \includegraphics[width=0.35\textwidth]{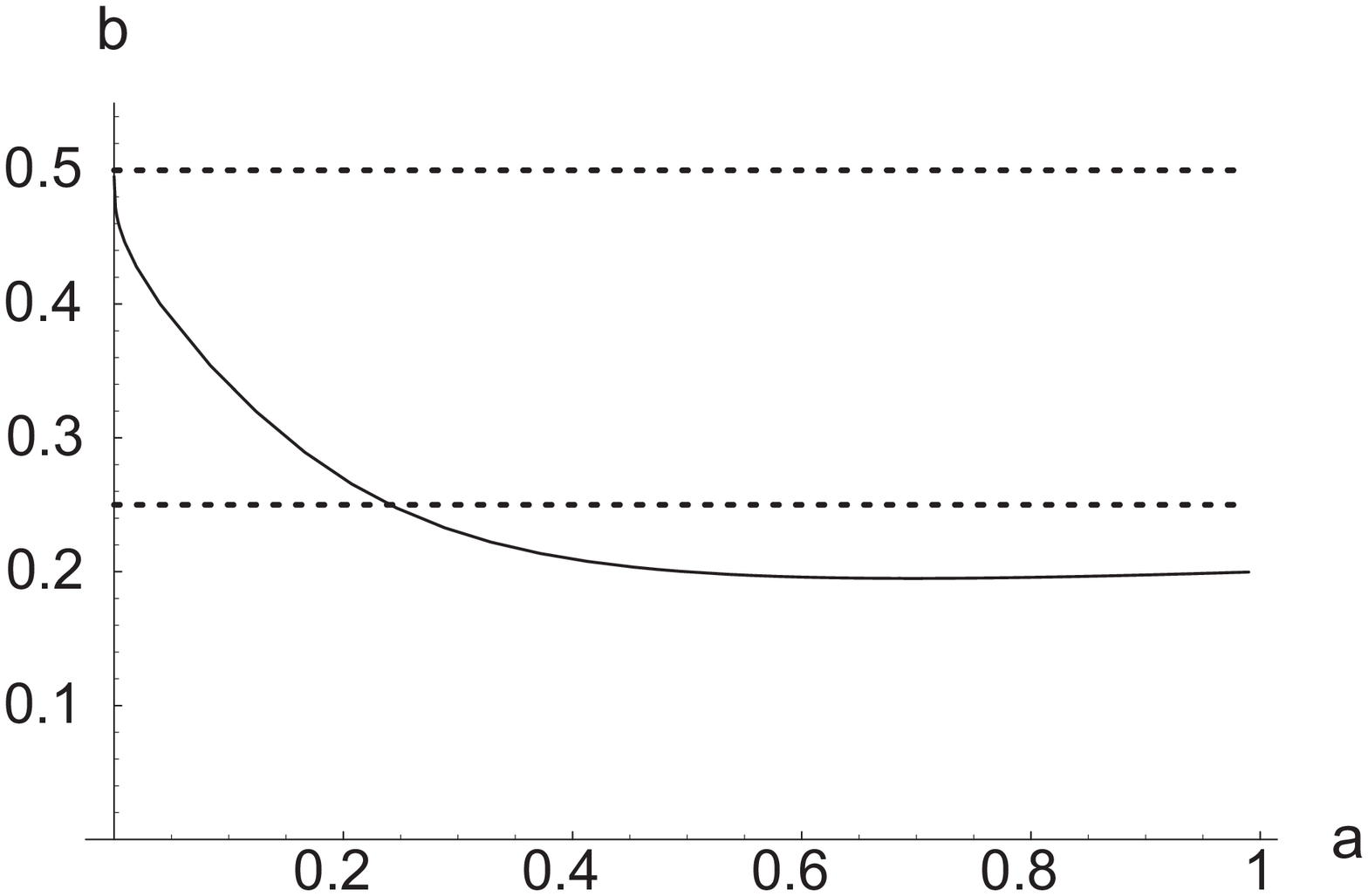}\label{fig:mrelv2}\\

}
\caption{}\label{fig:mrelv12}
\end{figure}

\begin{proof}
Since we are interested in the case where $c_u = c_v$ we will take $\mrelu = \mrelv$ from here on, which turns the conditions (\ref{eq:conditionsonchikappa}) into $\frac14 \leq \mrelv \leq \frac12$.
For the three eigenvalues of $\hat M_1(\mrelv, \mrelv, \mlength)$ we compute
\begin{align*}
E_1(\mrelv, \mlength) &:= \frac16 (3 - 3 \mlength^2 - 4 \mrelv \log^3\mlength),\\
E_{2, 3}(\mrelv, \mlength) &:= \frac1{12} \left( e_1(\mrelv, \mlength) \pm \sqrt{e_2(\mrelv, \mlength)}\right),
\end{align*}
where
\begin{align}
e_1(\mrelv, \mlength) &:= 15 + 3 \mlength^2 + (12 - 4 \log^2 \mlength + 4 \mrelv \log^2 \mlength) \log \mlength, \label{eq:e1}\\
e_2(\mrelv, \mlength) &:= 81 + 234 \mlength^2 + 9 \mlength^4 + 216 \log \mlength - 72 \mlength^2 \log \mlength + 144 \log^2 \mlength\nonumber\\
&\hspace{0.48cm} - 72 \log^3 \mlength + 24 \mlength^2 \log^3 \mlength - 96 \log^4 \mlength + 16 \log^6 \mlength\nonumber\\
&\hspace{0.48cm} + \left(216 \log^3 \mlength - 72 \mlength^2 \log^3 \mlength + 288 \log^4 \mlength - 96 \log^6 \mlength \right) \mrelv\nonumber\\
&\hspace{0.48cm} + \left(144 \log^6 \mlength\right) \mrelv^2\label{eq:e2}
\end{align}
and we choose the plus sign for $E_2$ and the minus sign for $E_3$.

First note that $\mlength \in(0,1)$ and $\chi\geq0$ imply that $E_1$ is always positive.
$E_{2,3}$ are real, since they are the eigenvalues of a symmetric matrix and thus $e_2(\mrelv, \mlength) \geq 0$ for all $\mrelv \in \vz{R}$ and for all $\mlength \in (0, 1)$.

Since for all $x>0$ and $\chi\leq 1/2$ we have $(1-\chi)x^3-3x\geq (1/2)x^3-3x\geq -2\surd 2$,
\[
e_1(\mrelv, \mlength) = 15 + 3 \mlength^2 + 4 \bigl[(1-\chi)|\log \mlength|^3 - 3|\log \mlength|\bigr] \geq 15 - 8 \sqrt{2} > 0.
\]
Combining this result with $e_2(\mrelv, \mlength) \geq 0$, we conclude that $E_2(\mrelv, \mlength) > 0$ for all admissible $\mrelv, \mlength$. Thus,
the only eigenvalue that might be negative (in all or part of parameter space) is $E_3$.

To prove the statements in (\ref{eq:signofE2conditions}) we compute
\begin{align*}
\frac1{16} \Bigl(e_1^2(\mrelv, \mlength) -e_2(\mrelv, \mlength)\Bigr) &= 9 (1-\mlength^2) + 9 (1+\mlength^2) \log\mlength - 3(1+\mlength^2)\log^3\mlength\\
&\hspace{0.4cm} + \Bigl(6 (-1+\mlength^2) \log^3\mlength + 4 (-3+\log^2\mlength)\log^4\mlength \Bigr)\mrelv\\
&\hspace{0.4cm} -\left(8 \log^6\mlength\right) \mrelv^2.
\end{align*}
This expression is negative on $(0, 1)$ if and only if $\mrelv \in \left[\frac14, \mrelv_1(\mlength)\right) \cup \left(\mrelv_2(\mlength), \frac12\right]$ and zero if and only if $\mrelv = \mrelv_1(\mlength)$ or $\mrelv = \mrelv_2(\mlength)$, where
\begin{equation}\label{eq:mrelv2}
\mrelv_{1,2}(\mlength) := \frac1{16 \log^6\mlength} \left(f(\mlength) \pm \sqrt{g(\mlength)}\right),
\end{equation}
with
\begin{align*}
f(\mlength) &:= \Bigl(6 (-1+\mlength^2) + 4 (-3+\log^2\mlength) \log\mlength\Bigr)\log^3\mlength;\\
g(\mlength) &:= 96 \log^6\mlength \Bigl( 3 (1-\mlength^2) + 3 (1+\mlength^2) \log\mlength - (1+\mlength^2) \log^3\mlength \Bigr)\\
 &\hspace{0.4cm}+ \Bigl( 6 (-1+\mlength^2) \log^3\mlength + 4 (\log^2\mlength-3)\log^4\mlength \Bigr)^2.
\end{align*}
The minus sign is chosen in $\mrelv_1$ while in $\mrelv_2$ we choose the plus sign. Plots of $\mrelv_1$ and $\mrelv_2$ are shown in Figure~\ref{fig:mrelv12}.

It is left to prove now that $\mrelv_1(\mlength) < 1/4$ for all $\mlength\in(0, 1)$. We will actually prove the stronger statement $\mrelv_1(\mlength) < 0$, which follows from
\begin{alignat*}4
&&&g(\mlength) > 0 & \qquad &\text{for $0<\mlength<1$},\\
&\Longleftarrow&\qquad &f(\mlength)^2 - g(\mlength) < 0 & \qquad &\text{for $0<\mlength<1$},\\
&\Longleftrightarrow&\qquad
  &3(1-\mlength^2)+3(1+\mlength^2)\log \mlength - (1+\mlength^2)\log^3 \mlength > 0 &&\text{for $0<\mlength<1$},\\
&\Longleftrightarrow&\qquad
  &3\frac{1-\mlength^2}{1+\mlength^2}+3\log \mlength - \log^3 \mlength > 0 &&\text{for $0<\mlength<1$},\\
&\hspace{-0.315cm}\stackrel{w = -\log \mlength}\Longleftrightarrow&\qquad
  &3\tanh w-3w + w^3  > 0 &&\text{for $w>0$}.
  \end{alignat*}
To prove that this last inequality holds, we define $h(w) := \tanh w-3w + w^3$ and use $\tanh'w=1-\tanh^2w$, to compute that $h'''(w) = 6 \tanh^2 w \, (-3 \tanh^4 w + 4) > 0$. From this it follows by integration that $h(w)>0$ for all $w>0$.
\end{proof}

\begin{remark}
For the excluded endpoints $0$ and $1$ we find
\begin{align*}
\underset{\mlength \downarrow 0}{\lim}\mrelv_1 &= 0, \qquad \underset{\mlength \uparrow 1}{\lim}\mrelv_1 = -\infty,\\
\underset{\mlength \downarrow 0}{\lim}\mrelv_2 &= \frac12, \qquad \underset{\mlength \uparrow 1}{\lim}\mrelv_2 = \frac15.
\end{align*}
The limits for $\mlength \uparrow 1$ were found by calculating the first terms in the Taylor expansion of $\mrelv_{1,2}$.
\end{remark}

\begin{figure}[ht]
\hspace{0.1\textwidth}
\subfloat[E3][Sign of $E_3$]
{
    \psfrag{a}{$\mlength$}
    \psfrag{b}{$\mrelv$}
    \psfrag{c}{\scriptsize $E_3>0$}
    \psfrag{d}{\color{white}$E_3<0$}
    \includegraphics[width=0.35\textwidth]{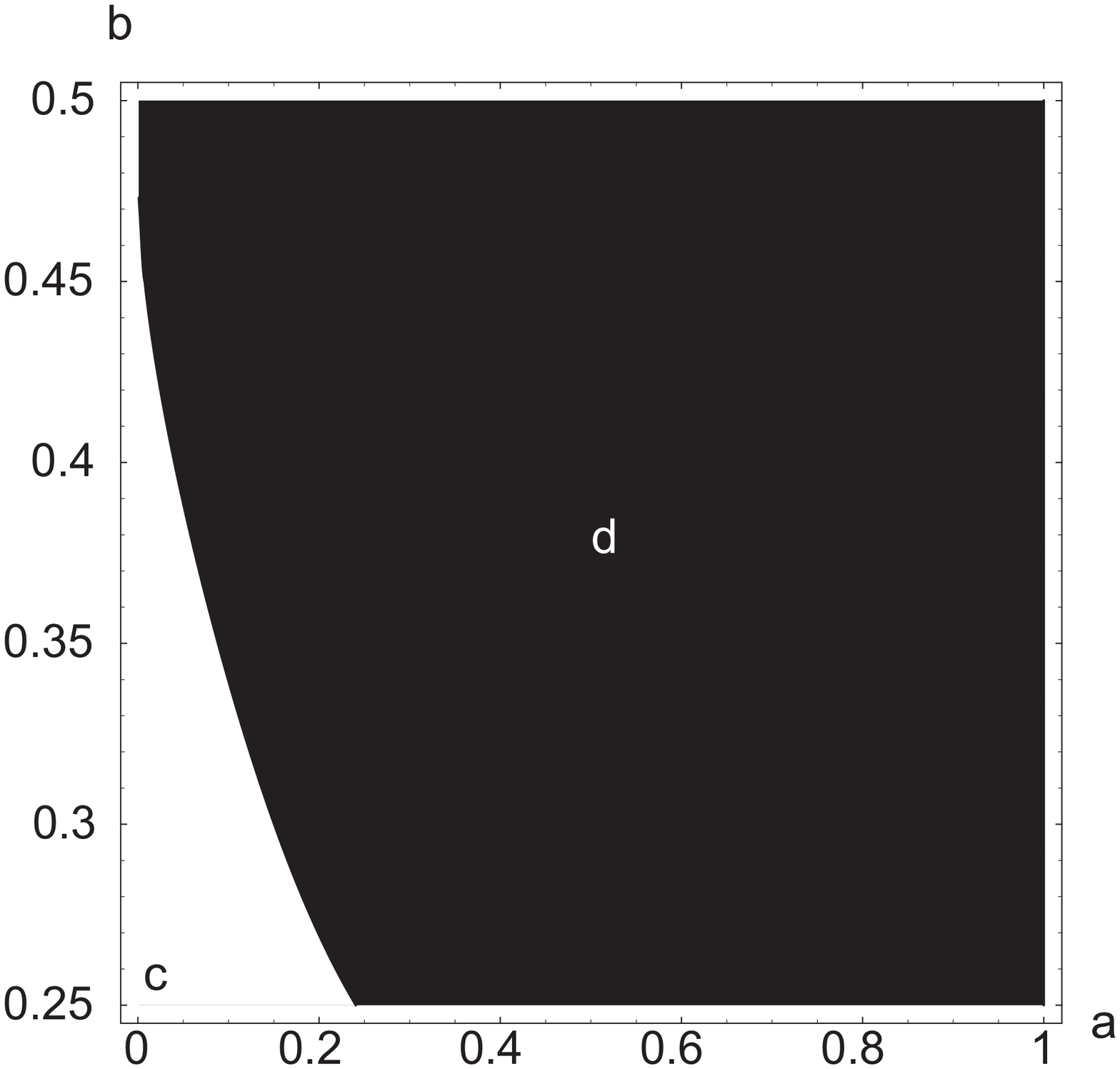}\label{fig:E2mrelva}\\

}
\hspace{0.1\textwidth}
\subfloat[E3extended][Sign of $E_3$ extended to $(0, 1)$ for $\mrelv$]
{
    \psfrag{a}{$\mlength$}
    \psfrag{b}{$\mrelv$}
    \psfrag{c}{$E_3>0$}
    \psfrag{d}{\color{white}$E_3<0$}
    \includegraphics[width=0.35\textwidth]{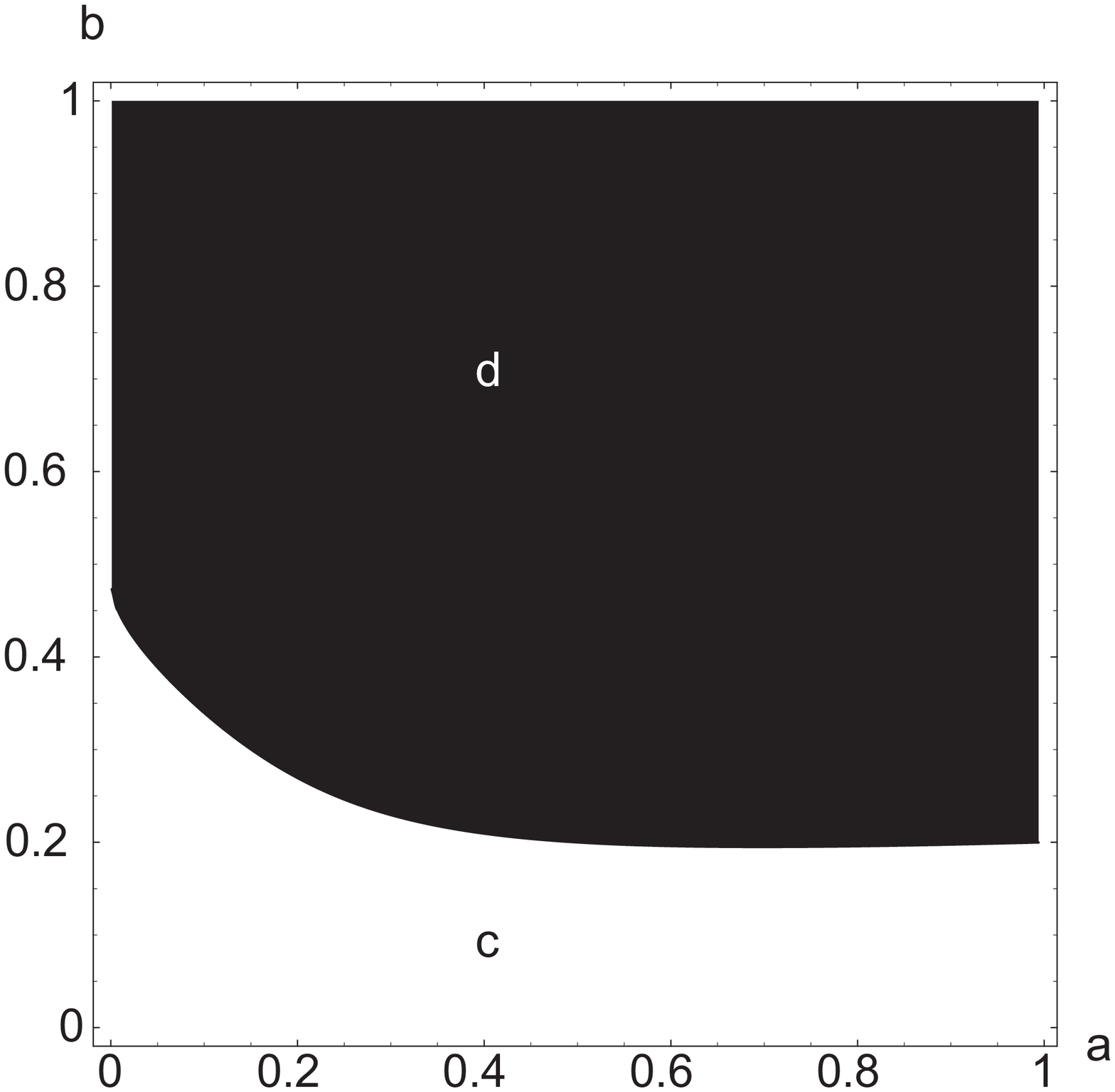}
    \label{fig:E2contour_cd}\\

}
\caption{}\label{fig:E2mrelv}
\end{figure}

Figure~\ref{fig:E2mrelv} shows the parts of parameter space where $E_3$ is positive and negative, both on the admissible domain $\left(\frac14, \frac12\right)$ for $\mrelv$ as well as extended to $(0, 1)$.

\begin{remark}
Expanding $E_3$ around $\mlength=1$ gives
\[
E_3(\mrelv, \mlength) = \frac4{45} (1 - 5 \mrelv) (1 - \mlength)^5 + \mathcal{O}\left(\left(1 - \mlength\right)^6\right),
\]
for $\mlength \uparrow 1$. Since $1-5\mrelv \leq -\frac14$ for $\mrelv\in\left[\frac14, \frac12\right]$ we can conclude that for $\mlength$ close to $1$ (or equivalently large $L$) the monolayer is unstable for all admissible values for the interfacial (surface tensions) coefficients $d_{ij}$ (or $c_i$). This corresponds to what is shown in Figure~\ref{fig:E2mrelva}.

Taking into account the assumption $d_{v0}=d_{u0}$, the condition $1-5\mrelv < 0$ for negativity of $E_3$ is equivalent to $d_{uv} < \frac32 (d_{u0} + d_{v0})$. In \cite[Theorem~8]{vanGennipPeletier07a} we show that for a circular two-dimensional monolayer the term in
in the energy per unit mass $\mathcal{F}/M$ that is quadratic in the curvature is given by
\[
m \left( -\frac12 (d_{u0}+d_{v0}) + \frac4{15} m^3 \right) \kappa^2,
\]
where $m$ is the thickness of the layers and $\kappa$ is the curvature. Taking $m=\monod$ we find that this term becomes negative exactly as $d_{uv} < \frac32 (d_{u0} + d_{v0})$, showing that the (large) circular monolayer loses stability at the same point as the flat monolayer on large domains. Note that conditions~(\ref{eq:ddemands}) imply $d_{uv} < \frac32 (d_{u0} + d_{v0})$. \end{remark}

In order to compare the results for the monolayer to those for the bilayer, we introduce the relative U-V interface penalisation
\begin{equation}\label{eq:yeahgoaheadanddefinemreluv}
\mreluv := 1-\mrelu-\mrelv = \frac{c_u + c_v}{2 (c_0 + c_u + c_v)},
\end{equation}
analogous to $\brel$ for the bilayer in Lemma~\ref{lem:B1negeig}. Note that conditions (\ref{eq:ddemands}) give $\mreluv \in [0, \frac12]$.
In terms of the surface tension coefficients,
\[
\mreluv = \frac{d_{uv}}{d_{u0} + d_{uv} + d_{v0}},
\]
$\mreluv$ is interpreted as the relative penalisation of the U-V interface.

\begin{theorem}\label{thm:stabilitymonolayer}
Let $d_{u0} = d_{v0}$ and let $\mrelv_2$ be as in Lemma~\ref{lem:M1negeig}. Define the functions $\underline{\mrelv}_j$ and $\tilde \mrelv$ by
\[
\underline{\mrelv}_j(\mlength) := \mrelv_2(\mlength^j), \quad \tilde \mrelv := \underset{j \geq 1}{\inf}\, \underline{\mrelv}_j, \quad \tilde \mreluv := 1 - 2 \tilde \mrelv.
\]
The monolayer of optimal width~\pref{def:monolayer-optimal} is stable with respect to perturbations in $\mathcal{P}_m^M$ if and only if $\mreluv \geq \tilde \mreluv(\mlength)$.
\end{theorem}
\begin{proof}
First we work with $\mrelv$ as in Lemma~\ref{lem:M1negeig} and afterwards we translate the results into conditions on $\mreluv$. By Definition~\ref{def:stability} and Theorem~\ref{thm:monolayersecond} in order to prove stability, we have to prove that
\[
M_0\left(\mathfrak{a}_0, \monod\right) + \sum_{j=1}^{\infty} M_j\left(\mathfrak{a}_j, \mathfrak{b}_j, d_{u0}, d_{uv}, d_{v0}, L\right) \geq 0,
\]
for all admissible perturbations.
Per definition we have $M_0\left(\mathfrak{a}_0, \monod\right) := \monod \left(a_{1,0} - a_{3,0}\right)^2 \geq 0$. By Lemma~\ref{lem:M1negeig} we know that if $(\mlength, \mrelv)$ is such that $\mrelv \in \left[\frac14, \mrelv_2(\mlength)\right]$ then $M_1\left(\mathfrak{a}_1, \zvec, d_{u0}, d_{uv}, d_{v0}, L\right) \geq 0$ for all $p \in \mathcal{P}_m$. By Lemma~\ref{lem:oneforall} we have for $j \geq 1$,
\[
\forall \mathfrak{a}_j, \mathfrak{b}_j, \, M_j\left(\mathfrak{a}_j, \mathfrak{b}_j, d_{u0}, d_{uv}, d_{v0}, L\right) \geq 0 \Longleftrightarrow \forall \mathfrak{a}_1, \, M    _1\left(\mathfrak{a}_1, \zvec, d_{u0}, d_{uv}, d_{v0}, L/j\right) \geq 0,
\]
thus we see that, if $\mrelv \in \left[\frac14, \tilde \mrelv(\mlength)\right]$ is satisfied, then for all $j \geq 1$, for all $\mathfrak{a}_j$, and for all $\mathfrak{b}_j$, $M_j\left(\mathfrak{a}_j, \mathfrak{b}_j, d_{u0}, d_{uv}, d_{v0}, L\right) \geq 0$.

Now note that $\mrelv = \frac{1-\mreluv}2$ and thus
\[
\mrelv \in \left[\frac14, \tilde \mrelv(\mlength)\right] \Longleftrightarrow \mreluv \in \left[1-2 \tilde \mrelv(\mlength), \frac12\right],
\]
which proves the statement of the theorem.
\end{proof}

To make the connection to Theorem~\ref{th:stab-monolayer} in the introduction, the function $f_1$ is defined by
\begin{equation}
\label{def:f1}
f_1(\ell) := 1-2\chi_2\left(e^{2\pi/\ell}\right),
\end{equation}
where $\chi_2$ is given in~\pref{eq:mrelv2}.

Figure~\ref{fig:signs-intro-mono} illustrates the stability properties of the monolayer of optimal width from Theorem~\ref{thm:stabilitymonolayer}.

\begin{remark}\label{rem:whichperturbationsmonolayer}
In Theorem~\ref{thm:stabilitymonolayer} we only consider perturbations in $\mathcal{P}_m^M$, i.e. perturbations that keep the total mass fixed. The statement about the positivity of the second variation still holds if we consider the larger set of perturbations $\mathcal{P}_m$, however, for these perturbations the monolayer of optimal width is not a stationary point, as was noted after Lemma~\ref{lem:monolayerstationary}.
\end{remark}

\begin{remark}\label{rem:stableunstablemodesmonolayer}
To find the stable and unstable first order Fourier modes of deformation we compute the eigenvectors belonging to the positive eigenvalues of $\tilde M_1\left(\mathfrak{a}_1, \mrelv, \mrelv, \mlength\right)$ and to the eigenvalues that are negative for some parameter choices. For the positive, stable directions we find
\begin{align*}
\mathfrak{a}_1^{s_1}(\mrelv, \mlength) &:= \left(-1, 0, 1 \right),\\
\mathfrak{a}_1^{s_2}(\mrelv, \mlength) &:= \left(1, \frac1{12 \mlength} \left(h_1(\mrelv, \mlength) + \sqrt{h_2(\mrelv, \mlength)}\right), 1\right),
\end{align*}
where
\begin{align*}
h_1(\mrelv, \mlength) &:= -9 + 3 \mlength^2 - 12 \log\mlength + (4-12 \mrelv) \log^3\mlength,\\
h_2(\mrelv, \mlength) &:= 9 \left( 9 + 26 \mlength^2 + \mlength^4 \right)\\
 &\hspace{0.5cm}+ 8 \log \mlength \left( 3+ (-1+3 \mrelv) \log^2\mlength\right) \left( 9 -3 \mlength^2 + 6 \log\mlength + (-2+6 \mrelv) \log^3\mlength\right).
\end{align*}
The direction belonging to the eigenvalues that can become negative, corresponding to the eigenvalue $E_3$ of in Lemma~\ref{lem:M1negeig}, is
\begin{align*}
\mathfrak{a}_1^{u}(\mrelv, \mlength) &:= \left(1, \frac1{12 \mlength} \left(h_1(\mrelv, \mlength) - \sqrt{h_2(\mrelv, \mlength)}\right), 1\right).
\end{align*}
Figure~\ref{fig:monostab1} shows the monolayer with a perturbation corresponding to $\mathfrak{a}_1^{s_1}$. Here we have chosen the values $d_{u0}=1$, $d_{uv}=0.7$, $d_{v0}=0.3$, $L=5$ and $\epsilon=0.25$. Similarly we get Figure~\ref{fig:monostab3} using
$\mathfrak{a}_1^{s_2}$, and Figure~\ref{fig:monounstab2} using $\mathfrak{a}_1^{u}$.
\end{remark}

\subsection{Discussion and comparison}

\begin{figure}[h]
\hspace{0.1\textwidth}
\subfloat[minmreluv][Plot of $\tilde \mreluv$ as a function of $L/\monod=2\pi (\log\mlength^{-1})^{-1}$]
{
    \psfrag{a}{$L/\monod$}
    \psfrag{b}{$\tilde \mreluv$}
    \includegraphics[width=0.35\textwidth]{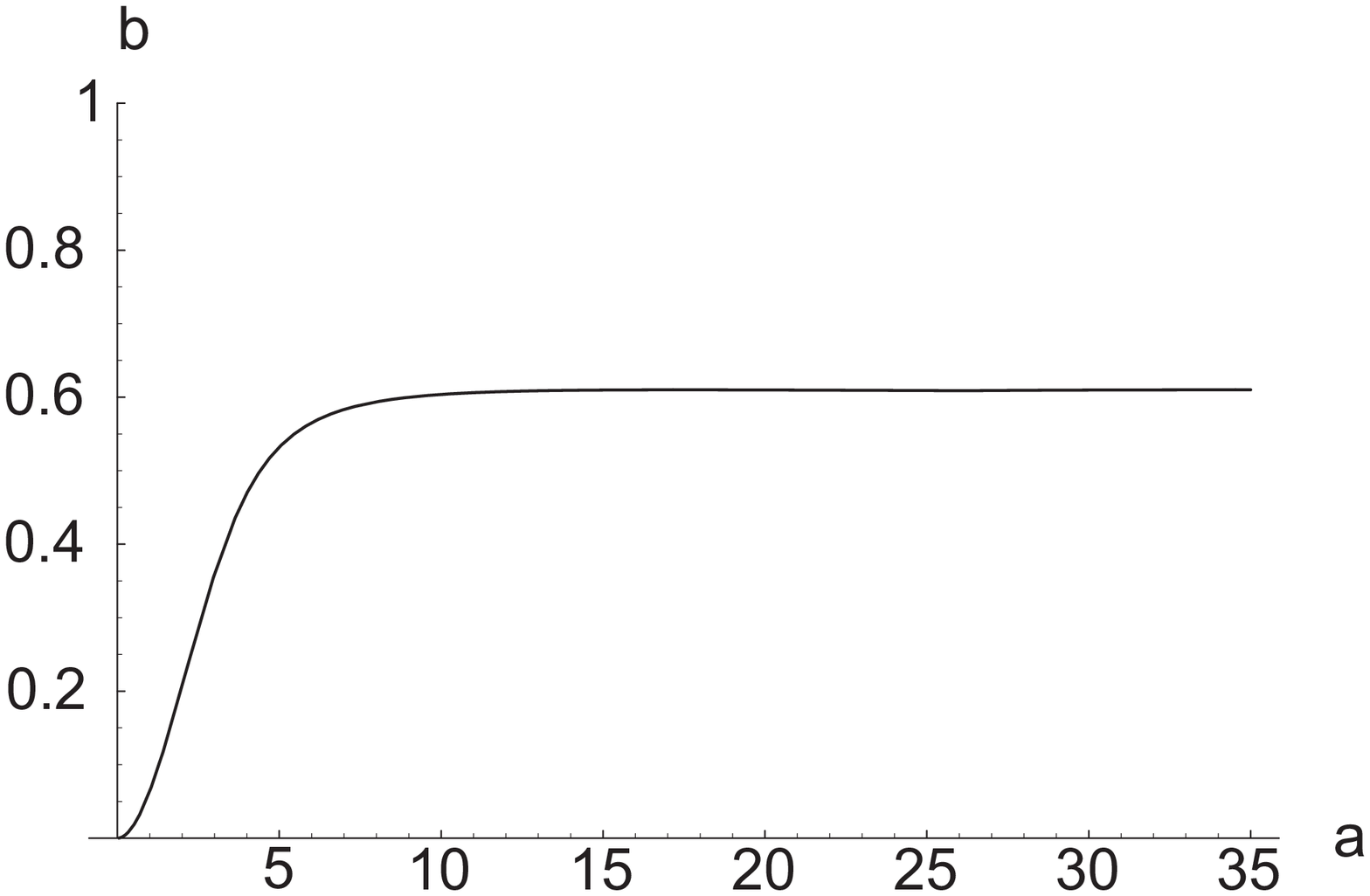}\\

}
\hspace{0.1\textwidth}
\subfloat[minmreluvrestricted][Plot of $\tilde \mreluv$ restricted to the admissible range $\tilde \mreluv \in \lbrack 0, \frac12 \rbrack$]
{
    \psfrag{a}{$L/\monod$}
    \psfrag{b}{$\tilde \mreluv$}
    \includegraphics[width=0.35\textwidth]{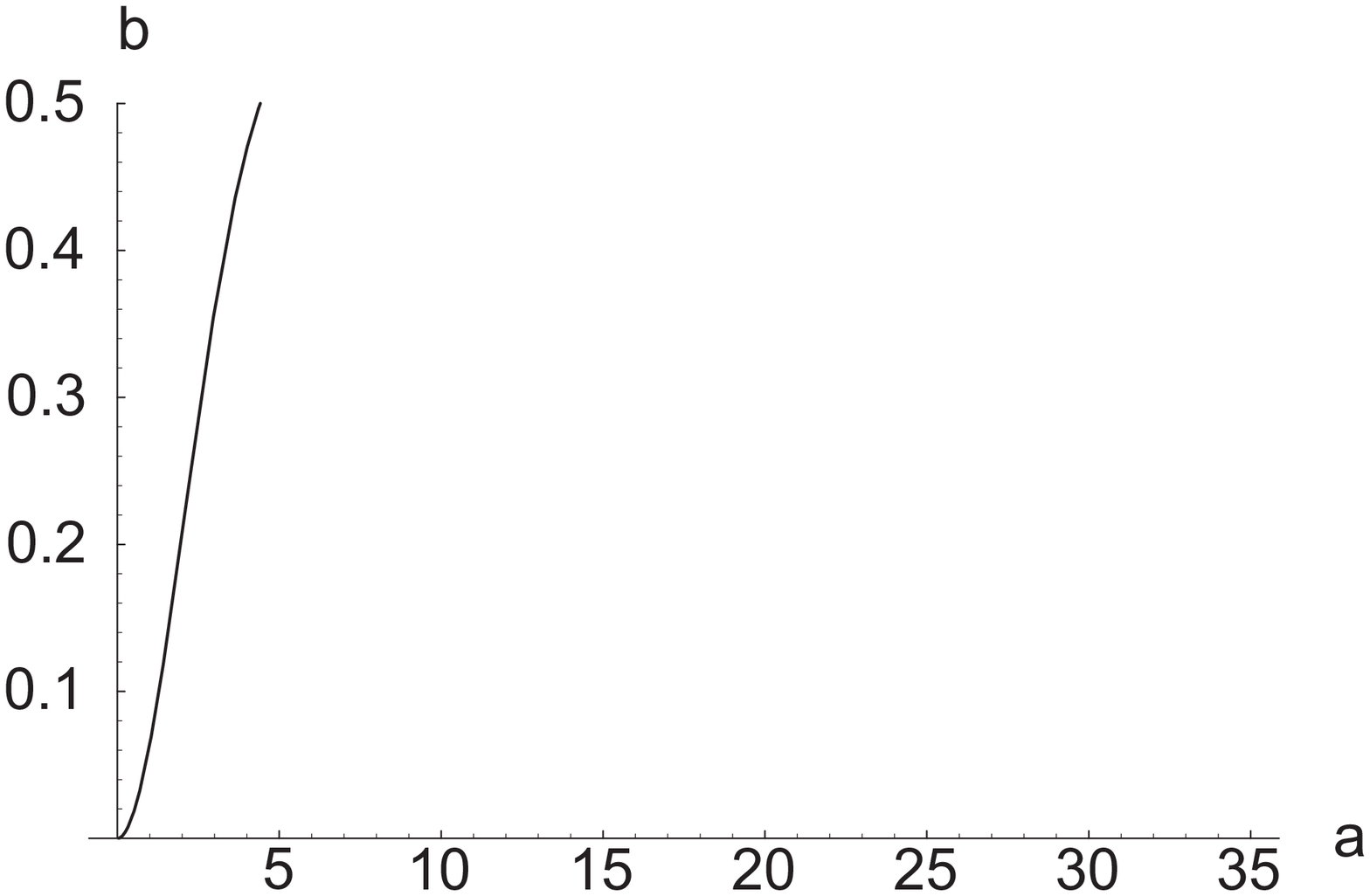}
    \label{fig:tildemreluvrestricted}
}

\caption{For the plots of $\tilde \mreluv$ from Theorem~\ref{thm:stabilitymonolayer} we have approximated $\tilde \mrelv$ by $\min_{1\leq j\leq 100} \underline{\mrelv}_j$} \label{fig:tildemreluv}
\end{figure}

\begin{figure}[h]
\centering
{
    \psfrag{a}{$L/\monod$}
    \psfrag{b}{$\tilde \mreluv$}
    \includegraphics[width=0.45\textwidth]{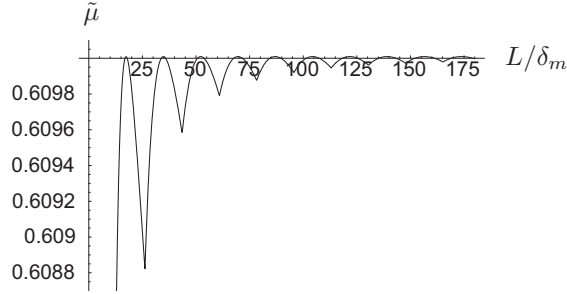}
}
\caption{Plot of $\tilde \mreluv$ from Theorem~\ref{thm:stabilitymonolayer} as a function of $L/\monod=2\pi (\log\mlength^{-1})^{-1}$, showing the small-scale oscillations where different Fourier orders become the dominant contributors.} \label{fig:tildemreluvsmallscale}
\end{figure}

\begin{figure}[h]

\hspace{0.1\textwidth}
\subfloat[tildebrelzoom][Plot of $\tilde \brel$ as a function of $L/\bid=2\pi (\log\blength^{-1})^{-1}$]
{
    \psfrag{a}{$L/\bid$}
    \psfrag{b}{$\tilde \brel$}
    \includegraphics[height=30mm]{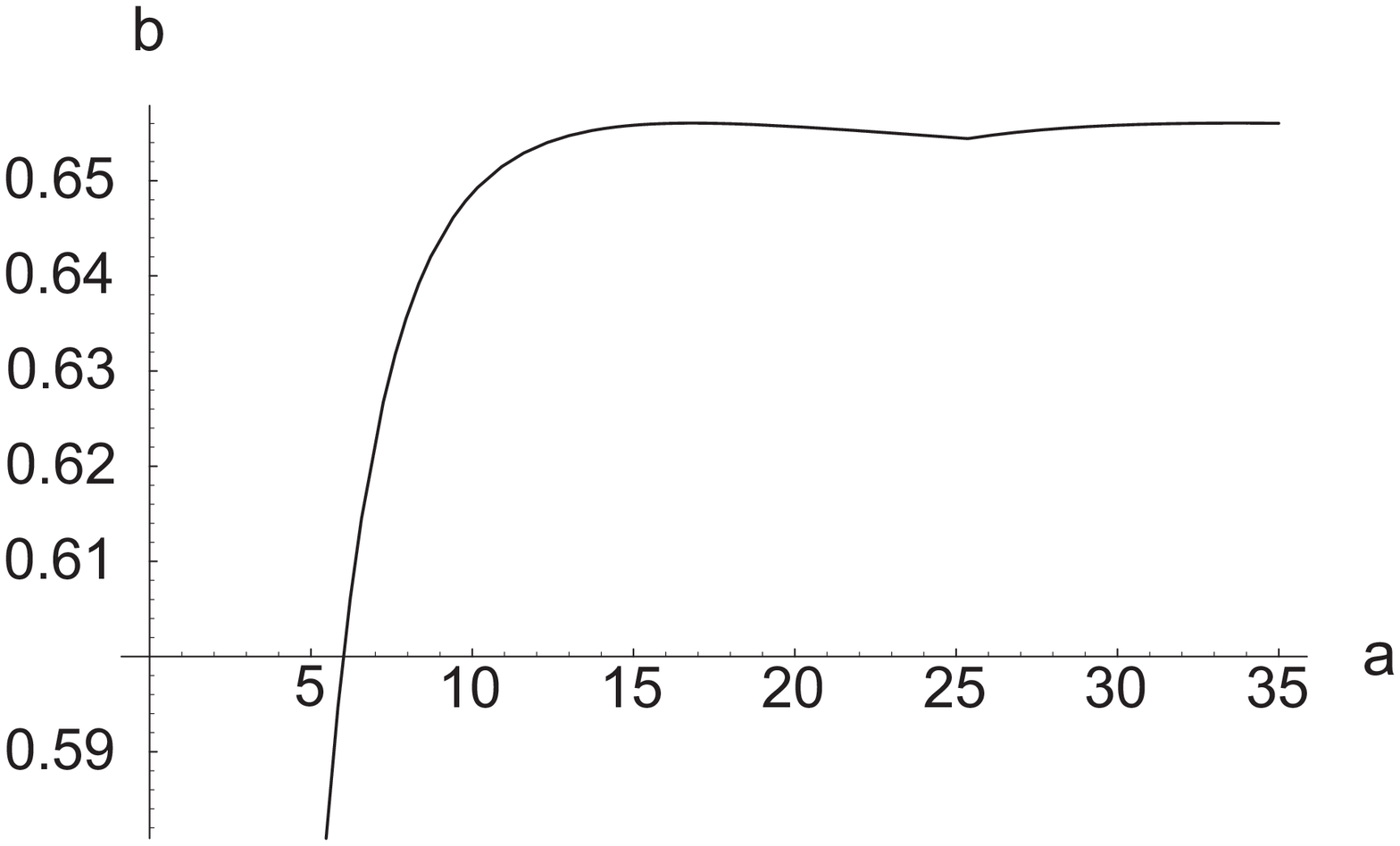}
    \label{fig:tildebrel}
}
\hspace{0.1\textwidth}
\subfloat[tildebreloscil][Plot of $\tilde \brel$ showing the small scale oscillations where different Fourier orders become the dominant contributors]
{
    \psfrag{a}{$L/\bid$}
    \psfrag{b}{$\tilde \brel$}
    \includegraphics[width=0.35\textwidth]{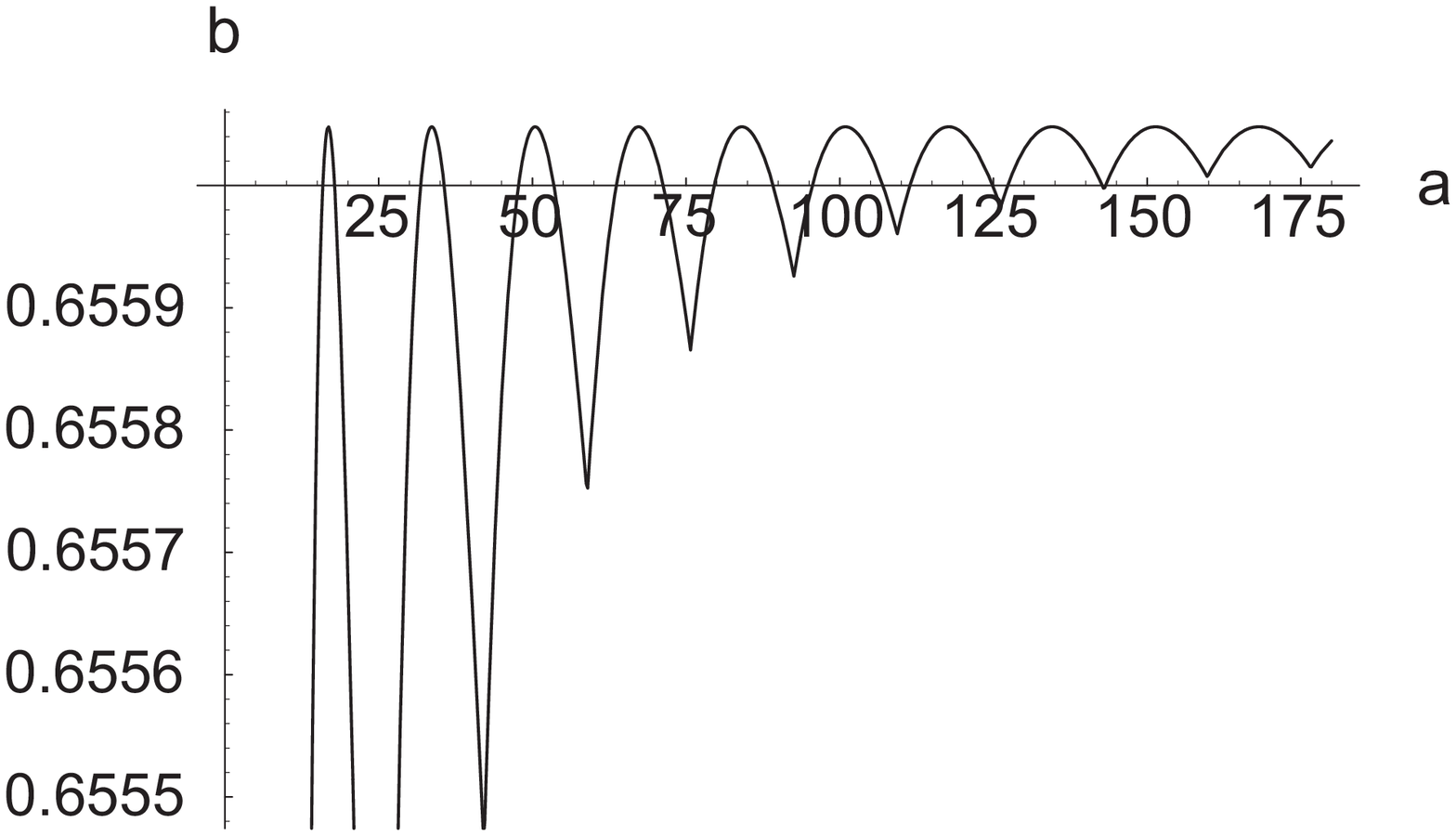}
}
\caption{For the plots of $\tilde \brel$ from Theorem~\ref{thm:stabilitybilayer} we have approximated $\tilde \brel$ by $\max_{1\leq j\leq 100}\underline{\brel}_j$} \label{fig:tildebrelsmallscale}
\end{figure}

In Sections~\ref{subsec:bilayerstability} and~\ref{subsec:monolayerstability} we found conditions for the stability of monolayers and bilayers with respect to some admissible perturbations. The main results are visualised in Figures~\ref{fig:tildemreluv} and~\ref{fig:tildemreluvsmallscale} for the monolayer and Figure~\ref{fig:tildebrelsmallscale} for the bilayer.

The monolayer is stable with respect to perturbations of the interface if $\mreluv \geq \tilde \mreluv$ (Theorem~\ref{thm:stabilitymonolayer}), and the bilayer is stable with respect to mass-preserving perturbations of the interface if $\brel \geq \tilde \brel$ (Theorem~\ref{thm:stabilitybilayer}).

$\tilde \mreluv$ and $\tilde \brel$ display very similar overall behaviour. They both rapidly increase for small values of $L/\monod$ or $L/\bid$ until they settle down around a value for $\mreluv$ or $\brel$ close to $0.6$. Around this value both $\tilde \mreluv$ and $\tilde \brel$ oscillate as with increasing $L$ different Fourier modes become dominant. The similarity is broken, however, by the restriction of $\mreluv$ to $\left[0, \frac12\right]$. Because of this the monolayer is unstable for all values of $L/\monod$ greater than about $6$ (see Figure~\ref{fig:tildemreluvrestricted}), while the bilayer can be stable for all values of $L/\bid$ (see Figure~\ref{fig:tildebrel}).

Remark that higher relative penalisation of the U-Vinterfaces, i.e. higher values of $\mreluv$ and $\brel$, improves stability. For the bilayer a sufficiently high value of $\brel$ even guarantees stability in the sense discussed here (Lemma~\ref{lem:brel1boundedawayfrom1}). This reinforces the notion that $d_{uv}$ plays a special role in the diblock copolymer-homopolymer problem, which was also encountered also in \cite[\S 2.2]{vanGennipPeletier07a}.

\section{Green's function on a periodic two-dimensional strip}\label{sec:greens}

When computing the first and second variation of $\mathcal{F}$ for monolayers and bilayers in Section~\ref{sec:perstrip} we required an explicit formula for the Green's function of $-\Delta$ on $S_L$. We present this Green's function here and prove that it satisfies the necessary conditions. For a heuristic derivation we refer to \cite[\S 6.3.1]{vanGennip08}.


\begin{theorem}\label{thm:gf}
Define $G: S_L\setminus\{(0, 0)\} \to \vz{R}$ in $L_{\text{loc}}^2(S_L)$ as follows:
\begin{equation}\label{eq:greensfunction}
G(x_1, x_2) := \frac{-1}{4 \pi} \log \left( 2 \cosh \left( \frac{2 \pi x_2}{L} \right) - 2 \cos \left( \frac{2 \pi x_1}{L} \right)\right).
\end{equation}
Then the equation $-\Delta G(x_1, x_2) = \delta(x_1, x_2)$ is satisfied with periodic boundary conditions $G(0, x_2) = G(L, x_2)$ and $\frac{\partial}{\partial x_1} G(0, x_2) = \frac{\partial}{\partial x_1} G(L, x_2)$. Writing the Fourier expansion of $G$ in $x_1$ gives
\begin{equation}\label{eq:greenfour}
G(x_1, x_2) = - \frac{1}{2 L} |x_2|  + \frac{1}{2 \pi} \sum_{q=1}^{\infty} \frac{1}{q} e^{-2 \pi |x_2| q / L} \cos\left( \frac{2 \pi x_1 q}{L} \right).
\end{equation}
\end{theorem}

\begin{proof}
We first prove that $G$, as given in equation~(\ref{eq:greensfunction}), satisfies the equation \mbox{$-\Delta G(x_1, x_2) = \delta(x_1, x_2)$} in the sense of distributions, i.e. we show that for all $\phi \in C_c^{\infty}(S_L)$,
\[
\int_{S_L} G(x_1, x_2) (-\Delta \phi)(x_1, x_2) \, dx_1 \, dx_2 = \phi(0, 0).
\]
Note that the constant term $- \frac{1}{4 \pi} \log 2$ implicitly present in (\ref{eq:greensfunction}) as the factor $2$ in the logarithm is of no importance here and so we will leave it out of subsequent calculations \footnote{The reason for adding it in (\ref{eq:greensfunction}) in the first place is to get a Fourier series without a term independent of $x_1$ and $x_2$.}.

We write
\begin{align*}
&\hspace{0.4cm}\int_{S_L} G (-\Delta \phi) \, d\mathcal{L}^2\\
  &= \lim_{\epsilon \to 0} \int_{S_L \setminus B(0,\epsilon)} G (-\Delta \phi) \, d\mathcal{L}^2\\
&= \lim_{\epsilon \to 0} \Biggl( -\int_{\partial B(0,\epsilon)} G \nabla \phi \cdot \nu \, d\mathcal{H}^1  - \int_{S_L \setminus B(0,\epsilon)} \Delta G \phi \, d\mathcal{L}^2 + \int_{\partial B(0,\epsilon)} \nabla G \cdot \nu \phi \, d\mathcal{H}^1 \Biggr),
\end{align*}
where $B(0, \epsilon)$ is the closed ball of radius $\epsilon$ and with the origin as center. $\nu$ is the unit outward normal to $S_L \setminus B(0,\epsilon)$, which means $\nu$ points into $B(0, \epsilon)$. Denote the three terms by $I_{\epsilon}, J_{\epsilon}$ and $K_{\epsilon}$ respectively. The integral $I_\e$ vanishes:
\begin{align*}
\lim_{\e\to0} |I_\e|
  &\leq \lim_{\e\to0}
   \|\nabla \phi\|_{L^\infty} \int_{\partial B(0,\epsilon)} |G|\, d\mathcal{H}^1\\
  &=\lim_{\e\to0}
   \|\nabla \phi\|_{L^\infty} \,2\pi\e\,
     \left|\log\biggl( \frac{2 \pi^2}{L^2} \Bigl(\epsilon^2 + \mathcal{O}(\epsilon^4)\Bigr) \biggr)\right| = 0.\\
\end{align*}

For $J_\e$ we calculate
\[
\nabla G(x_1, x_2) = -\frac{1}{2L} \left[ \cosh\left(\frac{2 \pi x_2}{L}\right) - \cos\left(\frac{2 \pi x_1}{L}\right) \right]^{-1} \left( \begin{array}{c} \sin \left(\frac{2 \pi x_1}{L}\right)\\ \sinh\left(\frac{2 \pi x_2}{L}\right)\end{array} \right).
\]
For notational convenience we will write $C(x_1, x_2) := \cosh\left(\frac{2 \pi x_2}{L}\right) - \cos\left(\frac{2 \pi x_1}{L}\right)$. Then we can compute that at $(x_1, x_2) \not=(0,0)$
\begin{align*}
\frac{\partial^2}{\partial x_1^2} G(x_1, x_2) &= \frac{\pi}{L^2} \left( C(x_1, x_2)^{-2} \sin^2\left(\frac{2 \pi x_1}{L}\right) - C(x_1, x_2)^{-1} \cos\left(\frac{2 \pi x_1}{L}\right) \right),\\
\frac{\partial^2}{\partial x_2^2} G(x_1, x_2) &= \frac{\pi}{L^2} \left( C(x_1, x_2)^{-2} \sinh^2\left(\frac{2 \pi x_2}{L}\right) - C(x_1, x_2)^{-1} \cosh\left(\frac{2 \pi x_2}{L}\right) \right),
\end{align*}
which gives $\Delta G(x_1, x_2) = 0$, from which it follows that $J_{\epsilon} = 0$ for all $\epsilon > 0$.

To determine $\lim_{\e\to0}K_\e$ we approximate $G$ by $G_{\R^2}(x_1,x_2) = -(4\pi)^{-1}\log(x_1^2+x_2^2)$, the Green's function of $-\Delta$ on $\R^2$. Estimating the difference on $\partial B(0,\e)$ by
\begin{align*}
\left|\nabla G(x_1,x_2) - \nabla G_{\R^2}(x_1,x_2)\right|
  &= \left| -\frac1{2L} \,
    \frac{\frac{2\pi x_1}L \vec e_1 + \frac{2\pi x_2}L \vec e_2
                              + \mathcal {O}\bigl((x_1^2+x_2^2)^{3/2}\bigr)}
          {\frac{2\pi^2}{L^2}(x_1^2+x_2^2) + \mathcal {O}\bigl((x_1^2+x_2^2)^2\bigr)}
     + \frac1{2\pi} \,
       \frac{x_1\vec e_1 + x_2 \vec e_2}{x_1^2+x_2^2}
      \right|\\
  &= \mathcal{O}\bigl((x_1^2+x_2^2)^{1/2}\bigr) \qquad\text{as }x_1^2+x_2^2 = \e^2 \to 0,
\end{align*}
we calculate
\begin{align*}
\lim_{\e\to0} K_\e
 &= \lim_{\e\to0} \int_{\partial B(0,\epsilon)}
    \nabla (G-G_{\R^2}) \cdot \nu \phi \, d\mathcal{H}^1 + \lim_{\e\to0} \int_{\partial B(0,\epsilon)}
    \nabla G_{\R^2} \cdot \nu \phi \, d\mathcal{H}^1 \\
 &= 0 + \lim_{\e\to0} \frac1{2\pi} \int_{\partial B(0,\epsilon)} \frac{x_1\vec e_1 + x_2 \vec e_2}{x_1^2+x_2^2}\cdot \frac{x_1\vec e_1 + x_2 \vec e_2}{(x_1^2+x_2^2)^{1/2}}\,\phi(x_1, x_2) \, d\mathcal{H}^1(x_1,x_2) = \phi(0,0).
\end{align*}
Taking these results together shows that $\lim_{\epsilon \downarrow 0}( I_{\epsilon} + J_{\epsilon} + K_{\epsilon}) = \phi(0, 0)$ and thus $-\Delta G = \delta$ holds in the sense of distributions.

\medskip

To prove that the Fourier series in (\ref{eq:greenfour}) corresponds to the Green's function (\ref{eq:greensfunction}), let $G$ be given by~(\ref{eq:greensfunction}) and $\tilde G$ by~(\ref{eq:greenfour}). Note that for every $x_2\not=0$ the series converges absolutely:
\[
\sum_{q=1}^{\infty} \left|  \frac1{q}e^{-\frac{2 \pi |x_2| q}L} \cos\left( \frac{2 \pi x_1 q}{L} \right)\right|
  \leq \sum_{q=1}^{\infty} \left(e^{-\frac{2 \pi |x_2|}L}\right)^q
  = \frac{e^{-\frac{2 \pi |x_2|}L}}{1-e^{-\frac{2 \pi |x_2|}L}}.
\]
If $x_2 \neq 0$ we then calculate
\begin{align}
\sum_{q=1}^{\infty} \frac1q e^{-\frac{2 \pi |x_2| q}L} \cos\left(\frac{2 \pi q x_1}L\right) &= \text{Re}\, \sum_{q=1}^{\infty} \frac1q e^{\frac{2 \pi}L (-|x_2| + i x_1)}\nonumber\\
&= -\text{Re}\log\left(1 - e^{\frac{2 \pi}L (-|x_2| + i x_1)}\right)\nonumber\\
&= -\log\big|1 - e^{\frac{2 \pi}L (-|x_2| + i x_1)}\big|\nonumber\\
&= -\frac12 \log\left( 1 - 2 e^{-\frac{2 \pi |x_2|}L} \cos\left(\frac{2 \pi x_1}L\right) +  e^{\frac{-4 \pi |x_2|}L} \right)\nonumber\\
&= -\frac12 \log\left(2 e^{-\frac{2 \pi |x_2|}L} \left( \frac12 e^{\frac{2 \pi |x_2|}L} - \cos\left(\frac{2 \pi x_1}L\right) + \frac12 e^{-\frac{2 \pi |x_2|}L} \right)\right)\nonumber\\
&= \frac{\pi}L |x_2| - \frac12 \log\left(2 \cosh\left(\frac{2 \pi |x_2|}L\right) - 2 \cos\left(\frac{2 \pi x_1}L\right)\right),\label{eq:greenfunctionandseries}
\end{align}
so that
the partial sums $\sum_{q=1}^{\ell} \frac1q e^{-\frac{2 \pi |x_2| q}L} \cos\left(\frac{2 \pi q x_1}L\right)$ converge pointwise to $2 \pi G(x_1,x_2) + \frac{\pi}L |x_2|$ for almost all $(x_1,x_2)\in S_L$. Since the partial sums  are all bounded by the $L_{\text{loc}}^2$-function on the right hand side of (\ref{eq:greenfunctionandseries}) the Dominated Convergence Theorem yields $\tilde G \in L^2(S_L)$. Together with $G = \tilde G$ a.e. on $S_L$ this shows that $G = \tilde G$ in $L_{\text{loc}}^2(S_L)$.
\end{proof}


\begin{corol}\label{cor:Greenproperty}
Let $G$ be as in (\ref{eq:greenfour}) and let $x_2 \in \vz{R}\setminus\{0\}$. Then
\[
\int_0^L G(x_1, x_2) \, dx_1 = -\frac12 |x_2|.
\]
\end{corol}
\begin{proof}
For all $q \geq 1$,
\[
\int_0^L \cos\left( \frac{2 \pi q x_1}{L} \right) \, dx_1 = 0.
\]
\end{proof}

Also note that $G(-x_1, x_2) = G(x_1, x_2)$ and $G(x_1, -x_2) = G(x_1, x_2)$.

\section{Discussion and conclusions}\label{sec:concdisc}

\subsection{Comparing mono- and bilayers} In this paper we showed that bilayers can be both stable and unstable, depending on the parameters: when the U-V interface penalty is strong enough relative to the penalties of the other interfaces, the bilayer is stable. On the other hand, monolayers are unstable as soon as the strip is wide enough to accommodate the unstable wavelengths, regardless of the values of the interface penalisation.

The bilayer can be thought of as two juxtaposed monolayers, and therefore the question presents itself how the unstable mode of the monolayer is prevented in the bilayer context. The correct answer seems to be that the unstable mode is actually not prevented at all; it continues to exist in the context of the bilayer, as can be witnessed in Figures~\ref{fig:monounstab2} and (especially)~\ref{fig:biunstab3}.

The reason why this unstable mode does not make every bilayer unstable lies in the admissible values of the coefficients, which are different in the two cases. For the VUV bilayer, for instance, the value of the U-0 interface penalty $d_{u0}$ is irrelevant; therefore, by choosing $d_{u0} := d_{v0}+d_{uv}$, every choice of $d_{uv}$ and $d_{v0}$ becomes admissible, and most importantly, the case of purely U-V penalisation ($\zeta \approx 1$, or $d_{v0}\approx 0$) is therefore allowed. For the monolayer, however, the conditions~\pref{eq:ddemands} imply that the two side interfaces (0-U and V-0) are necessarily penalised at least half as strongly as the central (U-V) interface. Most of the white (stable) region in Figure~\ref{fig:E2contour_cd} therefore is inaccessible, and only the unstable region remains as can be seen in Figure~\ref{fig:E2mrelva} (Figures~\ref{fig:E2mrelv} only show stability of the first Fourier mode, but the situation is similar for the higher modes).

\subsection{Comparison with~\cite{PeletierRoeger08}}

In previous work~\cite{PeletierRoeger08} one of the authors (Peletier) and R{\"o}ger studied a related functional, \begin{equation}\label{eq:lipidbil}
\mathcal{G}_{\epsilon}(u, v) := \left\{ \begin{array}{ll} \displaystyle \epsilon \int_{\R^2} |\nabla u| + \frac{1}{\epsilon} d_1(u,
v) & \mbox
{ if $(u, v) \in \mathcal{K}_{\epsilon}$,}\vspace{0.25cm}\\ \infty &\mbox{ otherwise.} \end {array} \right.
\end{equation}
Here
$d_1(\cdot,\cdot)$ is the Monge-Kantorovich distance with cost function $c(x,y) = |x-y|$, \cite{Villani08}, and
\[
\mathcal{K}_{\epsilon} := \left\{ (u, v) \in \text{BV}(\vz{R}^2; \{0,
1/\epsilon\}) \times L^1(\R^2; \{0, 1/\epsilon\}) : uv = 0 \text{ a.e., and }\int_{\R^2} u = \int_{\R^2} v = M \right\}.
\]
Apart from the choices $c_0 = c_v = 0$ and $c_u = 1$, the main difference between $\mathcal{F}$ and (\ref{eq:lipidbil}) is the different non-local term.

The scaling (constant mass but increasing amplitude $1/\e$) implies that the supports of $u$ and $v$ shrink to zero measure. The main goal in~\cite{PeletierRoeger08} was to investigate the limit $\e\to0$ and characterise the limiting structures and their energy.

The main result, a $\Gamma$-convergence theorem, can be interpreted as stating---in a very weak sense---that the limiting structures are VUV bilayers; in the limit $\e\to0$ these bilayers have a thickness equal to $4\e$ and their curvature is bounded in $L^2$. Most importantly, in connection with the present paper, the limit energy depends on the curvature in a stable way: the energy is minimal for straight bilayers and increases with curvature.

This result compares well with the results of this paper. The functional $\mathcal G_\e$ of~\cite{PeletierRoeger08} penalises only U-V and U-0 interfaces; the V-0 interface is free, or in terms of this paper $\zeta=1$. Both in~\cite{PeletierRoeger08} and in the present paper we therefore find that bilayers of optimal width are stable, although the precise results and their methods of proof are very different.

\subsection{Comparison with `wriggled lamellar' solutions}\label{sec:wriggledlamellar}

In a series of papers~\cite{Muratov02,RenWei03b,RenWei05} Muratov and Ren \& Wei investigate the stability of one-dimensional layered (lamellar) structures for copolymer melts---the case $u+v\equiv 1$. They find that for a critical value of the lamellar spacing the straight lamellar structures become unstable and a stable branch of curved, `wriggled'
lamellar structures bifurcates. Muratov considers unbounded domains and finds that the loss of stability happens at \emph{exactly} the optimal value of the width: for any larger value of the width unstable directions exist with  very large wavelength. Ren and Wei consider bounded domains, which provides a natural limit on the wavelength of perturbations, and consequently they find that at the optimal width the straight lamellar structures are stable, and the bifurcation occurs at slightly larger width.

The system studied in this paper is different in that there are three types of interfaces, not one; for comparison purposes one can identify the pure-melt case described above with the case of pure U-V interface penalisation for bilayers ($\brel=1$). In this case the bilayer of optimal width is stable, and this result mirrors the stability result of Ren and Wei for optimal-width lamellar structures.

\subsection{Generalizations and extensions} One might wonder whether the functional $\mathcal F$ depends in a smooth manner on the perturbations. The calculation of the second derivative of the functional in the melt case done by Choksi and Sternberg~\cite{ChoksiSternberg06} suggests that the second derivative of $\mathcal F$ depends continuously on $W^{1,2}$-regular perturbations of the interfaces. In that case the functional $\mathcal F$ is of class $C^2$, and the linear stability analysis of the current paper automatically implies the equivalent nonlinear stability properties.

One can also wonder whether the class of perturbations that are considered---those described by functions of the variable $x_1\in \mathbb T_L$---is not too restrictive. The class of all perturbations that are small in $L^1$, for instance, also includes many perturbations with small inclusions of one phase in another, which are not covered here. We believe that these will generally be less advantageous, since the results of this chapter show that perturbations with fast oscillations are energetically expensive (because the layers are stable with respect to the admissible perturbations for most values of the surface tension coefficients if $L$ is small). The same conclusion can be reached by a slightly different, heuristic argument as follows. Within the class of uniformly bounded functions the $H^{-1}$-norm is continuous with respect to the $L^1$-topology, as can be seen from
\[
\|f\|_{H^{-1}}^2 = \int f\varphi \leq \|f\|_{L^2} \|\varphi\|_{L^2} \leq C \|f\|_{L^2}^2 \leq C \|f\|_{L^1} \|f\|_{L^{\infty}},
\]
where $\varphi$ solves $-\Delta \varphi = f$. Therefore within that class of functions the $H^{-1}$-norm is also continuous with respect to the area of the inclusion; for small inclusions, with a large circumference-to-area ratio, a possible decrease in the $H^{-1}$-norm is thus dwarfed by the increase in interfacial length associated with such an inclusion.


Note that the problem has not completely been non-dimensionalised; it is possible to rescale the problem by the length scale $L$, resulting in a three-parameter problem (in the rescaled parameters $c_0$, $c_u$, and $c_v$). Instead we keep the length scale explicitly in the problem to illustrate the length-scale dependence of the stability properties.

\subsection{Diffuse interface model}

The functional $\mathcal{F}$ is the sharp interface limit (via $\Gamma$-convergence) of a well-known diffuse-interface functional~\cite{ModicaMortola77,Baldo90}
\[
\mathcal{F}_{\epsilon}(u, v) = \int \Bigl[ \frac{\epsilon}2 |\nabla u|^2 + \frac{\epsilon}2 |\nabla v|^2 + \frac{\epsilon}2 |\nabla (u+v)|^2 + \epsilon^{-1} W(u, v) \Bigr] \, dx + \frac12 \| u - v\|_{H^{-1}}^2.
\]
Here $W$ is a triple-well potential with wells at $(0,0)$, $(1,0)$, and $(0,1)$. The coefficients $d_{uv}$, $d_{u0}$, and $d_{v0}$ in the sharp interface limit depend on the specific form of $W$ via 
\[
d_{kl} := 2 \inf\left\{ \int_0^1 \sqrt{W(\gamma(t))} |\gamma'(t)|\,dt: \gamma\in C^1([0,1]; (\R_+)^n), \gamma(0)=\alpha_k, \gamma(1)=\alpha_l \right\},
\]
where $\alpha_u = (1,0)$, $\alpha_v=(0,1)$, and $\alpha_0=(0,0)$.

By the properties of $\Gamma$-convergence minimisers of $\mathcal{F}_{\epsilon}$ converge to minimisers of $\mathcal{F}$~\cite[Corollary~7.17]{DalMaso93}. Therefore our results indicate that in the regions of their respective instability monolayers and bilayers are not minimisers for $\mathcal{F}_{\epsilon}$ for small $\epsilon$.

\appendix

\section{Relevance of energy per unit mass for partial localisation}\label{sec:energy-per-unit-mass}

Throughout this paper we concentrate on layered structures with a specific width: the width that minimises the ratio of (one-dimensional) energy to (one-dimensional) mass. The origin for this choice lies in our interest in partially localised structures, as we now explain.

Since we are interested in long thin structures, we might first ask ourself the question what minimisers of $\mathcal{F}$ on the full domain $\R^2$ look like if we restrict the admissible functions to be rectangles with a fixed mass, oriented such that the long axis is parallel to the $x_1$-axis.

If the rectangle has a large aspect ratio, the structure is roughly constant in the $x_1$-direction. We can interpret the rectangle then as a one-dimensional structure in the $x_2$-direction, extended trivially in the $x_1$-direction and cut off at a certain length, $a$. In \cite{vanGennipPeletier07a} it is proven that for such a trivially extended one-dimensional structure the energy $\mathcal{F}$ per mass $M$ is approximately equal to the one-dimensional energy $F_{1\text{D}}$ per mass of the cross-section $M_{1\text{D}}$:
\[
\frac{\mathcal{F}}{M} = \frac{F_{1\text{D}}}{M_{1\text{D}}} + \mathcal{O}(1/a), \text{ for } a \to \infty.
\]
Put differently: although the energy depends on the structure in a nonlocal manner, for large mass (i.e. long rectangles) the energy is essentially equal to the one-dimensional energy of the cross-section times the length of the rectangle. Effects near the cut off points are less important.

This implies that the miminiser of $\mathcal{F}$ in the class of rectangles with large constrained mass should have a thickness $M_{1\text{D}}$ such that $F_{1\text{D}}/M_{1\text{D}}$ is minimal. Also when studying the stability of layered structures, it thus makes sense to concentrate on structures of optimal width, in the sense as described above.

In a monolayer of optimal width the U- and V-layers both have width~\cite{vanGennipPeletier07a}
\[
\monod := \left(\frac32\right)^{1/3} (c_0+c_u+c_v)^{1/3},
\]
while for the bilayer the thickness of the inner layer is
\[
2 \bid := 6^{1/3} (c_0+c_u+2c_v)^{1/3} \text{ (VUV)} \qquad \text{or} \qquad 2 \bid := 6^{1/3} (c_0+2c_u+c_v)^{1/3} \text{ (UVU)}.
\]

\section{Relevance of the choice $c_u = c_v$ for monolayers}\label{sec:cu=cv}

The choice $c_u=c_v$ for monolayers is similarly inspired by our interest in partial localisation and more or less forced upon us by the periodicity in the $x_1$-direction. If the U-0 and V-0 interfaces are penalised unequally, then a monolayer structure in $\R^2$ likely will tend to curve, in order to reduce the length of the `expensive' interface at the expense of the `cheap' interface.

When $c_u\not= c_v$, therefore, a straight monolayer is not even stationary under perturbations that allow for curving of the whole monolayer. The setup in the context of the strip $S_L$ disallows such curving over the whole length of the layer because of the periodicity in the $x_1$-direction. Therefore this instationarity is rendered invisible on $S_L$. However, with our interest in partial localisation in mind we make the choice $c_u=c_v$ in the case of monolayers throughout this paper.

\section{Proof of Theorem~\ref{thm:bilayersecondvar}}\label{app:proofbilayersecondvar}
For the interfacial terms we directly compute from (\ref{eq:inteps})
\begin{multline}\label{eq:secvar3}
\left. \frac{d^2}{d \epsilon^2} \left(c_0 \int_{S_L} |\nabla (u_\epsilon + v_\epsilon)| + c_u \int_{S_L} |\nabla u_\epsilon| + c_v \int_{S_L} |\nabla v_\epsilon| \right)\right|_{\epsilon = 0} = \\
\int_0^L \left( d_{uv} \left[ {p_1'}^2 + {p_3'}^2 \right] + d_{v0} \left[ {p_2'}^2 + {p_4'}^2 \right]\right)\, dx.
\end{multline}
In order to compute $\left. \frac{d^2}{d \epsilon^2} \|u_{\epsilon} - v_{\epsilon}\|_{H^{-1}(S_L)}^2\right|_{\epsilon=0}$ we split up the norm as follows:
\begin{equation}\label{eq:secondvarnormbilayer}
\|u_{\epsilon} - v_{\epsilon}\|_{H^{-1}(S_L)}^2 = \int_0^L \int_0^L f_{\epsilon}(x_1, \xi_1)\,d\xi_1\,dx_1,
\end{equation}
where
\begin{align}
&f_{\epsilon}(x_1, \xi_1) :=\nonumber\\
&\hspace{0.4cm}\int_{-2\bid-\epsilon p_4(x_1)}^{-\bid-\epsilon p_3(x_1)} \int_{-2\bid-\epsilon p_4(\xi_1)}^{-\bid-\epsilon p_3(\xi_1)} G(x-\xi)\,d\xi_2\,dx_2
  + \int_{-\bid-\epsilon p_3(x_1)}^{\bid+\epsilon p_1(x_1)} \int_{-\bid-\epsilon p_3(\xi_1)}^{\bid+\epsilon p_1(\xi_1)} G(x-\xi)\,d\xi_2\,dx_2\nonumber\\
&+ \int_{\bid+\epsilon p_1(x_1)}^{2\bid+\epsilon p_2(x_1)} \int_{\bid+\epsilon p_1(\xi_1)}^{2 \bid+\epsilon p_2(\xi_1)} G(x-\xi)\,d\xi_2\,dx_2\nonumber\\
&- 2\int_{-\bid-\epsilon p_3(x_1)}^{\bid+\epsilon p_1(x_1)} \int_{-2\bid-\epsilon p_4(\xi_1)}^{-\bid-\epsilon p_3(\xi_1)} G(x-\xi)\,d\xi_2\,dx_2 - 2\int_{\bid+\epsilon p_1(x_1)}^{2\bid+\epsilon p_2(x_1)} \int_{-\bid-\epsilon p_3(\xi_1)}^{\bid+\epsilon p_1(\xi_1)} G(x-\xi)\,d\xi_2\,dx_2\nonumber\\
&+ 2\int_{\bid+\epsilon p_1(x_1)}^{2\bid+\epsilon p_2(x_1)} \int_{-2\bid-\epsilon p_4(\xi_1)}^{-\bid-\epsilon p_3(\xi_1)} G(x-\xi)\,d\xi_2\,dx_2
\label{eq:fepsilon}
\end{align}
We compute now one of these terms in its general form. Let $n_1, n_2, n_3, n_4 \in \{-2, -1, 1, 2\}$, $r_1, r_2 \in \{p_1(x_1), p_2(x_1), -p_3(x_1), -p_4(x_1)\}$ and $r_3, r_4 \in \{p_1(\xi_1), p_2(\xi_1), -p_3(\xi_1), -p_4(\xi_1)\}$ (with $n_1<n_2<n_3<n_4$ and $r_1<r_2<r_3<r_4$), then we want to compute
\[
I = \left.\frac{d^2}{d\epsilon^2} \int_{n_1 \bid + \epsilon r_1}^{n_2 \bid + \epsilon r_2} \int_{n_3 \bid + \epsilon r_3}^{n_4 \bid + \epsilon r_4} G(\cdot, x_2-\xi_2)\,d\xi_2\,dx_2\right|_{\epsilon=0}.
\]
We can split up the integral over $[n_1 \bid + \epsilon r_1, n_2 \bid + \epsilon r_2] \times [n_3 \bid + \epsilon r_3, n_4 \bid + \epsilon r_4]$ into nine integrals over the domains
\begin{align*}
&[n_2 \bid, n_2 \bid + \epsilon r_2] \times [n_3 \bid + \epsilon r_3, n_3 \bid],&\quad [n_2 \bid, n_2 \bid + \epsilon r_2] \times [n_3 \bid, n_4 \bid],\\
&[n_2 \bid, n_2 \bid + \epsilon r_2] \times [n_4 \bid, n_4 \bid + \epsilon r_4],&\quad [n_1 \bid, n_2 \bid] \times [n_3 \bid + \epsilon r_3, n_3 \bid],\\
&[n_1 \bid, n_2 \bid] \times [n_3 \bid, n_4 \bid],&\quad [n_1 \bid, n_2 \bid] \times [n_4 \bid, n_4 \bid + \epsilon r_4],\\
&[n_1 \bid + \epsilon r_1, n_1 \bid] \times [n_3 \bid + \epsilon r_3, n_3 \bid],&\quad [n_1 \bid + \epsilon r_1, n_1 \bid] \times [n_3 \bid, n_4 \bid],\\
&[n_1 \bid + \epsilon r_1, n_1 \bid] \times [n_4 \bid, n_4 \bid + \epsilon r_4].&
\end{align*}
We compute two of these integrals. The others are computed in a similar vein. $G_{,2}$ denotes the partial derivative of $G$ with respect to its second argument.
\begin{align*}
&\hspace{0.4cm}\left.\frac{d^2}{d\epsilon^2} \int_{n_2 \bid}^{n_2 \bid + \epsilon r_2} \int_{n_3 \bid + \epsilon r_3}^{n_3 \bid} G(\cdot, x_2-\xi_2)\,d\xi_2\,dx_2\right|_{\epsilon=0}\\
&= \left.\frac{d^2}{d\epsilon^2} \int_0^{r_2} \int_{r_3}^0 \epsilon^2 G(\cdot, \epsilon (\tilde x_2 - \tilde \xi_2) + (n_2-n_3)\bid)\,d\tilde \xi_2\,d\tilde x_2\right|_{\epsilon=0} \\
&=\left.\frac{d}{d\epsilon} \int_0^{r_2} \int_{r_3}^0 \left[2 \epsilon G(\cdot, \epsilon (\tilde x_2 - \tilde \xi_2) + (n_2-n_3) \bid) + \epsilon^2 (\tilde x_2 - \tilde \xi_2) G_{,2}(\cdot, \epsilon (\tilde x_2 - \tilde \xi_2) + (n_2-n_3) \bid)\right]\,d\tilde \xi_2\,d\tilde x_2\right|_{\epsilon=0}\\
&= 2 \int_0^{r_2} \int_{r_3}^0 G(\cdot, (n_2-n_3) \bid) \,d\tilde \xi_2\,d\tilde x_2\\
&= -2 r_2 r_3 G\bigl(\cdot, (n_2-n_3) \bid\bigr).
\end{align*}
Another kind of integral we encounter is
\begin{align*}
&\hspace{0.4cm}\left.\frac{d^2}{d\epsilon^2} \int_{n_2 \bid}^{n_2 \bid + \epsilon r_2} \int_{n_3 \bid}^{n_4 \bid} G(\cdot, x_2-\xi_2)\,d\xi_2\,dx_2\right|_{\epsilon=0}\\
&= \left.-\frac{d^2}{d\epsilon^2} \int_0^{r_2} \int_{(n_2-n_3) \bid}^{(n_2-n_4) \bid} \epsilon G(\cdot, \epsilon \tilde x_2 + \tilde \xi_2)\,d\tilde \xi_2\,d\tilde x_2 \right|_{\epsilon=0}\\
&= -\left.\frac{d}{d\epsilon} \int_0^{r_2} \int_{(n_2-n_3) \bid}^{(n_2-n_4) \bid} \left[ G(\cdot, \epsilon \tilde x_2 + \tilde \xi_2) + \epsilon \tilde x_2 G_{,2}(\cdot, \epsilon \tilde x_2 + \tilde \xi_2)\right]\,d\tilde \xi_2\,d\tilde x_2\right|_{\epsilon=0}\\
&= -\int_0^{r_2} \int_{(n_2-n_3) \bid}^{(n_2-n_4) \bid} 2 \tilde x_2 G_{,2}(\cdot, \tilde \xi_2)\,d\tilde \xi_2\,d\tilde x_2\\
&= r_2^2 \Bigl( G\bigl((n_2-n_3) \bid\bigr) - G\bigl((n_2-n_4) \bid\bigr) \Bigr).
\end{align*}
Combining all integrals we find
\begin{align*}
I &= -2 r_2 r_3 G\bigl(\cdot, (n_2-n_3) \bid\bigr) + r_2^2 \Bigl( G\bigl(\cdot, (n_2-n_3) \bid\bigr) - G\bigl(\cdot, (n_2-n_4) \bid\bigr)\Bigr) + 2 r_2 r_4 G\bigl(\cdot, (n_2-n_4) \bid\bigr)\\
&+ r_3^2 \Bigl( G\bigl(\cdot, (n_2-n_3) \bid\bigr) - G\bigl(\cdot, (n_1-n_3) \bid\bigr)\Bigr) - r_4^2 \Bigl( G\bigl(\cdot, (n_2-n_4) \bid\bigr) - G\bigl(\cdot, (n_1-n_4) \bid\bigr)\Bigr)\\
&+ 2 r_1 r_3 G\bigl(\cdot, (n_1-n_3) \bid\bigr) + r_1^2 \Bigl( G\bigl(\cdot, (n_1-n_4) \bid\bigr) - G\bigl(\cdot, (n_1-n_3) \bid\bigr)\Bigr) - 2 r_1 r_4 G\bigl(\cdot, (n_1-n_4) \bid\bigr).
\end{align*}
Applying this result to (\ref{eq:fepsilon}) while keeping in mind that $G(-x_1, \cdot) = G(x_1, \cdot)$ and $G(\cdot, -x_2) = G(\cdot, x_2)$, we find
\begin{align*}
f_{\epsilon}(x_1, \xi_1) &= \Bigl[ -8 p_1^2(x_1) + 8 p_1(x_1) p_1(\xi_1) - 2 p_2^2(x_1) + 2 p_2(x_1) p_2(\xi_1)\Bigr.\\
 &\hspace{0.65cm}\Bigl.- 8 p_3^2(x_1) + 8 p_3(x_1) p_3(\xi_1) - 2 p_4^2(x_1) + 2 p_4(x_1) p_4(\xi_1)\Bigr] G(x_1-\xi_1, 0)\\
 &\hspace{0.3cm} + \Bigl[ 4 p_1(x_1) - 8 p_1(x_1) p_2(\xi_1) + 4 p_2^2(x_1) + 4 p_3^2(x_1) - 8 p_3(x_1) p_4(\xi_1) + 4 p_4^2(x_1)\Bigr] G(x_1-\xi_1, \bid)\\
 &\hspace{0.3cm} + \Bigl[ 8 p_1^2(x_1) + 16 p_1(x_1) p_3(\xi_1) + 8 p_3^2(x_1)\Bigr] G(x_1-\xi_1, 2\bid)\\
 &\hspace{0.3cm} - \Bigl[ 4 p_1^2(x_1) + 4 p_2^2(x_1) + 8 p_2(x_1) p_3(\xi_1) + 4 p_3^2(x_1) + 8 p_1(x_1) p_4(\xi_1) + 4 p_4^2(\xi_1)\Bigr] G(x_1-\xi_1, 3\bid)\\
 &\hspace{0.3cm} + \Bigl[ 2 p_2^2(x_1) + 4 p_2(x_1) p_4(\xi_1) + 2 p_4^2(x_1)\Bigr] G(x_1-\xi_1, 4 \bid),
\end{align*}
where we have used that in (\ref{eq:secondvarnormbilayer}) the integrations over $x_1$ and $\xi_1$ are indistinguishable.

Note that, by Corollary~\ref{cor:Greenproperty}, for $\xi \in \vz{T}_L, r \in \vz{R}$
\[
\int_0^L G(x - \xi, r) \, dx = \int_0^L G(x, r) \, dx = - \frac12 |r|.
\]
Using this, as well as
equations (\ref{eq:Parseval}--\ref{eq:parsconv}) and the equality $\overline{\hat G(q, r)} = \hat G(q, r)$ for $r \in \vz{R}$, we find
\begin{align}
\left. \frac{d^2}{d \epsilon^2} \|u_{\epsilon} - v_{\epsilon}\|_{H^{-1}(S_L)}^2\right|_{\epsilon=0} &= -4 \bid \sum_{q \in \vz{Z}} \left( |\hat p_1(q)|^2 + |\hat p_3(q)|^2 \right)\nonumber\\
&\hspace{0.4cm} + L^{\frac12} \sum_{q \in \vz{Z}} \left[ \bigl\{8 |\hat p_1(q)|^2  + 2 |\hat p_2(q)|^2 + 8 |\hat p_3(q)|^2 + 2 |\hat p_4(q)|^2\bigr\} \hat G(q, 0)\right.\nonumber\\
&\hspace{2cm} - 8 \bigl\{ \hat p_1(q) \overline{\hat p_2}(q) + \hat p_3(q) \overline{\hat p_4}(q)\bigr\} \hat G(q, \bid)\nonumber\\
&\hspace{2cm} + 16 \hat p_1(q) \overline{\hat p_3}(q) \hat G(q, 2 \bid)\nonumber\\
&\hspace{2cm} - 8 \bigl\{ \hat p_2(q) \overline{\hat p_3}(q) + \hat p_1(q) \overline{\hat p_4}(q) \bigr\} \hat G(q, 3 \bid)\nonumber\\
&\hspace{2cm} \left.+ 4 \hat p_2(q) \overline{\hat p_4}(q) \hat G(q, 4 \bid)\right].\label{eq:svarofnormblyr}
\end{align}

Adding the results (\ref{eq:secvar3}) and (\ref{eq:svarofnormblyr}), we get
\begin{align*}
\left.\frac{d^2}{d \epsilon^2} \mathcal{F}(u_{\epsilon}, v_{\epsilon})\right\vert_{\epsilon=0} &= \int_0^L \left( d_{uv} \left[ {p_1'}^2 + {p_3'}^2 \right] + d_{v0} \left[ {p_2'}^2 + {p_4'}^2 \right]\right)\, dx\\
&+ L^{\frac12}  \sum_{q \in \vz{Z}} \left[ \bigl\{8 |\hat p_1(q)|^2  + 2 |\hat p_2(q)|^2 + 8 |\hat p_3(q)|^2 + 2 |\hat p_4(q)|^2\bigr\} \hat G(q, 0)\right.\\
&\quad\quad\quad\quad\quad\quad\quad- 8 \bigl\{ \hat p_1(q) \overline{\hat p_2}(q) + \hat p_3(q) \overline{\hat p_4}(q)\bigr\} \hat G(q, \bid)\\
&\quad\quad\quad\quad\quad\quad\quad+ 16 \hat p_1(q) \overline{\hat p_3}(q) \hat G(q, 2 \bid)\\
&\quad\quad\quad\quad\quad\quad\quad- 8 \bigl\{ \hat p_2(q) \overline{\hat p_3}(q) + \hat p_1(q) \overline{\hat p_4}(q) \bigr\} \hat G(q, 3 \bid)\\
&\quad\quad\quad\quad\quad\quad\quad\left.+ 4 \hat p_2(q) \overline{\hat p_4}(q) \hat G(q, 4 \bid)\right]\\
&-4\bid \sum_{q \in \vz{Z}} \left( |\hat p_1(q)|^2 + |\hat p_3(q)|^2 \right).
\end{align*}

Because we have, for all $q, \tilde q \in \vz{N}$,
\begin{align*}
\frac2L \int_0^L \sin \left( \frac{2 \pi x q}{L} \right) \sin \left( \frac{2 \pi x \tilde q}{L} \right) \, dx &= \frac2L \int_0^L \cos \left( \frac{2 \pi x q}{L} \right) \cos \left( \frac{2 \pi x \tilde q}{L} \right) \, dx = \delta_{q \tilde q},\\
\frac2L \int_0^L \sin \left( \frac{2 \pi x q}{L} \right) \cos \left( \frac{2 \pi x \tilde q}{L} \right) \, dx &= 0,
\end{align*}
the integral over the derivatives  in the second variation gives us
\begin{align*}
\sum_{j=1}^{\infty} \left(\frac{2 \pi j}{L}\right)^2 &\left[ d_{uv} \left\{  \left( a_{1,j} \right)^2 + \left( a_{3,j} \right)^2 + \left( b_{1,j} \right)^2 + \left( b_{3,j} \right)^2 \right\}\right.\\
&\left.\hspace{0.2cm}+ d_{v0} \left\{ \left( a_{2,j} \right)^2 + \left( a_{4,j} \right)^2 + \left( b_{2,j} \right)^2 + \left( b_{4,j} \right)^2 \right\} \right].
\end{align*}

Because $p$ is $\vz{R}^4$-valued, $\overline{\hat p(q)} = \hat p(-q)$. Furthermore $\hat G(-q, x_2) = \hat G(q, x_2)$ by equation (\ref{eq:greensfunction}). This enables us to write terms as follows, for $k, l \in \{1, 2, 3, 4\}$:
\[
\sum_{q \in \vz{Z}} \hat p_k(q) \overline{\hat p_l(q)} \hat G(q, x_2) = \hat p_k(0) \hat p_l(0) \hat G(0, x_2) + 2 \text{Re}\, \sum_{q=1}^{\infty} \hat p_k(q) \overline{\hat p_l(q)} \hat G(q, x_2).
\]
Note that, for $k \in \{1, 2, 3, 4\}$, $q \in \vz{Z}\setminus\{0\}$, we have $\hat p_k(q) = \frac{1}{\sqrt{2}} \left(a_q^{(k)} - i b_q^{(k)}\right)$ and $\overline{\hat p_k(q)} = \frac{1}{\sqrt{2}} \left(a_q^{(k)} + i b_q^{(k)}\right)$ and thus, for $l \in \{1, 2, 3, 4\}$,
\[
\text{Re}\, \hat p_k(q) \overline{\hat p_l(q)} = \frac12 \left( a_q^{(k)} a_q^{(l)} + b_q^{(k)} b_q^{(l)} \right).
\]
From the Fourier series (\ref{eq:greenfour}) we get furthermore that
\begin{align*}
\hat G(0, x_2) &=  - \frac{1}{2 \sqrt{L}} |x_2|,\\
\hat G(q, x_2) &=  \frac{\sqrt{L}}{4 \pi q} e^{- 2 \pi |x_2| q / L}, \text{ for } q \geq 1.
\end{align*}
Using these results in the expression for the second variation yields the desired result.

\section{Detailed calculations in the proof of Lemma~\ref{lem:B1negeig}}\label{app:details}

\begin{figure}[ht]
\hspace{0.1\textwidth}
\subfloat[brel1][Plot of $\brel_1$]
{
    \psfrag{a}{$\blength$}
    \psfrag{b}{$\brel_1$}
    \includegraphics[width=0.35\textwidth]{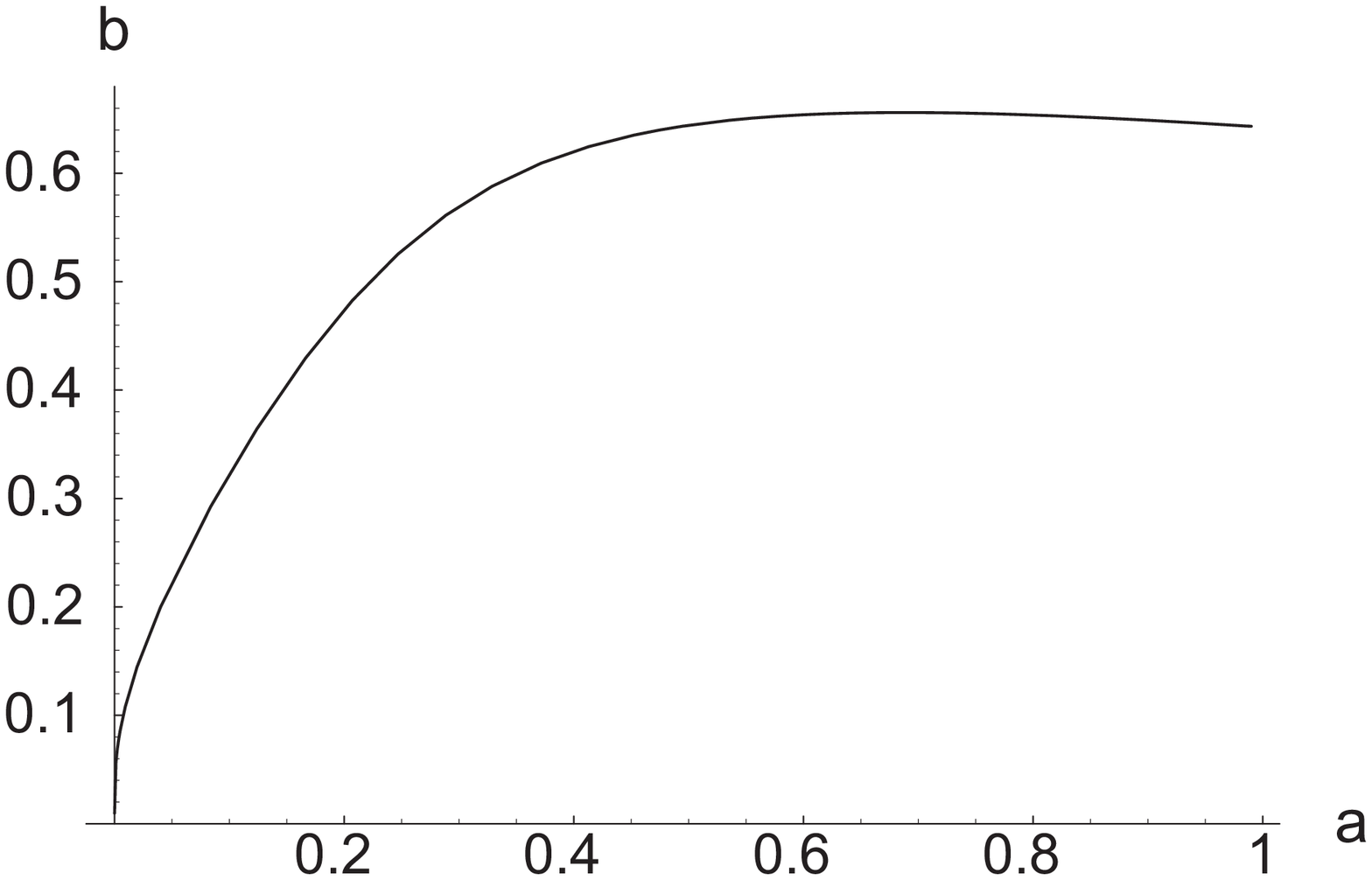}\label{fig:brel1}\\

}
\hspace{0.1\textwidth}
\subfloat[brel2][Plot of $\brel_2$]
{
    \psfrag{a}{$\blength$}
    \psfrag{b}{$\brel_2$}
    \includegraphics[width=0.35\textwidth]{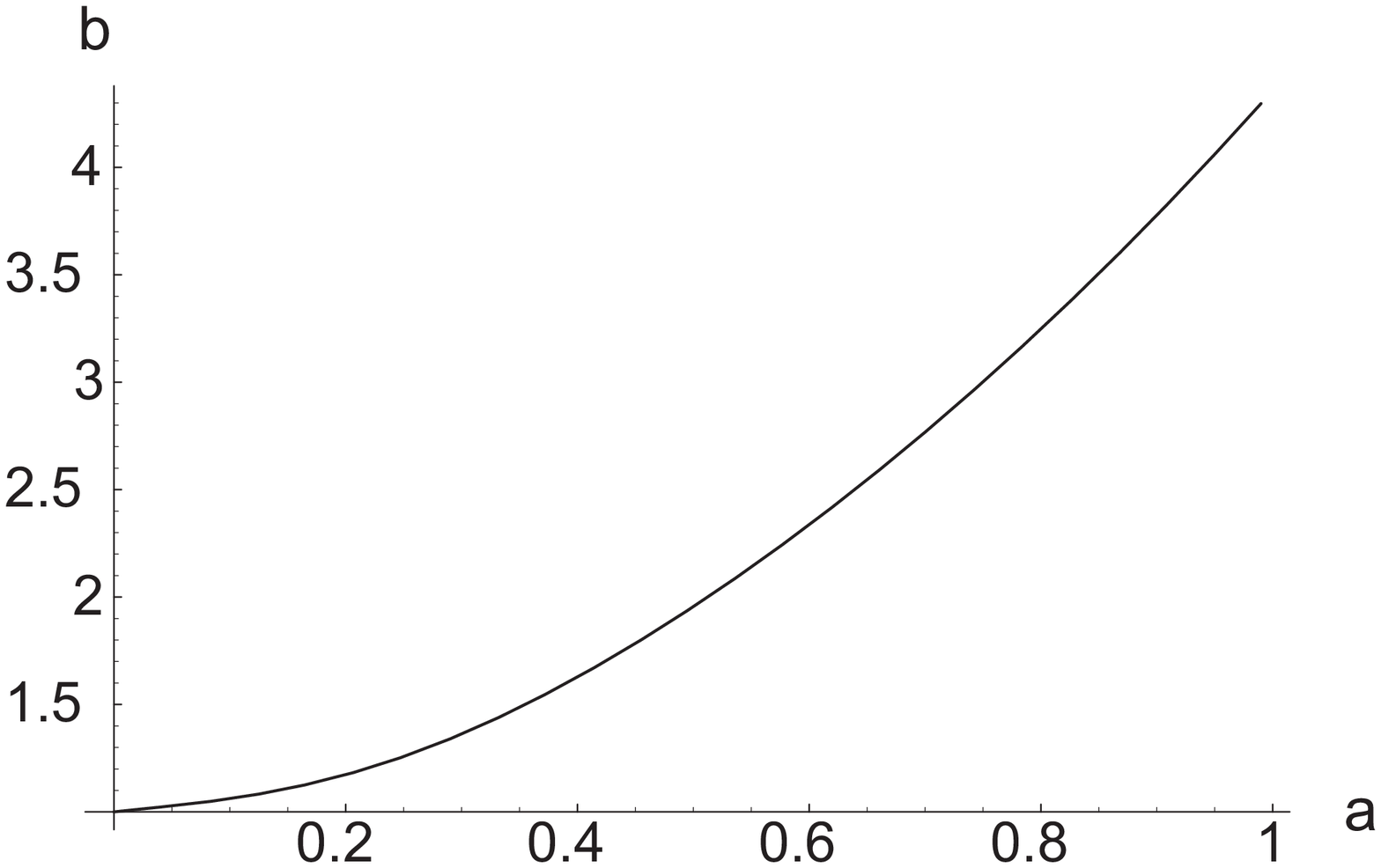}\\

}
\caption{}\label{fig:brel1brel2}
\end{figure}

In this appendix we prove~\pref{eq:F1neg}. Since $0 < \blength < 1$ we have $3 (-1 + \blength^4) - 4 (-1 + \brel) \log^3 \blength < 0$ and thus $G_- < 0 \Longleftrightarrow h_- > 0$. Because $\frac43 \log^6 \blength$, the coefficient in front of $\brel^2$ in $h_-$, is positive, we know that $h_-$ is positive for $\brel \in [0, \brel_1(\blength)) \cup (\brel_2(\blength), 1]$, where $\brel_{1,2}$ are the $\blength$-dependent zeros of $h_-$, with $\brel_1 \leq \brel_2$. These zeroes are given by
\begin{align*}
\brel_{1,2}(\blength) &= (8 \log^3 \blength)^{-1} \Bigl(9 - 12 \blength^2 +  3 \blength^4 + (4 \log \blength) (3 + \log^2 \blength)\Bigr.\\
    &\hspace{1.9cm}\left.\pm \bigl\{ 225 - 504 \blength^2 + 342 \blength^4 - 72 \blength^6 + 9 \blength^8 + (360 - 288 \blength^2 - 72 \blength^4) \log \blength \right.\\
    &\hspace{2.3cm}\Bigl. + 144 \log^2 \blength + (-120 + 96 \blength^2 + 24 \blength^4) \log^3 \blength - 96 \log^4 \blength + 16 \log^6 \blength \bigr\}^{\frac12} \Bigr).
\end{align*}
We take the plus sign in $\brel_1$ and the minus sign in $\brel_2$. In this way the negativity of $(8 \log^3 \blength)^{-1}$ ensures that $\brel_1 \leq \brel_2$. Plots of $\brel_1$ and $\brel_2$ are given in Figure~\ref{fig:brel1brel2}.

We start by proving that $9 -12 \blength^2 + 3 \blength^4 + (4 \log \blength) (3 + \log^2 \blength) < 0$ on $(0, 1)$. The equation
\[
\blength \frac{d}{d\blength} \left(2 \blength^4 - 2 \blength^2 + \log\blength\right) = 8 \blength^4 - 4 \blength^2 + 1 = 0
\]
has no real solutions on $(0, 1)$ and so $\frac12 \blength \frac{d}{d\blength} \left(\blength^4 - 2 \blength^2 + \log^2\blength + 1\right) = 2 \blength^4 - 2 \blength^2 + \log\blength \leq 0$ on $(0, 1]$, with equality iff $\blength=1$. This in turn shows that $12 \blength \frac{d}{d\blength}\left(9 -12 \blength^2 + 3 \blength^4 + (4 \log \blength) (3 + \log^2 \blength)\right) = \blength^4 - 2 \blength^2 + \log^2\blength + 1 \geq 0$ on $(0, 1]$, with equality iff $\blength=1$, from which
$9 -12 \blength^2 + 3 \blength^4 + (4 \log \blength) (3 + \log^2 \blength) < 0$ follows. Consequently, since $(8 \log^3 \blength)^{-1} < 0$, we have that $\brel_2 > 0$.

Next we calculate
\[
\frac{d}{d \blength} \left(3 \log \blength - 3 + \frac6{\blength^2 + 1} - \log^3 \blength\right) = \frac3{\blength} - \frac{12 \blength}{(\blength^2 + 1)^2} - \frac3{\blength} \log^2 \blength.
\]
This is equal to zero if and only if $\log^2 \blength = \frac{(1 - \blength^2)^2}{(1 + \blength^2)^2}$, which leads to $\blength = e^{-\frac{1 - \blength^2}{1 + \blength^2}}$.
We will now prove
\[
\blength\in[0,1] \wedge \blength = e^{\frac{-1+\blength^2}{1+\blength^2}} \Longleftrightarrow \blength = 1.
\]
Since $\blength=1$ clearly satisfies the equation on the left, it remains to show that there are not more solutions.
We start by computing
\begin{align*}
\frac{d}{d\blength} \left( e^{\frac{-1+\blength^2}{1+\blength^2}} - \blength \right) &= \frac{4\blength}{\left(1+\blength^2\right)^2} e^{\frac{-1+\blength^2}{1+\blength^2}} - 1;\\
\frac{d^2}{d\blength^2 } \left( e^{\frac{-1+\blength^2}{1+\blength^2}} - \blength \right) &= \frac1{\left(1+\blength^2\right)^4} \left(-12 \blength^4 + 8 \blength^2+4 \right) e^{\frac{-1+\blength^2}{1+\blength^2}}.
\end{align*}
On $[0, 1]$ we have
\[
\frac{d^2}{d\blength^2 } \left( e^{\frac{-1+\blength^2}{1+\blength^2}} - \blength \right) = 0 \Longleftrightarrow -12 \blength^4 + 8 \blength^2+4 = 0 \Longleftrightarrow \blength = 1,
\]
showing that $\frac{d}{d\blength} \left( e^{\frac{-1+\blength^2}{1+\blength^2}} - \blength \right)$ has at most one zero on $[0, 1]$ and thus its only zero is at $\blength=1$, which in turn shows that also $e^{\frac{-1+\blength^2}{1+\blength^2}} - \blength$ has at most one zero on $[0, 1]$, which is what we set out to prove. This now leads us to conclude
\[
\frac{d}{d \blength} \left(3 \log \blength - 3 + \frac6{\blength^2 + 1} - \log^3 \blength\right) = 0 \Longleftrightarrow \blength = 1.
\]
This means that $3 \log \blength - 3 + \frac6{\blength^2 + 1} - \log^3 \blength$ has a minimum at $\blength = 1$ and thus this expression is positive on $(0, 1)$. Then
\begin{align}
&\Bigl\{\left(9 - 12 \blength^2 +  3 \blength^4 + (4 \log \blength) (3 + \log^2 \blength)\right) - 8 \log^3 \blength\Bigr\}^2\nonumber\\
&- \Bigl\{ 225 - 504 \blength^2 + 342 \blength^4 - 72 \blength^6 + 9 \blength^8 + (360 - 288 \blength^2 - 72 \blength^4) \log \blength \nonumber\\
&\hspace{0.5cm} + 144 \log^2 \blength + (-120 + 96 \blength^2 + 24 \blength^4) \log^3 \blength - 96 \log^4 \blength + 16 \log^6 \blength \Bigr\}\nonumber\\
&= -144 (\blength^2 - 1)^2 + 144 (\blength^4 - 1) \log \blength - 48 (\blength^4 - 1) \log^3 \blength\nonumber\\
&= 48 (\blength^4 - 1) \left(3 \log \blength - 3 \left( 1 - \frac2{\blength^2 + 1}\right) - \log^3 \blength \right)\nonumber\\
&< 0.\label{eq:lotsofinfoinhere}
\end{align}
Note that this also proves that the expression in the square root in $\brel_{1,2}$ is positive. Together with $8 \log^3 \blength < 0$ these inequalities give us $\brel_2(\blength) > 1$.
These results lead to the conclusion that
\[
G_-(\blength, \brel) < 0 \Longleftrightarrow \brel \in [0, \brel_1(\blength)).
\]
The other sign possibilities for $G_-$ follow immediately.

\begin{remark}\label{rem:limitsbrel}
For the excluded endpoints $0$ and $1$ we find
\begin{align*}
\underset{\blength \downarrow 0}{\lim}\,\brel_1 &= 0, \qquad \underset{\blength \uparrow 1}{\lim}\,\brel_1 = \frac52 - \frac12 \sqrt{\frac{69}5},\\
\underset{\blength \downarrow 0}{\lim}\,\brel_2 &= 1, \qquad \underset{\blength \uparrow 1}{\lim}\,\brel_2 = \frac52 + \frac12 \sqrt{\frac{69}5}.
\end{align*}
The limits for $\blength \uparrow 1$ were found by calculating the first terms in the Taylor expansion of $\brel_{1,2}$.
\end{remark}

\bibliographystyle{acm}
\bibliography{bibliography}

\end{document}